\documentclass[fleqn, 10pt, 3p]{elsarticle}

\usepackage[overload]{empheq}

\usepackage{amssymb}

\usepackage{mathrsfs}

\usepackage{amsthm}

\usepackage{siunitx}

\usepackage{graphicx}

\usepackage{overpic}

\usepackage{grffile}

\usepackage[dvipsnames]{xcolor}

\usepackage{mdframed}

\usepackage{paracol}
\columnratio{0.5}

\usepackage[linesnumbered,ruled]{algorithm2e}

\usepackage[title,titletoc]{appendix}

\usepackage[font=small,labelfont=bf,labelsep=period,width=.9\textwidth]{caption}

\usepackage{subcaption}

\usepackage{hyperref}
\hypersetup{
  colorlinks=true,
  linkcolor=blue,
  filecolor=blue,
  citecolor=black,
  urlcolor=cyan,
}

\usepackage{cleveref}

\usepackage{array}
\usepackage{booktabs}

\usepackage{enumitem}

\usepackage{stringstrings}
\usepackage{xstring}

\makeatletter
\def\ps@pprintTitle{%
  \let\@oddhead\@empty
  \let\@evenhead\@empty
  \def\@oddfoot{\reset@font\hfil\thepage\hfil}
  \let\@evenfoot\@oddfoot
}
\makeatother

\usepackage{pgfplots}

\usepgfplotslibrary{colorbrewer}
\usepgfplotslibrary{colormaps}
\usepgfplotslibrary{fillbetween}
\usepgfplotslibrary{groupplots}
\usetikzlibrary{patterns}

\pgfplotsset{select coords between index/.style 2 args={
      x filter/.code={
          \ifnum\coordindex<#1\fi
          \ifnum\coordindex>#2\fi
        }
    }}

\usepackage{tikz}

\usetikzlibrary{external}
\newcommand{\tikzsetnextfilenamesafe}[1]{
  \StrSubstitute{#1}{/}{-}[\temp]
  \tikzsetnextfilename{\temp}
}

\newcommand{\bfu}{\boldsymbol{u}}

\newcommand{\bfw}{\boldsymbol{w}}
\newcommand{\bfx}{\boldsymbol{x}}

\newcommand{\bfB}{\boldsymbol{B}}
\newcommand{\bfC}{\boldsymbol{C}}

\newcommand{\bfF}{\boldsymbol{F}}

\newcommand{\bfI}{\boldsymbol{I}}

\newcommand{\bfN}{\boldsymbol{N}}

\newcommand{\bfP}{\boldsymbol{P}}

\newcommand{\bfT}{\boldsymbol{T}}

\newcommand{\bfX}{\boldsymbol{X}}

\DeclareMathOperator{\tr}{tr}
\DeclareMathOperator{\dev}{dev}
\DeclareMathOperator*{\argmin}{arg\,min}

\newcommand{\diff}[1]{\text{ d}#1}

\newcommand{\grad}{\bs{\nabla}}
\newcommand{\divergence}{{\grad \cdot}}

\newcommand{\norm}[1]{\left\lVert#1\right\rVert}
\newcommand{\macaulay}[1]{\left<#1\right>}
\newcommand{\bs}[1]{\boldsymbol{#1}}
\newcommand{\abs}[1]{\left\vert{#1}\right\vert}
\newcommand{\overbar}[1]{\mkern 1.5mu\overline{\mkern-1.5mu#1\mkern-1.5mu}\mkern 1.5mu}

\newcommand{\strain}{\boldsymbol{\varepsilon}}

\newcommand{\crackset}{\Gamma}
\newcommand{\Gc}{{\mathcal{G}_c}}
\newcommand{\activeenergy}{{\left< A \right>}}
\newcommand{\inactiveenergy}{{\left< I \right>}}

\newcommand{\plasticstrain}{{\overbar{\varepsilon}^p}}

\newcommand{\set}[1]{#1}
\newcommand{\fspace}[1]{\mathcal{#1}}
\newcommand{\body}{\set{\Omega}}
\newcommand{\bodyboundary}{{\partial \body}}

\theoremstyle{definition}
\newtheorem*{mdalternative}{Alternative}
\definecolor{alternativegray}{gray}{0.98}
\newenvironment{alternative}%
{\begin{mdframed}[backgroundcolor=alternativegray,nobreak=true]\begin{mdalternative}}
      {\end{mdalternative}\end{mdframed}}

\theoremstyle{definition}
\newtheorem*{mdexample}{Example}
\definecolor{examplegray}{gray}{0.98}
\newenvironment{example}%
{\begin{mdframed}[backgroundcolor=examplegray,nobreak=true]\begin{mdexample}}
      {\end{mdexample}\end{mdframed}}

\theoremstyle{definition}
\newtheorem*{mdremark}{Remark}
\definecolor{remarkgray}{gray}{0.95}
\newenvironment{remark}%
{\begin{mdframed}[backgroundcolor=remarkgray,nobreak=true]\begin{mdremark}}
      {\end{mdremark}\end{mdframed}}

\begin{document}

\begin{frontmatter}
  \title{A Variational Phase-Field Model For Ductile Fracture \\ with Coalescence Dissipation }

  \author[1]{Tianchen~Hu}
  \author[3]{Brandon~Talamini}
  \author[3]{Andrew~J.~Stershic}
  \author[4]{Michael~R.~Tupek}
  \author[1,2]{John~E.~Dolbow}
  \ead{jdolbow@duke.edu}

  \address[1]{Department of Mechanical Engineering and Materials Science, Duke University, Durham, NC 27708, USA}
  \address[2]{Department of Civil and Environmental Engineering, Duke University, Durham, NC 27708, USA}
  \address[3]{Sandia National Laboratories, Livermore, CA 94551, USA}
  \address[4]{Sandia National Laboratories, Albuquerque, NM 87185, USA}

  \begin{abstract}

    A novel phase-field for ductile fracture model is presented.
    The model is developed within a consistent variational framework in the context of finite-deformation kinematics.
    A novel coalescence dissipation introduces a new coupling mechanism between plasticity and fracture by degrading the fracture toughness as the equivalent plastic strain increases.
    The proposed model is compared with a recent alternative where plasticity and fracture are strongly coupled.
    Several representative numerical examples motivate specific modeling choices.
    In particular, a linear crack geometric function provides an ``unperturbed'' ductile response prior to crack initiation, and Lorentz-type degradation functions ensure that the critical fracture strength remains independent of the phase-field regularization length.
    In addition, the response of the model is demonstrated to converge with a vanishing phase-field regularization length.
    The model is then applied to calibrate and simulate a three-point bending experiment of an aluminum specimen with a complex geometry.
    The effect of the proposed coalescence dissipation coupling on simulations of the experiment is first investigated in a two-dimensional plane strain setting. The calibrated model is then applied to a three-dimensional calculation, where the calculated load-deflection curves and the crack trajectory show  excellent agreement with experimental observations.  Finally, the model is applied to simulate crack nucleation and growth in a specimen from a recent Sandia Fracture Challenge.

  \end{abstract}

\end{frontmatter}

\section{Introduction}
\label{s: intro}

The phase-field method for fracture has demonstrated advantages in unifying crack initiation, propagation, merging and branching, over a wide range of applications. Recent efforts have extended the thermodynamic framework to couple phase-field for fracture with plasticity, and demonstrated how different modeling choices can lead to versatile model behavior. In this work, we cast a family of phase-field models for large-deformation ductile fracture as a minimization problem and demonstrate how several key constitutive and modeling choices can be made to yield favorable model behavior. In particular, the proposed model manifests plastic hardening, critical fracture strength, and softening response that are largely insensitive to the choice of regularization length.
Importantly, a novel coalescence dissipation term is proposed in a variational setting to characterize the decrease of fracture toughness as plastic flow occurs, which serves to effectively delay crack initiation in a computationally efficient manner.

The variational approach to fracture was originally introduced by \citet{Francfort98} and \citet{Bourdin2000}. Since its introduction, the accompanying phase-field approach to fracture has proven to be successful for modeling brittle fracture \cite{karma_2001, karma_2004, amor_2009, AMOR20091209, miehe_2010_p1, miehe_2010_p2}. The approach introduces a phase-field and accompanying length scale to regularize the sharp discontinuities associated with fracture surfaces.  Gamma convergence of the phase-field for fracture model to a Griffith model of fracture has been established in certain cases \cite{may2015numerical,negri2020gamma}. In practice, many researchers employ phase-field for fracture models as effectively cohesive models of fracture, setting the regularization length according to the critical fracture strength of the material  (see e.g.\  \citet{borden2012isogeometric} for a detailed analysis).
More recently, phase-field models of fracture that asymptote towards cohesive models of fracture (with finite strength as the regularization length decreases) have been proposed by \citet{lorentz2011convergence} and \citet{lorentz2017nonlocal}.    This formulation leads to  phase-field for fracture models that exhibit a reduced sensitivity of the effective fracture response to the particular choice of regularization length.  For additional developments and discussions of the advantages of the cohesive approach, see the recent contributions by \citet{wu2017unified}, \citet{geelen2019phase}, \citet{HuGary2020}, and \citet{brandon2020cohesive}.

The thermodynamic framework employed by the phase-field approach provides a convenient extension to incorporate plasticity by incorporating a plastic potential. However, how best to design the form of that plastic potential and its coupling with a phase-field that regularizes a fracture surface remains an open question.
There exist many approaches to modeling plasticity in the context of phase-field fracture \cite{alessi_gradient_2014, alessi_gradient_2015, alessi_coupling_2018, ambati_phase-field_2015, ambati_phase-field_2016, miehe_phase_2016, borden_phase-field_2016, borden_phase-field_2017}.  For a recent review of the various approaches that have been proposed, see  \citet{alessi_comparison_2017}. In this work, we extend the work by \citet{Francfort98} which formulates brittle fracture as a minimization problem.  In particular, we propose a framework to formulate the phase-field model for ductile fracture as a minimization problem much in the spirit of \citet{simo1988framework} and \citet{ortiz_variational_1999}. The proposed framework is flexible in the sense that a wide range of plastic flow rules and hardening behavior can be cast in a variational setting, as illustrated in \cite{ortiz_variational_1999}.  The consistent variational approach does impart certain restrictions, such as the degradation of the yield surface with the phase-field implying a corresponding plastic contribution to crack growth.  While other models that break the variational structure may be of utility to effectively capture certain responses, they are beyond the scope of the current work.

By adopting such a variational framework to model ductile fracture, the remaining major modeling choices are the form of the local term in the fracture energy (the crack geometric function)\footnote{Also sometimes referred to as the \textit{local fracture dissipation function} in the phase-field for fracture literature, despite its energetic nature in the thermodynamic framework.}, the plastic degradation function (the functional dependence of the plastic potential on the phase-field variable), and the power expenditure. \citet{pham2013onset} used a crack geometric function that is linear in the phase-field variable, resulting in a purely elastic response prior to crack initiation.  This fracture energy functional was subsequently referred to as the \texttt{AT-1} functional by Bourdin and co-workers (meant to evoke \textit{Ambrosio} and \textit{Tortorelli}, who initially introduced this and related functionals in the context of image segmentation \cite{ambrosio1990approximation}). \citet{wu2017unified} generalized the \textit{Ambrosio-Tortorelli} functional \cite{ambrosio1990approximation} to a family of second-order polynomials with various phase-field profiles, with specific choices of recovering the widely adopted \texttt{AT-1} and \texttt{AT-2} models. In this work, we extend the results to the regime of ductile fracture and demonstrate that an unperturbed elastic-plastic response can be obtained in certain cases, i.e.\ the yield stress and the plastic hardening law remain unmodified prior to crack initiation even when the plastic potential is degraded as a function of the phase-field variable.

In the case of brittle and quasi-brittle fracture, several studies have demonstrated that a regularization-length-independent critical fracture strength and softening response can be obtained by a family of rational degradation functions in combination with the appropriate choice of the crack geometric function \cite{lorentz2011convergence, wu2017unified, geelen2019phase}. These are referred to as Lorentz-type degradation functions hereinafter.  In another work \cite{brandon2020cohesive}, we showed that this decoupling of critical fracture strength and regularization length is crucial in elastic-plastic materials to ensure the crack growth resistance is convergent with respect to the regularization. This independence or lack of sensitivity is important both in the study of the $\Gamma$-convergence of the model behavior and in engineering applications where the specimen used for calibration and the actual structure of interest span vastly different spatial scales. In this manuscript, we extend the use of these degradation functions to the coupling of plastic potential so that the model response converges as the phase-field regularization length diminishes.

Although the phase-field regularization length can be divorced from the set of material properties by choosing the Lorentz-type degradation functions, an upper bound still exists, stemming from either polyconvexity requirements or physical constraints (see e.g. \cite{wu2017unified, geelen2019phase} for detailed analysis). With regard to phase-field models for ductile fracture, various choices of the coupling between damage and plasticity can result in the need for surprisingly small regularization lengths.  In particular, some models require the need for smaller regularization lengths as the transition from plastic localization to the onset of damage is delayed.
There have been several attempts to address this issue. \citet{ambati_phase-field_2015} and \citet{ambati_phase-field_2016} scaled the plastic degradation function by a critical plastic strain to be calibrated. \citet{miehe2015phase} introduced barrier functions related to critical values of the state variables. Most recently, \citet{borden_phase-field_2016} devised a threshold for the plastic work contributions to crack initiation. Although these are all valid, thermodynamically consistent approaches to introduce an arbitrary delay to crack initiation, the hybrid nature of these formulations makes it difficult to fit them into a variational framework.

Coupling the phase field to the plasticity tends to promote cracks in regions of extensive plasticity, which is often the case observed experimentally, but the observed damage process resulting from the model can be far more gradual than expected. On the other hand, when the phase field is coupled only to the elastic strain energy, the damage process is more abrupt, but the predicted crack growth directions tend not to agree with observations in ductile alloys.  In particular, they tend to propagate orthogonally to the maximum tensile stress as in brittle materials, rather than in response to intense plastic straining. Here, we have identified a plausible micro-mechanical argument to reconcile these problems, and used this to formulate a model that fractures in regions with large plastic deformation even without the strong coupling between plasticity and fracture. Motivated by findings from several recent attempts to homogenize the fracture toughness, in this work we decouple the plastic disspation from fracture degradation and propose a novel coalescence dissipation to model the coupling between plasticity and fracture. \citet{rodriguez2016silica}, \citet{chowdhury2019effects}, and \citet{vo2020molecular} found that ligament length, shape, and orientation of defects at the micro-scale influence the Mode-I fracture toughness. We postulate that the presence of plastic flow (or dislocations at the micro-scale) alters those properties related to defects, and at the continuum level, some configurational energy has been dissipated prior to crack initiation, effectively reducing the fracture toughness. We note that the idea of degrading the fracture toughness is not necessarily new, as a similar approach was recently proposed by \citet{yin2020ductile}. Importantly, by virtue of considering the coalescence dissipation in a variational framework, we show that fracture occurs in regions with large plastic deformation even without the strong coupling between plasticity and fracture.  An upshot is a significant reduction in the fracture driving energy and a corresponding relaxation of the upper bound for the regularization length.

This paper is organized as follows: \Cref{s: theory} outlines the variational framework. \Cref{s: theory/potential} describes a specific form of the functional to be minimized. \Cref{s: theory/minimizer} describes the minimization problem and the resulting governing equations and constitutive constraints. \Cref{s: theory/discretization} presents the initial boundary value problem, the strong form, the weak form and the discrete Galerkin form.
\Cref{s: example/homogenized} motivates the choices of the crack geometric function and the degradation function by examining a homogeneous problem in detail. \Cref{s: example/nonhomogeneous} demonstrates the regularization-length-independent softening behavior as well as the effect of the coalescence dissipation term. \Cref{s: example/3pb} applies the proposed model to simulate a three-point bending problem, and \Cref{s: example/SFC} presents simulation results for a recent Sandia Fracture Challenge problem. Finally, \Cref{s: conclusion} provides a summary and some concluding remarks.

\section{Theory}
\label{s: theory}

Let $\bs{\phi}(\bfX, t)$ be the deformation map at time $t$ mapping a material point $\bfX \in \body$ to a point in the current configuration $\bfx \in \bs{\phi}(\body)$. The deformation tensor is defined as $\bfF = \partial \bfx/\partial \bfX$, and $J = \det(\bfF)$ is the associated Jacobian determinant.

The state of the system is characterized by the set of variables $\mathcal{S} = \{ \bfF, \bfF^p, \plasticstrain, d, \grad d \}$, where $\plasticstrain$ is the effective plastic strain, and $d$ is the damage with its nonlocal counterpart $\grad d$.  Plasticity is modeled through the framework of multiplicative decomposition, i.e.\ $\bfF = \bfF^e \bfF^p$.

Consider the total energy $\Psi(\mathcal{S})$ for the system which can be decomposed as
\begin{align}
  \Psi(\mathcal{S}) = W^e(\bfF, \bfF^p, d) + W^p(\plasticstrain, d) + F(d, \grad d), \label{eq: total potential}
\end{align}
where  $W^e(\bs{\phi}, \bfF^p, d)$ is the strain energy, $W^p(\plasticstrain, d)$ is the plastic hardening potential, and $F(d, \grad d)$ is the fracture energy. We write $\psi$, $w^e$, $w^p$, and $f$ for their accompanying volumetric densities, respectively. The corresponding dissipation potentials are denoted as
\begin{align}
  \Pi(\mathcal{V}, \mathcal{S}) = V(\dot{\bs{\phi}}, \mathcal{S}) + \Theta(\dot{\plasticstrain}, \mathcal{S}) + \Lambda(\dot{d}, \mathcal{S}), \label{eq: total dual kinetic potential}
\end{align}
where $\mathcal{V} = \{ \dot{\bs{\phi}}, \dot{\bfF}^p, \dot{\plasticstrain}, \dot{d} \}$ is the set of generalized velocities, $V$ is the viscoelastic dissipation, $\Theta$ is the viscoplastic dissipation, and $\Lambda$ is the fracture dissipation. Again, $\pi$, $v$, $\theta$, and $\lambda$ denote their volumetric densities.

The initial boundary value problem describing an elastic-plastic-fracture coupled system is obtained by minimizing the sum of the rate of the total energy and its dissipative counterpart, i.e.
\begin{align*}
  \mathcal{V} =\  & \argmin_{\mathcal{V}} \dot{\Psi}(\mathcal{V}) + \Pi(\mathcal{V}, \mathcal{S}) - P(\mathcal{V}),
\end{align*}

\noindent subject to appropriate constraints. In the above, $P(\mathcal{V})$ is the external power expenditure. Mass balance and angular momentum balance are satisfied by construction, i.e.
\begin{align}
  J \rho = \rho_0, \quad \forall \bfX \in \body, \\
  \bfP \bfF^T = \bfF \bfP^T, \quad \forall \bfX \in \body,
\end{align}
where $\rho_0$ and $\rho$ denote the density in the reference and current configuration, respectively, and $\bfP$ is the first Piola-Kirchhoff stress.
The balance of linear momentum is satisfied at the minimizer of the objective function. Other constitutive constraints including the plastic flow rule, the plastic yield surface (together with the hardening law), and the fracture evolution law are obtained by minimizing the objective. In this manner, the  Clausius-Duhem inequality is satisfied according to the principle of maximum dissipation.

\subsection{Postulation of the objective function}
\label{s: theory/potential}

In this work, with reference to \eqref{eq: total potential}, $W^e(\bfC, \bfF^p, d)$ is constructed to represent a compressible Neo-Hookean hyperelastic material with tension-compression asymmetry coupled with fracture, $W^p(\plasticstrain, d)$ models J$_2$ plasticity (and optionally consistent viscoplasticity) with a linear hardening law, and $F(d, \grad d)$ models cohesive fracture with a finite critical fracture strength. The elastic-fracture coupling is modeled using an elastic degradation $g^e(d)$, and the plastic-fracture coupling is effected through a plastic degradation function $g^p(d)$. The specific form of each potential, the degradation functions, the power expenditure, and the resulting initial boundary value problem are given below. We note that this approach is sufficiently general and can accommodate many other choices of hyperelastic materials, plastic models, and fracture models.

Following \cite{borden2016phase,ambati2016phase}, the strain energy is defined as
\begin{subequations}\label{eq: strain energy potential}
  \begin{align}
    W^e(\bfC, \bfC^p, d)            & = \int\limits_\body w^e(\bfC, \bfC^p, d) \diff{V} = \int\limits_\body g^e(d) w^e_\activeenergy(\bfC, \bfC^p) + w^e_\inactiveenergy(\bfC) \diff{V},         \\
    w^e_\activeenergy(\bfC, \bfC^p) & = \mathbb{H}_1(J)\left\{ \dfrac{1}{2}K\left( \dfrac{1}{2}(J^2-1) - \ln{J} \right) \right\} + \dfrac{1}{2}G\left( \overbar{\bfC}:{\bfC^p}^{-1} - 3 \right), \\
    w^e_\inactiveenergy(\bfC)       & = \left( 1-\mathbb{H}_1(J) \right) \left\{ \dfrac{1}{2}K\left[ \dfrac{1}{2}(J^2-1) - \ln{J} \right] \right\},
  \end{align}
\end{subequations}
where $K$ is the bulk modulus, $G$ is the shear modulus, $\mathbb{H}_a(x)$ is the Heaviside function with the jump located at $x = a$, $\bfC = \bfF^T \bfF$ is the right Cauchy-Green strain tensor, $\overbar{\bfC} = J^{-2/3}\bfC$ is the volume-preserving part of the right Cauchy-Green strain tensor, $\bfC^p = {\bfF^p}^T \bfF^p$ is the plastic counterpart of the right Cauchy-Green strain, and $g^e(d)$ is the elastic degradation function (to be defined). We note that the kinematic variable $\bfF$ enters the potential through $\bfC$ and $\bfC^p$ only, according to the principle of frame indifference.
The strain energy is split into an active part (with the subscript $\activeenergy$) and an inactive part (with the subscript $\inactiveenergy$) such that fracture is associated with only volumetric expansion and deviatoric deformation. In this work, the viscoelastic dissipation $V = 0$, but in principle most forms of damping and Newtonian viscosity can be modeled through this term.

\begin{alternative}[Hencky hyperelastic model]
  \vspace{-0.5em}
  Another suitable hyperelastic material based on spectral decomposition is given in \cite{brandon2020cohesive} as
  \begin{align*}                     \\
    w^e_\activeenergy(\bfC^e)   & = \dfrac{1}{2} K \macaulay{\tr(\strain^e)}_+^2 + G \dev{\strain^e} : \dev{\strain^e}, \\
    w^e_\inactiveenergy(\bfC^e) & = \dfrac{1}{2} K \macaulay{\tr(\strain^e)}_-^2,
  \end{align*}
  where $\strain^e = \frac{1}{2} \ln(\bfC^e)$ is the logarithmic elastic strain based on the elastic right Cauchy-Green strain, and $\macaulay{\cdot}_\pm$ is the signed version of the standard Macaulay bracket.
\end{alternative}

The plastic hardening potential is defined as
\begin{subequations}\label{eq: plastic hardening potential}
  \begin{align}
    W^p(\plasticstrain, d) & = \int\limits_\body w^p(\plasticstrain, d) \diff{V},                               \\
    w^p(\plasticstrain, d) & = g^p(d) \left( \sigma_y \plasticstrain + \dfrac{1}{2} H \plasticstrain^2 \right),
  \end{align}
\end{subequations}
where the yield stress $\sigma_y$ and the hardening modulus $H$ are parameters of the hardening law. In this work, the viscoplastic dissipation ${\Theta^p} = 0$, resulting in a rate-independent yield surface.

\begin{alternative}[power-law hardening and power-law rate-sensitivity]
  \vspace{-0.5em}
  As shown in \cite{ortiz_variational_1999, brandon2020cohesive}, a widely used power-law hardening model can be cast within the current framework with
  \begin{align*}                     \\
    w^p(\plasticstrain, d) = g^p(d) \dfrac{n}{n+1} \sigma_y \epsilon_0 \left[ \left( 1 + \dfrac{\plasticstrain}{\epsilon_0} \right)^{(n+1)/n} - 1 \right],
  \end{align*}
  and a power-law rate-sensitivity can be modeled using a viscoplastic dissipation function of the form:
  \begin{align*}
    \Theta(\dot{\plasticstrain}, d) = \int\limits_\body \theta(\dot{\plasticstrain}, d) \diff{V}, \quad \theta(\dot{\plasticstrain}, d) = g^p(d) \dfrac{m}{m+1} \sigma_y \dot{\epsilon}_0 \left( \dfrac{\dot{\plasticstrain}}{\dot{\epsilon_0}} \right)^{(m+1)/m}.
  \end{align*}
  Note that including viscoplastic dissipation results in a consistent viscoplasticity model.
\end{alternative}

The fracture potential is the surface energy released by the formation of an irreversible crack surface. The phase-field fracture potential regularizes the fracture potential using a standard Allen-Cahn approximation:
\begin{subequations}\label{eq: fracture potential}
  \begin{align}
    F(d, \grad d; l, C)      & = \int\limits_\body f(d, \grad d; l, C) \diff{V} = \int\limits_\body \Gc \gamma(d, \grad d; l, C) \diff{V} ,                                                \\
    \gamma(d, \grad d; l, C) & = \dfrac{1}{c_0l} \left( C \alpha(d) + l^2 \grad d \cdot \grad d \right), \quad c_0 = 4 \int\limits_0^1 \sqrt{\alpha(d)} \diff{d}, \quad 0 < C \leqslant 1,
  \end{align}
\end{subequations}
where $\Gc$ is the fracture toughness\footnote{It can also be interpreted as the critical energy release rate depending on the context.}, $l$ is the phase-field regularization length, and $C$ represents the fraction of the fracture process that is energetic in the Griffith's sense. The dissipative portion of the fracture process will be defined by the fracture dissipation momentarily.
The normalization constant $c_0$ is chosen such that $\lim_{l \to 0} \int_\body \Gc \gamma(d, \grad d; l, C = 1) \diff{V} = \int_{\crackset} \Gc \diff{A}$ where $\crackset$ is the sharp crack surface represented by the phase-field. In this work, we consider the family of crack geometric functions\footnote{We refer to $\alpha(d)$ as the crack geometric function to keep it consistent with previous works. This should not be confused with the energetic processes considered in this work.} $\alpha(d)$ that take the quadratic form:
\begin{align}
  \alpha(d; \xi) = (1 - \xi) d^2 + \xi d, \quad \xi \geqslant 0
\end{align}

\begin{example}[various crack geometric functions]
  \vspace{-0.5em}
  The parameter $\xi$ in the crack geometric function determines the form of the critical fracture energy, the support of the phase-field, and the upper bound for the phase-field regularization length.
  \begin{itemize}[nosep]
    \item $\xi = 1$ recovers the so-called \texttt{AT-1} model, with
          \begin{align*}
            c_0 = \dfrac{8}{3}, \quad D_u = 2l, \quad d_u(x) = \left(1-\dfrac{\abs{x}}{2l}\right)^2.
          \end{align*}
    \item $\xi = 2$ recovers the model advocated by \citet{wu2017unified}, with
          \begin{align*}
            c_0 = \pi, \quad D_u = \dfrac{\pi}{2}l, \quad d_u(x) = 1-\sin\left(\dfrac{\abs{x}}{l}\right).
          \end{align*}
    \item $\xi = 0$ recovers the so-called \texttt{AT-2} model, with
          \begin{align*}
            c_0 = 2, \quad D_u = \infty, \quad d_u(x) = \exp\left(-\dfrac{\abs{x}}{l}\right).
          \end{align*}
  \end{itemize}
\end{example}

We will demonstrate through numerical examples (\Cref{s: example/homogenized}) that a strictly positive $\xi$ is necessary to obtain an unperturbed elastic-plastic response. For practical purposes, $\xi = 1$ and $\xi = 2$ (and their resulting models) give qualitatively similar response for all numerical examples considered in this manuscript. For the sake of simplicity, only results for the case of $\xi = 1$ are shown in \Cref{s: example/nonhomogeneous,s: example/3pb}, and similar results can be obtained using $\xi = 2$ with slightly different parameters.

A variety of damage softening response (or equivalently cohesive traction-separation laws) may be achieved by different choices for the degradation functions $g^e(d)$ and $g^p(d)$.
A family of degradation functions originally proposed by Lorentz et  al.~\cite{lorentz2011convergence} ensures a regularization-length independent critical fracture strength in phase-field models with $\xi > 0$. In this work, we adopt the two-parameter Lorentz-type degradation function
\begin{align}\label{eq: lorentz degradation}
  g^\text{Lorentz}(d; m, p) = \dfrac{(1-d)^2}{(1-d)^2 + md (1+pd)}
\end{align}
as the elastic and (optionally) the plastic degradation functions. A regularization-length independent critical fracture strength can be obtained with $m = \dfrac{\xi \Gc}{c_0 l \psi_c}$, where $\psi_c$ is the critical fracture energy regarded as a material property. The parameter $p \geqslant 1$ can be used to calibrate the softening response.

\begin{alternative}[other Lorentz-type degradation functions]
  \vspace{-0.5em}
  A one-parameter degradation function is given as
  \begin{align*}
    g(d; m) = \dfrac{(1-d)^2}{(1+md)^2}, \quad m = \dfrac{\xi \Gc}{2 c_0 l \psi_c} - 1,
  \end{align*}
  which is useful when the structural response at the macro-scale is not strongly affected by the damage softening response. \citet{wu2017unified} also derived a family of degradation functions that give equivalent responses of commonly used traction-separation laws for quasi-brittle fracture. An extension of such analysis to ductile fracture has yet to be performed.
\end{alternative}

The fracture dissipation is defined as
\begin{subequations}\label{eq: dual fracture potential}
  \begin{align}
    \Lambda(\dot{d}, \plasticstrain; \beta, \varepsilon_0) & = \int\limits_\body \lambda(\dot{d}, \plasticstrain; \beta, \varepsilon_0) \diff{V},                                                                                                \\
    \lambda(\dot{d}, \plasticstrain; \beta, \varepsilon_0) & = \dfrac{1}{2} \eta \dot{d}^2 - (1-\beta)\dfrac{\Gc}{c_0l}\alpha_{,d}(d)\left( 1-e^{-\plasticstrain/\varepsilon_0} \right)\dot{d} + (1-C) \dfrac{\Gc}{c_0 l}\alpha_{, d}(d)\dot{d},
  \end{align}
\end{subequations}
where the first term describes the viscous dissipation of crack propagation and $\eta$ is the kinetic crack propagation viscosity.
The second term introduces a novel representation of the dissipation (hereinafter referred to as the \textit{coalescence dissipation}) associated with the development of the fracture process zone in ductile materials, such as through the evolution of voids, defects and dislocations. It models an exponential decay in the effective fracture toughness as defects increase, with the parameter $0 < \beta \leqslant 1 $ controlling the ratio between the final fracture toughness and the initial fracture toughness.
The third term represents the dissipative portion of the fracture process.

\begin{remark}[regarding the second law of thermodynamics]
  \vspace{-0.5em}
  Consider the more strict form of the dissipation inequality where each of the thermodynamic processes is dissipative. For the process related to the fracture dissipation, that is
  \begin{align*}
    \lambda_{, \dot{d}}(\dot{d}, \plasticstrain; \beta, \varepsilon_0) \dot{d}                                                                      & \geqslant 0  \\
    \eta \dot{d}^2 + \left[ (1-C) - (1-\beta) \left( 1-e^{-\plasticstrain/\varepsilon_0} \right) \right] \dfrac{\Gc}{c_0 l} \alpha_{, d}(d) \dot{d} & \geqslant 0.
  \end{align*}
  Since $\dot{d} \geqslant 0$ and $\alpha_{, d}(d) \geqslant 0, \forall d \in [0, 1]$, the inequality requires $\beta - C \geqslant 0$.  In principle, $C$ can be chosen to calibrate the energetic and dissipative portions of the fracture potential in a specific model given sufficient experimental data, which would in turn set the lower bound for $\beta$.
\end{remark}

The external power expenditure can be expressed as
\begin{subequations}\label{eq: power expenditure}
  \begin{align}
    P(\mathcal{V})       & = \int\limits_\body p_b \diff{V} + \int\limits_\bodyboundary p_t \diff{A}, \\
    p_b(\dot{\bs{\phi}}) & = \rho_0 \bs{B} \cdot \dot{\bs{\phi}},                                     \\
    p_t(\dot{\bs{\phi}}) & = \bs{T} \cdot \dot{\bs{\phi}},
  \end{align}
\end{subequations}
where $\bfB$ is the referential body force and $\bfT$ is the traction acting on the Neumann boundaries of the body in the reference configuration. We assume that the damage field $d$ cannot be driven directly by external agencies.

\subsection{Minimization of the objective function}
\label{s: theory/minimizer}

At this point, given a geometry $\body$, and the set of hyperparameters $\mathscr{H} = \{ \xi, l, m, p, \eta, C, \beta, \varepsilon_0\}$ and material constants $\mathscr{M} = \{ K, G, \sigma_y, H, \Gc, \psi_c \}$, the postulated potential is completely defined. In the following derivations, the functional dependence on $\mathscr{H}$ and $\mathscr{M}$ is dropped notation-wise for simplicity. Substituting \eqref{eq: strain energy potential}, \eqref{eq: plastic hardening potential}, \eqref{eq: fracture potential} and \eqref{eq: dual fracture potential} into \eqref{eq: total potential} and \eqref{eq: total dual kinetic potential}, along with \eqref{eq: power expenditure}, we obtain the objective function to be minimized:
\begin{equation}
  \begin{aligned}
    \   & \dot{\Psi}(\mathcal{V}) + \Pi(\mathcal{V}, \mathcal{S}) - P(\mathcal{V})                                                                                                                                                                                                                          \\
    =\  & \int\limits_\body w^e_{, \bfF}(\bfF, \bfF^p, d) : \dot{\bfF} \diff{V} + \int\limits_\body w^e_{, \bfF^p}(\bfF, \bfF^p, d) : \dot{\bfF}^p \diff{V} + \int\limits_\body w^e_{, d}(\bfF, \bfF^p, d) \dot{d} \diff{V}                                                                                 \\
    \   & + \int\limits_\body w^p_{, \plasticstrain}(\plasticstrain, d) \dot{\plasticstrain} \diff{V} + \int\limits_\body w^p_{, d}(\plasticstrain, d) \dot{d} \diff{V} + \int\limits_\body f_{, d}(d, \grad d) \dot{d} \diff{V} + \int\limits_\body f_{, \grad d}(d, \grad d) \cdot \grad \dot{d} \diff{V} \\
    \   & + \int\limits_\body \lambda(\dot{d}, \plasticstrain) \diff{V} - \int\limits_\body p_b(\dot{\bs{\phi}}) \diff{V} - \int\limits_\bodyboundary p_t(\dot{\bs{\phi}}) \diff{A}.
  \end{aligned}
\end{equation}

We find the minimizer of the objective with respect to the generalized velocity $\mathcal{V} = \{ \dot{\bs{\phi}}, \dot{\bfF}^p, \dot{\plasticstrain}, \dot{d} \}$:
\begin{subequations}\label{eq: minimization}
  \begin{align}
    \mathcal{V} =\        & \argmin_{\mathcal{V}} \dot{\Psi}(\mathcal{V}) + \Pi(\mathcal{V}, \mathcal{S}) - P(\mathcal{V}) \\
    \text{subject to } \  & \dot{\plasticstrain} \geqslant 0, \label{eq: irreversible plastic deformation}                 \\
                          & \dot{d} \geqslant 0, \label{eq: damage irreversibility}                                        \\
                          & \det{\bfF^p} = 1, \label{eq: plastic isochoricity}
  \end{align}
\end{subequations}

\noindent where  \eqref{eq: irreversible plastic deformation} and \eqref{eq: damage irreversibility} enforce the irreversibility of the plastic deformation and damage, respectively, and \eqref{eq: plastic isochoricity} enforces the isochoric nature of plastic flow in J$_2$ plasticity.

Let us define $(\cdot, \cdot)$ as the inner product over the undeformed body $\body$ and $\left<\cdot, \cdot\right>$ as the inner product over the Neumann boundary. Let us also define the variation operator $\delta (\cdot)$ as the variation of the generalized velocity (at a fixed time) while the generalized displacement \emph{is kept constant}. The unconstrained minimization with respect to $\dot{\bs{\phi}}$ given by the variation of the objective yields the balance of linear momentum:
\begin{subequations}
  \begin{align}
               & \left( \dot{\psi}_{, \dot{\bs{\phi}}}(\mathcal{V}), \delta \dot{\bs{\phi}} \right) - \left( {p_b}_{, \dot{\bs{\phi}}}(\dot{\bs{\phi}}), \delta \dot{\bs{\phi}} \right) - \left< {p_t}_{, \dot{\bs{\phi}}}(\dot{\bs{\phi}}), \delta \dot{\bs{\phi}} \right> \\
    =\         &
    \begin{aligned}[t]
       & - \int\limits_\body \divergence w^e_{, \bfF}(\bfF, \bfF^p, d) \cdot \delta \dot{\bs{\phi}} \diff{V} + \int\limits_\bodyboundary \left( w^e_{, \bfF}(\bfF, \bfF^p, d) \cdot \bfN \right) \cdot \delta \dot{\bs{\phi}} \diff{A}   \\
       & - \int\limits_\body \rho_0 \bs{B} \cdot \delta \dot{\bs{\phi}} \diff{V} - \int\limits_\bodyboundary \bs{T} \cdot \delta \dot{\bs{\phi}} \diff{A} = 0, \quad \forall \body' \subset \body, \bodyboundary' \subset \bodyboundary,
    \end{aligned}                                                                                                                                                                                                                                              \\
    \implies\  & \left\{
    \begin{aligned}
      \divergence \bfP + \rho_0 \bs{B} & = \bs{0}, \quad \forall \bs{X} \in \body,         \\
      \bfP \cdot \bfN                  & = \bs{T}, \quad \forall \bs{X} \in \bodyboundary,
    \end{aligned} \right.
  \end{align}
\end{subequations}
with the constitutive constraints (following the Coleman-Noll procedure):
\begin{subequations}\label{eq: stress-strain}
  \begin{align}
    \bfP                      & = w^e_{, \bfF}(\bfF, \bfF^p, d) = \bs{\tau}\bs{F}^{-T},        \\
    \bs{\tau}                 & = g^e(d)\bs{\tau}^\activeenergy + \bs{\tau}^\inactiveenergy,   \\
    \bs{\tau}^\activeenergy   & = \dfrac{1}{2}\mathbb{H}_1(J)K(J^2-1)\bs{I} + G\dev(\bs{b}^e), \\
    \bs{\tau}^\inactiveenergy & = \dfrac{1}{2}(1-\mathbb{H}_1(J))K(J^2-1)\bs{I}.
  \end{align}
\end{subequations}

In keeping with a variational framework, here we determine the plastic flow rule following the principle of maximum dissipation.  As a result, the plastic flow rule and the plastic yield surface are derived from a pair of coupled partial differential equations. To simplify the derivation, we follow \citet{simo1988framework} and assume, subject to verification, that the plastic yield surface (serving as the Lagrange multiplier for the constraint \eqref{eq: irreversible plastic deformation}) takes the following form:
\begin{align}
  \phi^p(\bfF, \bfF^p, \plasticstrain, d) = \norm{\dev(\bs{\tau})} - c_1 w^p_{, \plasticstrain}(\plasticstrain, d) \leqslant 0 \label{eq: yield surface guess}.
\end{align}
It can be shown that, by minimizing the Lagrangian with respect to $\bfF^p$ or $\bfC^p$, the plastic flow rule constrains the spin of plastic deformation up to a constant $c_1$:
\begin{align}
  \dot{\bfF}^p {\bfF^p}^{-1} & = \dfrac{2}{3} c_1 \dot{\plasticstrain} \dfrac{\dev(\bs{\tau})}{\norm{\dev(\bs{\tau})}}.
\end{align}

Next, by minimizing the Lagrangian with respect to $\dot{\plasticstrain}$, the plastic yield surface is recovered:
\begin{subequations}
  \begin{align}
               & \left( \dot{\psi}_{, \dot{\plasticstrain}}(\mathcal{S}), \delta \dot{\plasticstrain} \right) + \left( c_2 \phi^p, \delta \dot{\plasticstrain} \right) = 0, \\
    =\         &
    \begin{aligned}[t]
       & \int\limits_\body w^e_{, \bfF^p}(\bfF, \bfF^p, d) : \dot{\bfF}^p_{, \dot{\plasticstrain}} \delta \dot{\plasticstrain} \diff{V} + \int\limits_\body w^p_{, \plasticstrain}(\plasticstrain, d) \delta \dot{\plasticstrain} \diff{V} + \int\limits_\body c_2 \phi^p \delta \dot{\plasticstrain} \diff{V},
    \end{aligned}                                                                                                                                              \\
    \implies\  & \left\{
    \begin{aligned}
      \phi^p \leqslant 0, \quad \forall \bs{X} \in \body,               \\
      \dot{\plasticstrain} \geqslant 0, \quad \forall \bs{X} \in \body, \\
      \phi^p \dot{\plasticstrain} = 0, \quad \forall \bs{X} \in \body,
    \end{aligned} \right.
  \end{align}
\end{subequations}
where, by comparing with the assumption \eqref{eq: yield surface guess}, the yield surface can be written as
\begin{subequations}
  \begin{align}\label{eq: plastic yield surface}
    \phi^p(\bfF, \bfF^p, \plasticstrain, d) & = \norm{\dev(\bs{\tau})} - \sqrt{\dfrac{2}{3}} w^p_{, \plasticstrain}(\plasticstrain, d),         \\
                                            & = \norm{\dev(\bs{\tau})} - \sqrt{\dfrac{2}{3}} g^p(d) \left( \sigma_y + H \plasticstrain \right),
  \end{align}
\end{subequations}
which resembles a linear hardening law. The corresponding plastic flow rule is given by
\begin{align} \label{eq: flow rule}
  \dot{\bfF}^p {\bfF^p}^{-1} = \dot{\plasticstrain} \bfN^p, \quad \bfN^p = \sqrt{\dfrac{3}{2}} \dfrac{\dev(\bs{\tau})}{\norm{\dev(\bs{\tau})}}, \quad \det(\bfF^p) = 1.
\end{align}

Finally, the constrained minimization with respect to $\dot{d}$ (i.e.\  through Lagrange multipliers) is given as
\begin{subequations}
  \begin{align}
               & \left( \dot{\psi}_{, \dot{d}}(\mathcal{S}), \delta \dot{d} \right) + \left( \pi_{, \dot{d}}(\mathcal{V}, \mathcal{S}), \delta \dot{d} \right) + \left( \phi^f, \delta \dot{d} \right) \\
    =\         &
    \begin{aligned}[t]
       & \int\limits_\body f_{, d}(d, \grad d) \delta \dot{d} \diff{V} - \int\limits_\body \divergence f_{, \grad d}(d, \grad d) \delta \dot{d} \diff{V} + \int\limits_\bodyboundary f_{, \grad d}(d, \grad d) \cdot \bfN \delta \dot{d} \diff{A}                                                         \\
       & + \int\limits_\body w^e_{, d}(\bfF, \bfF^p, d) \delta \dot{d} \diff{V} + \int\limits_\body w^p_{, d}(\plasticstrain, d) \delta \dot{d} \diff{V} + \int\limits_\body \lambda_{, \dot{d}}(\dot{d}, \plasticstrain) \delta \dot{d} \diff{V} + \int\limits_\body \phi^f \delta \dot{d} \diff{V} = 0,
    \end{aligned}                                                                                                                                                                         \\
    \implies\  & \left\{
    \begin{aligned}
      f_{, \grad d}(d, \grad d) \cdot \bfN = 0 , \quad \forall \bs{X} \in \bodyboundary, \\
      \phi^f \leqslant 0, \quad \forall \bs{X} \in \body,                                \\
      \dot{d} \geqslant 0, \quad \forall \bs{X} \in \body,                               \\
      \phi^f \dot{d} = 0, \quad \forall \bs{X} \in \body,
    \end{aligned} \right.
  \end{align}
\end{subequations}
with
\begin{subequations}\label{eq: fracture yield surface}
  \begin{align}
    -\phi^f                               & = \eta \dot{d} - \divergence \left( \dfrac{2 \Gc l}{c_0} \grad d \right) + \dfrac{\widehat{\Gc}}{c_0 l}\alpha_{,d}(d) + \chi(\bfF, \bfF^p, \plasticstrain, d),              \\
    \widehat{\Gc}                         & = g^c(\plasticstrain; \beta, \varepsilon_0) \Gc, \quad g^c(\plasticstrain; \beta, \varepsilon_0) = 1-(1-\beta)\left( 1-e^{-\plasticstrain/\varepsilon_0} \right)            \\
    \chi(\bfF, \bfF^p, \plasticstrain, d) & = g^e_{, d}(d) w^e_\activeenergy(\bfF, \bfF^p) + g^p_{, d}(d) \left( \sigma_y \plasticstrain + \dfrac{1}{2} H \plasticstrain^2 \right), \label{eq: fracture driving energy}
  \end{align}
\end{subequations}
where $\chi(\bfF, \bfF^p, \plasticstrain, d)$ may be interpreted as the generalized fracture driving energy which consists of contributions from the active strain energy and the plastic work.  Importantly, the fracture toughness $\Gc$ is ``degraded'' by the function $g^c(\plasticstrain; \beta, \varepsilon_0)$, resulting in an effective fracture toughness $\widehat{\Gc}$ that decreases as the plastic deformation increases. The underlying effective fracture dissipation is
\begin{align}\label{eq: effective fracture dissipation}
  \widehat{f}(d,\grad d) = \dfrac{\Gc}{c_0 l} (g^c(\plasticstrain) \alpha(d) + l^2 \grad d \cdot \grad d).
\end{align}

\subsection{Discretization}
\label{s: theory/discretization}

We begin this section by summarizing the strong form of the initial boundary value problem of interest:
\begin{mdframed}[
    frametitle={The initial boundary value problem},
    frametitlerule=true,
    frametitlebackgroundcolor=gray!20,
    linewidth=1pt,
    nobreak=true
  ]
  \vspace{-1em}
  \begin{align*}
    \text{linear momentum balance: }   & \divergence \bfP + \rho_0 \bfB = \bs{0},                  &  & \forall \bfX \in \body,           \\
                                       & \bfP \cdot \bfN = \bfT,                                   &  & \forall \bfX \in \bodyboundary_t, \\
                                       & \bfu = \bfu_g,                                            &  & \forall \bfX \in \bodyboundary_u, \\
    \text{stress-strain relations: }   & \bfP = \bs{\tau}\bs{F}^{-T},                              &  & \forall \bfX \in \body,           \\
    \text{KKT system for plasticity: } & \phi^p \leqslant 0,                                       &  & \forall \bfX \in \body,           \\
                                       & \dot{\plasticstrain} \geqslant 0,                         &  & \forall \bfX \in \body,           \\
                                       & \phi^p \dot{\plasticstrain} = 0,                          &  & \forall \bfX \in \body,           \\
    \text{plastic flow rule: }         & \dot{\bfF}^p {\bfF^p}^{-1} = \dot{\plasticstrain} \bfN^p, &  & \forall \bfX \in \body,           \\
                                       & \det(\bfF^p) = 1,                                         &  & \forall \bfX \in \body,           \\
    \text{KKT system for fracture: }   & f_{, \grad d}(d, \grad d) \cdot \bfN = 0 ,                &  & \forall \bfX \in \bodyboundary,   \\
                                       & \phi^f \leqslant 0,                                       &  & \forall \bfX \in \body,           \\
                                       & \dot{d} \geqslant 0,                                      &  & \forall \bfX \in \body,           \\
                                       & \phi^f \dot{d} = 0,                                       &  & \forall \bfX \in \body,           \\
    \text{initial conditions: }        & \bfF^p(t=0) = \bfI,                                       &  & \forall \bfX \in \body,           \\
                                       & \plasticstrain(t=0) = 0,                                  &  & \forall \bfX \in \body,           \\
                                       & d(t=0) = 0,                                               &  & \forall \bfX \in \body,
  \end{align*}
\end{mdframed}
supplemented by detailed definitions of the Kirchhoff stress $\bs{\tau}$ \eqref{eq: stress-strain}, the plastic yield surface $\phi^p$ \eqref{eq: plastic yield surface}, the plastic flow direction $\bfN^p$ \eqref{eq: flow rule}, and the fracture envelope $\phi^f$ \eqref{eq: fracture yield surface}.

The stress-strain relations, the KKT system for plasticity and the plastic flow rule are satisfied during constitutive updates within the finite element method, consistent with standard practice. Therefore, the field equations (together with their initial and boundary conditions) to be discretized are:

\begin{mdframed}[
    frametitle={The strong form},
    frametitlerule=true,
    frametitlebackgroundcolor=gray!20,
    linewidth=1pt,
    nobreak=true
  ]
  \vspace{-1em}
  \begin{align*}
    \text{governing equations: } & \divergence \bfP + \rho_0 \bfB = \bs{0},                                                                                                       &  & \forall \bfX \in \body,           \\
                                 & -\phi^f = \eta \dot{d} - \divergence \left( \dfrac{2 \Gc l}{c_0} \grad d \right) + \dfrac{\widehat{\Gc}}{c_0 l}\alpha_{,d} + \chi \geqslant 0, &  & \forall \bfX \in \body,           \\
    \text{boundary conditions: } & \bfP \cdot \bfN = \bfT,                                                                                                                        &  & \forall \bfX \in \bodyboundary_t, \\
                                 & \bfu = \bfu_g,                                                                                                                                 &  & \forall \bfX \in \bodyboundary_u, \\
                                 & f_{, \grad d} \cdot \bfN = 0 ,                                                                                                                 &  & \forall \bfX \in \bodyboundary,   \\
    \text{initial condition: }   & d(t=0) = 0,                                                                                                                                    &  & \forall \bfX \in \body,           \\
    \text{constraints: }         & \dot{d} \geqslant 0,                                                                                                                           &  & \forall \bfX \in \body,           \\
                                 & \phi^f \dot{d} = 0,                                                                                                                            &  & \forall \bfX \in \body.
  \end{align*}
\end{mdframed}

To develop the weak form, we begin by introducing the trial spaces $\bs{\fspace{U}}_t$ and $\fspace{D}_t$ as:
\begin{align}
  \bs{\fspace{U}}_t = \{ \bfu \in \fspace{H}^1(\body)^d \mid \bfu(t) = \bfu_g \text{ on } \bodyboundary_u \}, \quad \fspace{D}_t = \{ d \in \fspace{H}^1(\body) \mid \dot{d} \geqslant 0, \phi^f \dot{d} = 0 \},
\end{align}
and the weighting spaces $\bs{\fspace{V}}$ and $\fspace{Q}$ as:
\begin{align}
  \bs{\fspace{V}} = \{ \bfw \in \fspace{H}^1(\body)^d \mid \bfw(t) = \bs{0} \text{ on } \bodyboundary_u \}, \quad \fspace{Q} = \{ q \in \fspace{H}^1(\body) \}.
\end{align}

The weak form can be derived as:
\begin{mdframed}[
    frametitle={The weak form},
    frametitlerule=true,
    frametitlebackgroundcolor=gray!20,
    linewidth=1pt,
    nobreak=true
  ]
  Given $\bfu_g$, $\bfT$ and $d_0$, find $\bfu(t) \in \bs{\fspace{U}}_t$ and $d(t) \in \fspace{D}_t$, $t \in [0, t']$, such that $\forall \bfw \in \bs{\fspace{V}}$ and $\forall q \in \fspace{Q}$,
  \begin{subequations}
    \begin{align}
      \left< \bfw, \bfT \right>_{\bodyboundary_t} - \left( \grad \bfw, \bfP \right) + \left( \bfw, \rho_0 \bfB \right)                                                             & = \bs{0},    \\
      \left( q, \eta \dot{d} \right) + \left( \grad q, \dfrac{2 \Gc l}{c_0} \grad d \right) + \left( q, \dfrac{\widehat{\Gc}}{c_0 l} \alpha_{, d} \right) + \left( q, \chi \right) & \geqslant 0, \\
      \left( q, d(0) \right) - \left( q, d_0 \right)                                                                                                                               & = 0.
    \end{align}
  \end{subequations}
\end{mdframed}

Using the Galerkin method, with finite dimensional function spaces $\widetilde{\bs{\fspace{U}}}_t \subset \bs{\fspace{U}}_t$, $\widetilde{\bs{\fspace{V}}} \subset \bs{\fspace{V}}$, $\widetilde{\fspace{D}}_t \subset \fspace{D}_t$, $\widetilde{\fspace{Q}} \subset \fspace{Q}$, we arrive at the spatially discrete form of the problem:
\begin{mdframed}[
    frametitle={The Galerkin form},
    frametitlerule=true,
    frametitlebackgroundcolor=gray!20,
    linewidth=1pt,
    nobreak=true
  ]
  Given $\bfu_g$, $\bfT$ and $d_0$, find $\bfu^h(t) \in \widetilde{\bs{\fspace{U}}}_t$ and $d^h(t) \in \widetilde{\fspace{D}}_t$, $t \in [0, t']$, such that $\forall \bfw^h \in \widetilde{\bs{\fspace{V}}}$ and $\forall q^h \in \widetilde{\fspace{Q}}$,
  \begin{subequations}
    \begin{align}
      \left< \bfw^h, \bfT \right>_{\bodyboundary_t} - \left( \grad \bfw^h, \bfP \right) + \left( \bfw^h, \rho_0 \bfB \right)                                                                 & = \bs{0},    \label{eq: galerkin mechanics} \\
      \left( q^h, \eta \dot{d} \right) + \left( \grad q^h, \dfrac{2 \Gc l}{c_0} \grad d^h \right) + \left( q^h, \dfrac{\widehat{\Gc}}{c_0 l} \alpha_{, d} \right) + \left( q^h, \chi \right) & \geqslant 0, \label{eq: galerkin damage}    \\
      \left( q^h, d^h(0) \right) - \left( q^h, d_0 \right)                                                                                                                                   & = 0. \label{eq: galerkin bound}
    \end{align}
  \end{subequations}
\end{mdframed}

The discrete inequality for crack irreversibility is satisfied node-wise with a primal-dual active set strategy. See e.g.\  \citet{heister2015primal} for implementational details of such a solver. The solver is also generally available in numerical toolboxes, e.g.\  PETSc \cite{petsc-web-page}. The discrete approximation is calculated using a fixed-point iterative solution scheme outlined from \cite{HuGary2020}. The KKT system for plasticity with the plastic flow rule constraints are enforced using the return mapping algorithm from \cite{borden2016phase}, where the F-bar approach (e.g. \cite{neto2005f}) is adopted to circumvent volumetric locking.

\section{Numerical examples}
\label{s: example}

The numerical results shown in this section are generated using RACCOON \cite{raccoon, raccoon_doc}, a massively parallel finite element code specializing in phase-field fracture problems. RACCOON is built upon the MOOSE framework \cite{permann2020moose} developed at the Idaho National Laboratory.
The mechanics sub-problem \eqref{eq: galerkin mechanics} is solved using the Newton-Raphson method. The phase-field sub-problem \eqref{eq: galerkin damage}\&\eqref{eq: galerkin bound} is solved using PETSc's variational inequality (VI) solver \cite{benson2006flexible}.
We choose $C = \beta$ for the fracture energy \eqref{eq: fracture potential} and the fracture dissipation \eqref{eq: dual fracture potential} in all calculations to satisfy the second law of thermodynamics. All other model parameters and material constants are provided separately for each of the problems in this section, below.

\begin{figure}[htb!]
  \centering
  \begin{subfigure}{0.4\textwidth}
    \centering
    \tikzsetnextfilenamesafe{example/terminology/EPD}
    \includegraphics{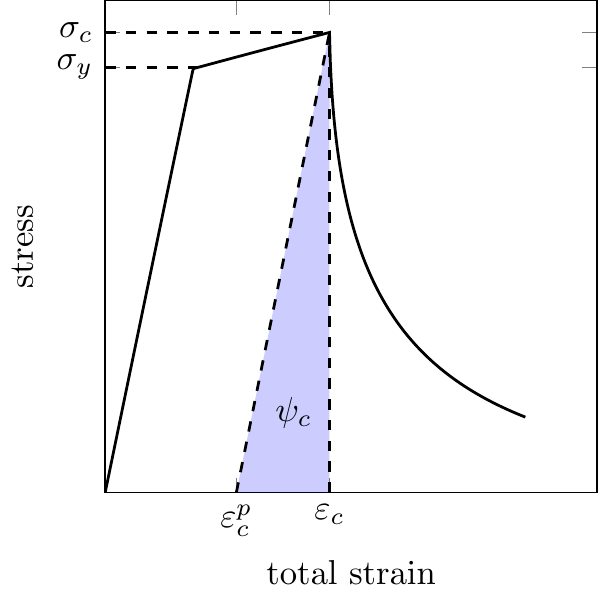}
    \caption{}
    \label{fig: example/terminology/EPD}
  \end{subfigure}
  \begin{subfigure}{0.4\textwidth}
    \centering
    \tikzsetnextfilenamesafe{example/terminology/EPPD}
    \includegraphics{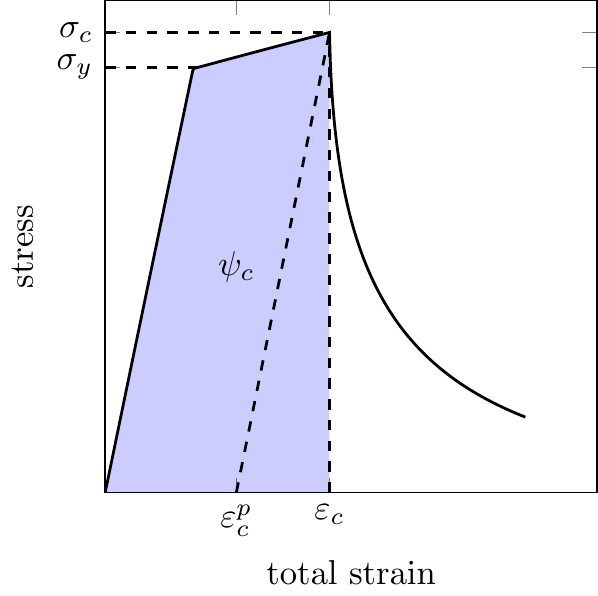}
    \caption{}
    \label{fig: example/terminology/EPPD}
  \end{subfigure}
  \caption{A prototypical stress-strain curve. The critical fracture energy $\psi_c$ is represented by the shaded area. The compositions of the fracture driving energy are different for (a) an E-P-D model and (b) an E-P-PD model.}
  \label{fig: example/terminology}
\end{figure}

Prototypical stress-strain curves (\Cref{fig: example/terminology}) obtained with the model presented in \Cref{s: theory} serve to illustrate several terms that are used throughout this section. Following \cite{alessi_coupling_2018}, the stress-strain response is categorized into \textrm{E}, \textrm{P}, \textrm{D}, and \textrm{PD} stages. First, during the \emph{elastic} (\textrm{E}) loading stage, the stress increases with strain with a slope governed by the elastic moduli. Once the stress (in a measure defined by the plastic flow rule) reaches the yield stress $\sigma_y$, it enters the \emph{plastic} (\textrm{P}) hardening stage, plastic flow occurs, and the plastic strain begins to accumulate. In this stage, the slope of the stress-strain curve is governed by the plastic flow rule and the plastic hardening law.
Once the generalized fracture driving energy $\chi$ reaches the critical fracture energy $\psi_c$, a crack initiates and the stress-strain curve enters the softening stage. If the plastic strain keeps increasing during the softening stage, it is referred to as \emph{plastic-damaging} (\textrm{PD}), otherwise it is simply \emph{damaging} (\textrm{D}). The stress, the total strain, and the equivalent plastic strain values attained just before the curve enters the softening stage are referred to as the critical fracture strength $\sigma_c$, the critical strain $\varepsilon_c$, and the critical plastic strain $\varepsilon^p_c$, respectively.

Although the variational framework presented in \Cref{s: theory} allows the use of a variety of fracture models, here we focus on those obtained via particular choices for the crack geometric function $\alpha(d)$, the elastic degradation function $g^e(d)$, and the plastic degradation function $g^p(d)$. In particular, we propose a new model (hereinafter referred to as the E-P-D model) by selecting
\begin{align*}
  \alpha(d) = d, \quad g^e(d) = g^\text{Lorentz}(d), \quad g^p(d) = 1, \quad \beta \in (0,1], \quad \varepsilon_0 > 0,
\end{align*}
where plasticity and fracture are not strongly coupled, but the evolution of plastic deformation results in a ``degraded'' fracture toughness. For this E-P-D model, the corresponding generalized fracture driving energy $\chi$ consists of only the active elastic energy $w^e_\activeenergy$, and the critical fracture energy $\psi_c$ is pictorially depicted in \Cref{fig: example/terminology/EPD}.

In what follows, we will compare the behavior of the E-P-D model to an existing phase-field for ductile fracture model presented in \cite{brandon2020cohesive}.  The existing model,
which we refer to as the E-P-PD model,  is recovered by selecting
\begin{align*}
  \alpha(d) = d, \quad g^e(d) = g^p(d) = g^\text{Lorentz}(d), \quad \beta = 1.
\end{align*}
These choices give rise to a model in which plasticity and fracture are strongly coupled, i.e.\ the yield surface is degraded by the phase-field and the plastic energy contributes to fracture. For this E-P-PD model, the corresponding generalized fracture driving energy $\chi$ consists of both the active elastic energy $w^e_\activeenergy$ and the plastic energy $w^p$.  The corresponding critical fracture energy $\psi_c$ is pictorially depicted in \Cref{fig: example/terminology/EPPD}.

In this section, we demonstrate through numerical examples the motivations behind the various modeling choices. The naming of each model is explained \Cref{s: example/homogenized/characteristics} by investigating their characteristic loading and unloading behavior. \Cref{s: example/homogenized/alpha} motivates the use of a linear crack geometric function, and \Cref{s: example/homogenized/degradation} motivates the use of the Lorentz-type degradation functions.
The effect of the coalescence dissipation in a homogenized setting is demonstrated in \Cref{s: example/homogenized/coalescence}.
\Cref{s: example/nonhomogeneous} presents an example demonstrating that the nonhomogeneous model response is independent of the regularization length for different values of $\beta$. Finally, both the proposed E-P-D model and the E-P-PD model are applied to simulate a three-point bending experiment in \Cref{s: example/3pb} and the experiment adopted for the Sandia Fracture Challenge in \Cref{s: example/SFC}.

\subsection{A homogeneous example: uniaxial constitutive response}
\label{s: example/homogenized}

We begin by considering a problem in which the damage is spatially constant and so nonlocal features are effectively suppressed.  This is effected by considering the response of a single element subjected to uniaxial tension.  Examining this relatively simple system allows us to introduce the effect of different modeling choices.

For this study and the sake of comparison, the material parameters are largely taken in accordance with an analogous numerical example described in \citet{borden2016phase}. The domain is a one-element (\texttt{HEX8}) cube with side length $a$. The material and fracture parameters are: Young's modulus $E = \SI{68.8}{\giga\pascal}$;  Poisson's ratio $\nu = 0.33$; yield stress $\sigma_y = \SI{320}{\mega\pascal}$; hardening modulus $h = \SI{688}{\mega\pascal}$, and; the fracture toughness $\Gc = \SI{1.38e5}{\milli\joule\per\square\milli\meter}$.

The single element is loaded in uniaxial tension with displacement control. Representative loading and unloading behaviors are illustrated using the E-P-D model and the E-P-PD model. In what follows,  motivations for  particular modeling choices of the crack geometric function $\alpha(d)$ and the degradation functions $g^e(d)$ and $g^p(d)$ are provided.

\subsubsection{Characteristic loading and unloading behavior}
\label{s: example/homogenized/characteristics}

\begin{figure}[htb!]
  \centering
  \begin{subfigure}[b]{0.45\textwidth}
    \centering
    \tikzsetnextfilenamesafe{example/homogenized/lu/EPD}
    \includegraphics{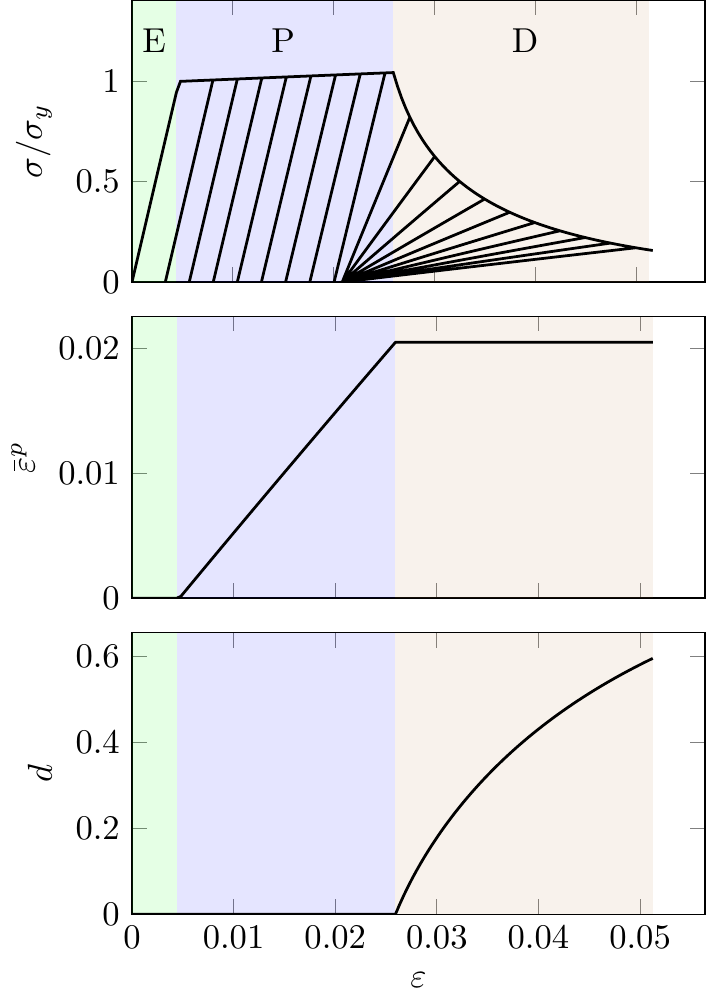}
    \caption{}
    \label{fig: example/homogenized/lu/EPD}
  \end{subfigure}
  \hspace{0.04\textwidth}
  \begin{subfigure}[b]{0.45\textwidth}
    \centering
    \tikzsetnextfilenamesafe{example/homogenized/lu/EPPD}
    \includegraphics{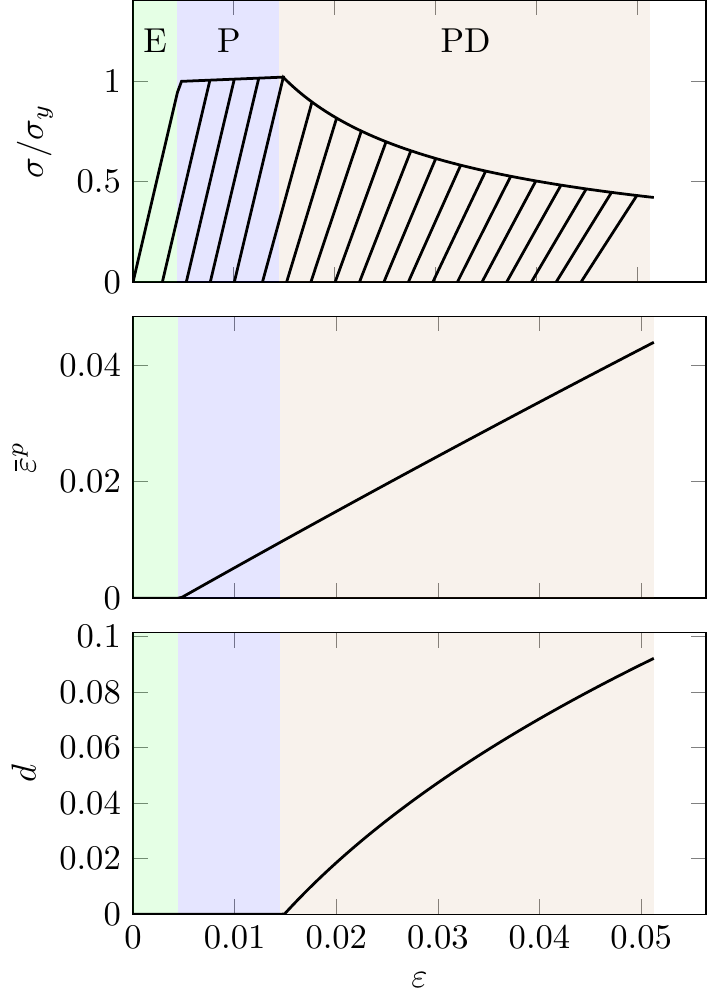}
    \caption{}
    \label{fig: example/homogenized/lu/EPPD}
  \end{subfigure}
  \caption{Characteristic loading and unloading curves of (a) the E-P-D model and (b) the E-P-PD model. }
  \label{fig: example/homogenized/lu}
\end{figure}

Consider $\beta = 1$, the proposed E-P-D model yields an E-P-D loading and unloading response (\Cref{fig: example/homogenized/lu/EPD}), and the E-P-PD model yields an E-P-PD loading and unloading response (\Cref{fig: example/homogenized/lu/EPPD}).

We note that if the critical fracture strength is chosen to be smaller than the yield stress, e.g.\  when the critical fracture energy is sufficiently small, the E-P-D response becomes an E-D response, and the E-P-PD response becomes an E-D-PD response, recovering the other two representative cases for the stress-strain response following \citet{alessi_coupling_2018}.

\subsubsection{The crack geometric function}
\label{s: example/homogenized/alpha}

\begin{figure}[htb!]
  \centering
  \begin{subfigure}[b]{0.29\textwidth}
    \centering
    \tikzsetnextfilenamesafe{example/homogenized/compare_alpha/EPPD_xi_0}
    \includegraphics{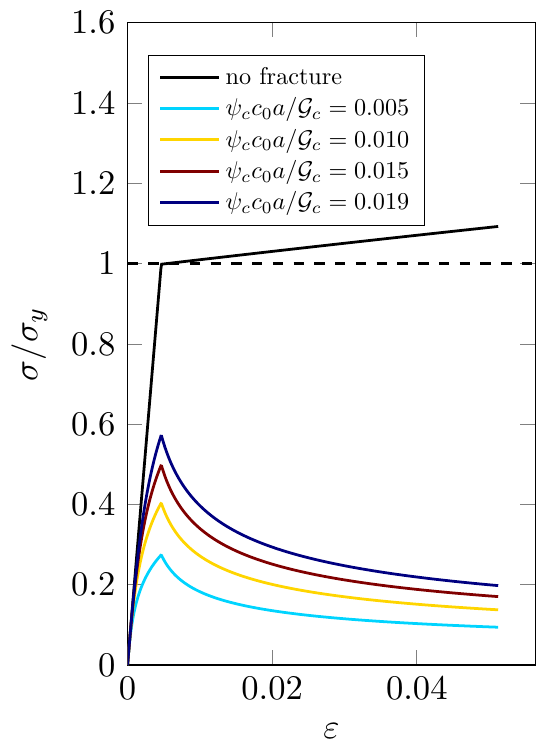}
    \vspace{-12pt}
    \caption{}
    \label{fig: example/homogenized/compare_alpha/EPPD_xi_0}
  \end{subfigure}
  \hspace{0.04\textwidth}
  \begin{subfigure}[b]{0.29\textwidth}
    \centering
    \tikzsetnextfilenamesafe{example/homogenized/compare_alpha/EPPD_xi_1}
    \includegraphics{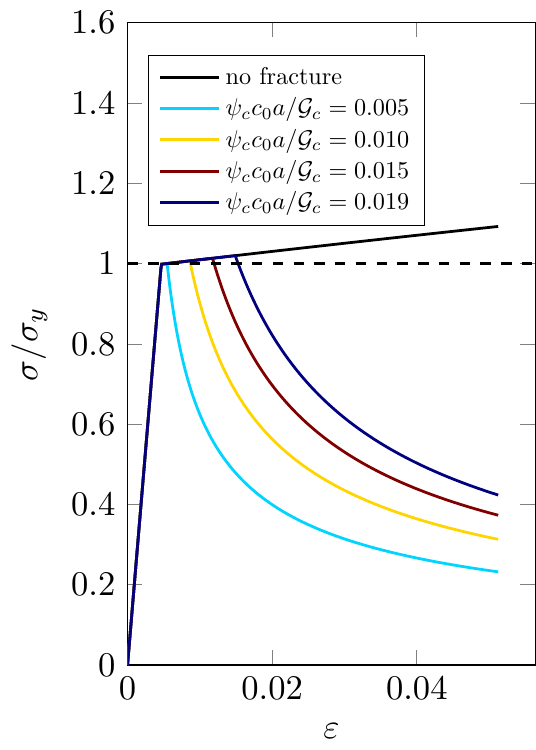}
    \vspace{-12pt}
    \caption{}
    \label{fig: example/homogenized/compare_alpha/EPPD_xi_1}
  \end{subfigure}
  \hspace{0.04\textwidth}
  \begin{subfigure}[b]{0.29\textwidth}
    \centering
    \tikzsetnextfilenamesafe{example/homogenized/compare_alpha/EPD_xi_1}
    \includegraphics{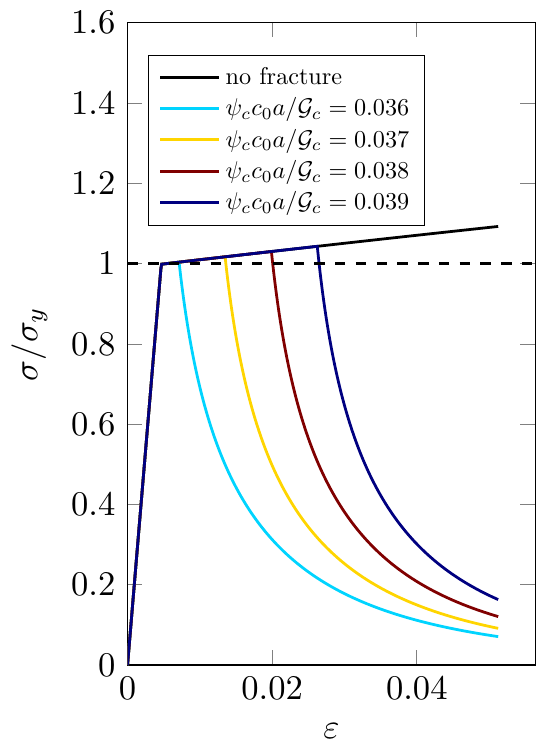}
    \vspace{-12pt}
    \caption{}
    \label{fig: example/homogenized/compare_alpha/EPD_xi_1}
  \end{subfigure}
  \caption{Stress-strain curves obtained with (a) the quadratic crack geometric function $\alpha(d;\xi=0) = d^2$ and the Lorentz-type degradation function $g^e(d) = g^p(d) = g^\text{Lorentz}(d;m,p=1)$, (b) the E-P-PD model with $\xi = 1$, and (c) the E-P-D model with $\xi = 1$.}
  \label{fig: example/homogenized/compare_alpha}
\end{figure}

Next, two crack geometric functions, $\alpha(d; \xi = 0) = d^2$ and $\alpha(d; \xi = 1) = d$, are compared. The corresponding stress-strain curves are shown in \Cref{fig: example/homogenized/compare_alpha}.
In \Cref{fig: example/homogenized/compare_alpha/EPPD_xi_0}, we observe that  the stress-strain response enters a damage softening stage \textit{before} the stress reaches the yield stress. In essence, what happens is that a small amount of damage is accumulated during the elastic loading stage due to the nature of the specific crack geometric function $\alpha(d; \xi=0) = d^2$.
The plastic yield surface is consequently shrunk according to \eqref{eq: plastic yield surface}, and a fair amount of plastic work contributes to the total fracture driving energy $\chi$ according to \eqref{eq: fracture driving energy}. As a result, the response enters the damage softening stage much earlier than expected.

By adopting a crack geometric potential with a linear term, e.g. $\xi > 0$, a finite mobility is attained in the absence of damage, and an unperturbed elastic-plastic response can be recovered. The results shown correspond to the use of  $\xi = 1$.  The stress-strain curves for the E-P-PD model and the E-P-D model are shown in \Cref{fig: example/homogenized/compare_alpha/EPPD_xi_1,fig: example/homogenized/compare_alpha/EPD_xi_1}. Prior to crack initiation, the stress-strain curves with $\xi = 1$ align perfectly with the one where fracture is not considered. Such behavior is favorable in calibrations with experiments because it allows for decoupled hyperparameter tuning for plasticity and fracture: the elastic moduli, the yield stress, and the hardening parameters can be tuned without considering fracture. Accordingly, from this point forward we rely only on the use of a linear crack geometric function $\alpha(d; \xi = 1) = d$ for both the E-P-D model and the E-P-PD model.

\subsubsection{The degradation functions}
\label{s: example/homogenized/degradation}

\begin{figure}[htb!]
  \centering
  \begin{subfigure}[b]{0.29\textwidth}
    \centering
    \tikzsetnextfilenamesafe{example/homogenized/compare_degradation/EPPD_quad}
    \includegraphics{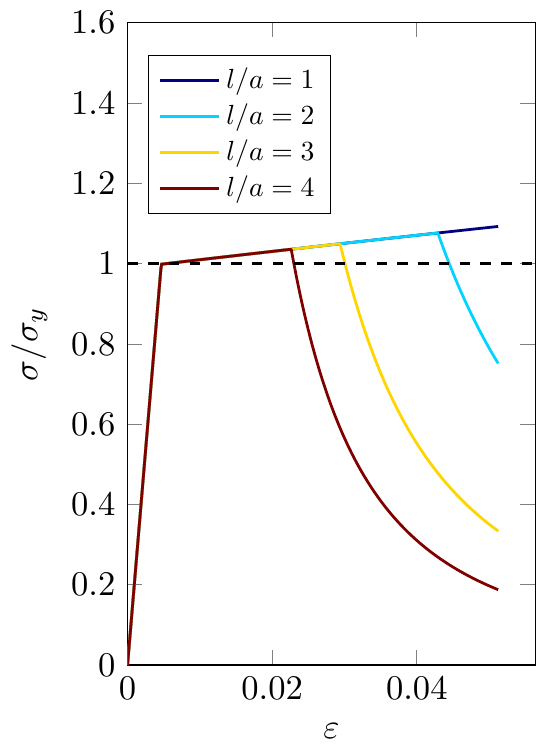}
    \vspace{-12pt}
    \caption{}
    \label{fig: example/homogenized/compare_degradation/EPPD_quad}
  \end{subfigure}
  \hspace{0.04\textwidth}
  \begin{subfigure}[b]{0.29\textwidth}
    \centering
    \tikzsetnextfilenamesafe{example/homogenized/compare_degradation/EPPD_lorentz}
    \includegraphics{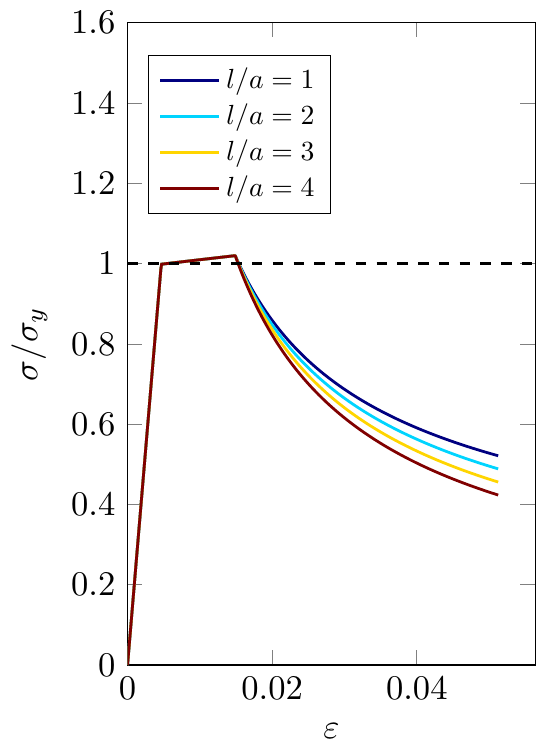}
    \vspace{-12pt}
    \caption{}
    \label{fig: example/homogenized/compare_degradation/EPPD_lorentz}
  \end{subfigure}
  \hspace{0.04\textwidth}
  \begin{subfigure}[b]{0.29\textwidth}
    \centering
    \tikzsetnextfilenamesafe{example/homogenized/compare_degradation/EPD_lorentz}
    \includegraphics{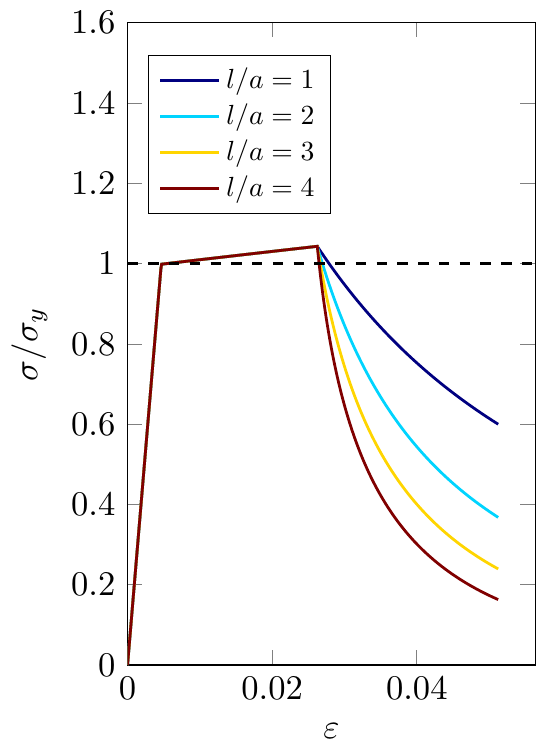}
    \vspace{-12pt}
    \caption{}
    \label{fig: example/homogenized/compare_degradation/EPD_lorentz}
  \end{subfigure}
  \caption{Stress-strain curves obtained with the linear crack geometric function $\alpha(d) = d$ and (a) the quadratic degradation function (b-c) a Lorentz-type degradation function. Both (b) the E-P-PD model ($\psi_c c_0 a / \Gc = 0.019$) and (c) the E-P-D model ($\psi_c c_0 a / \Gc = 0.039$) show a regularization-length-independent critical fracture strength $\sigma_c$.}
  \label{fig: example/homogenized/compare_degradation}
\end{figure}

Next, we focus attention on the critical fracture strength $\sigma_c$. The Lorentz-type degradation is compared with the widely used quadratic degradation function $g^\text{quad}(d) = (1-d)^2$.
As shown in \Cref{fig: example/homogenized/compare_degradation/EPPD_quad}, in conjunction with $\alpha(d) = d$, the quadratic degradation function still relates the critical fracture energy (or equivalently the critical fracture strength) to the phase-field regularization length.  As a result, as the regularization length is decreased, the stress at which the softening begins increases.
By contrast, the use of a Lorentz-type degradation function \eqref{eq: lorentz degradation} removes this correlation between the critical fracture strength and the regularization length.  As shown in \Cref{fig: example/homogenized/compare_degradation/EPPD_lorentz,fig: example/homogenized/compare_degradation/EPD_lorentz}, the use of a Lorentz-type degradation function thus gives rise to softening that begins at a stress that is insensitive to the regularization length, which confirms the observations of \cite{brandon2020cohesive}.
The influence of the regularization length is instead largely exhibited in the rate of decay in the stress during the softening portion of the response.  Going forward, we therefore confine attention to models using only the Lorentz-type degradation functions.

We conclude this subsection by noting that while the observations regarding the critical fracture strength in this homogeneous setting can be generalized to a non-homogeneous setting, the softening response and its convergence have to be investigated separately in a non-homogeneous setting.

\subsubsection{The coalescence dissipation}
\label{s: example/homogenized/coalescence}

With clear justification for the modeling choices of the crack geometric function (\Cref{s: example/homogenized/alpha}) and the degradation functions (\Cref{s: example/homogenized/degradation}), we now examine the effect of the coalescence dissipation in this homogenized setting.

\textit{Without} fracture dissipation, i.e.\ $C = \beta = 1$, the E-P-PD model and the E-P-D model have some shortcomings.
First, the E-P-PD model introduces a strong coupling between plasticity and fracture, as both the active part of the strain energy and the plastic energy contribute to damage evolution (according to \eqref{eq: fracture driving energy}, or see \Cref{fig: example/terminology/EPPD} for a graphical illustration). As a result, the use of a large critical fracture energy $\psi_c$ may be required to calibrate the E-P-PD model against experimental measurements. However, due to convexity concerns, the Lorentz-type degradation functions typically impose an upper-bound on the regularization length that scales inversely with the critical fracture energy:
\begin{align}
  l \leqslant l_u \propto \dfrac{1}{\psi_c}.
  \label{eq: upper bound}
\end{align}
As a result, large values of $\psi_c$ can give rise to a regularization length that is prohibitively small to resolve computationally.

\begin{figure}[htb!]
  \centering
  \begin{subfigure}[b]{0.4\textwidth}
    \centering
    \tikzsetnextfilenamesafe{example/homogenized/compare_beta/beta_1/stress_strain}
    \includegraphics{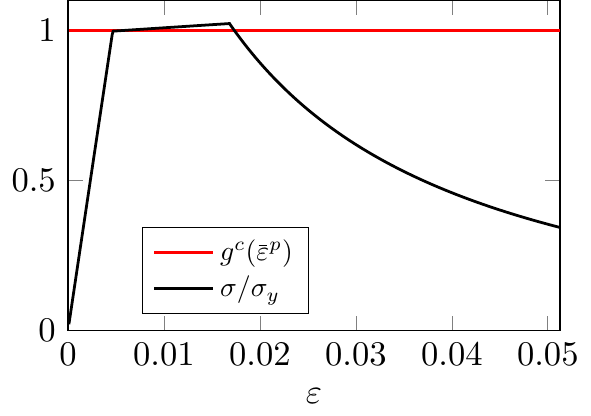}
    \caption{}
    \label{fig: example/homogenized/compare_beta/beta_1/stress_strain}
  \end{subfigure}
  \begin{subfigure}[b]{0.4\textwidth}
    \centering
    \tikzsetnextfilenamesafe{example/homogenized/compare_beta/beta_1/energy}
    \includegraphics{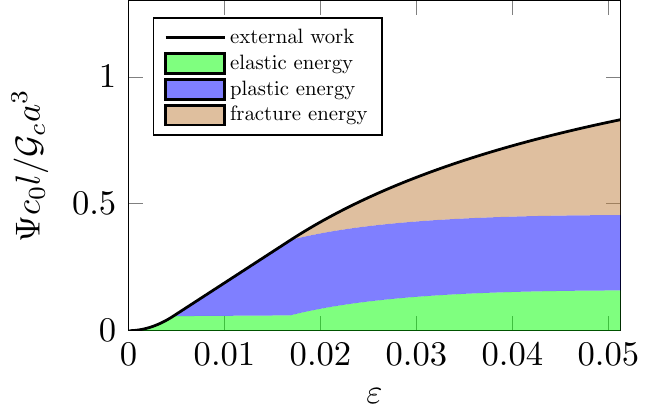}
    \vspace{-12pt}
    \caption{}
    \label{fig: example/homogenized/compare_beta/beta_1/energy}
  \end{subfigure}

  \begin{subfigure}[b]{0.4\textwidth}
    \centering
    \tikzsetnextfilenamesafe{example/homogenized/compare_beta/beta_0.5/stress_strain}
    \includegraphics{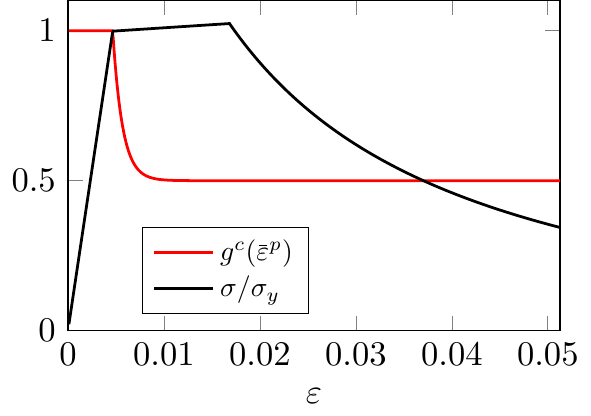}
    \caption{}
    \label{fig: example/homogenized/compare_beta/beta_0.5/stress_strain}
  \end{subfigure}
  \begin{subfigure}[b]{0.4\textwidth}
    \centering
    \tikzsetnextfilenamesafe{example/homogenized/compare_beta/beta_0.5/energy}
    \includegraphics{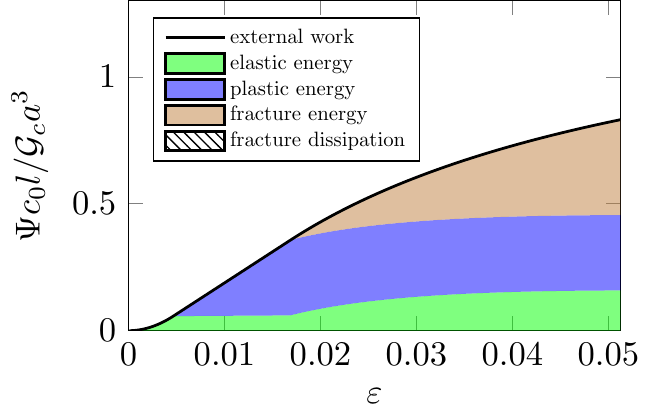}
    \vspace{-12pt}
    \caption{}
    \label{fig: example/homogenized/compare_beta/beta_0.5/energy}
  \end{subfigure}
  \caption{(a, c) the stress strain curve and the degradation for the fracture toughness and (b, d) the corresponding energy conservation. In (a-b), $C=\beta = 1$ and $\dfrac{\psi_cc_0a}{\Gc} = 0.0021$. In (c-d), $C=\beta=0.5$, $\varepsilon_0 = 0.001$, and $\dfrac{\psi_cc_0a}{\Gc} = 0.0021$. }
  \label{fig: example/homogenized/compare_beta}
\end{figure}

Second, the plastic energy contributes to the fracture driving energy in the E-P-PD model. Hence, by construction, the E-P-PD model results in fracture in regions with high plastic deformation, which agrees reasonably well with experimental observations. However, in many plasticity models, plastic flow occurs under both tension and compression. As a result the E-P-PD model can permit damage growth under compression. In contrast, with the E-P-D model, plastic energy does not contribute to the fracture driving energy.  This both precludes crack growth in regions that are flowing plastically under compression and also tends to permit the use of larger regularization lengths compared to the E-P-PD model.  The drawback is that, in tension, the E-P-D model tends to favor crack growth in regions with large elastic strains as opposed to large plastic strains.

In this work, the coalescence dissipation is proposed to address the aforementioned issues. According to \eqref{eq: fracture yield surface}, the coalescence dissipation introduces a dependency of the effective fracture toughness on the plastic strain.
To demonstrate the influence, the stress-strain curve and the corresponding energy balance when $C = \beta = 1$ (i.e.\ no coalesence dissipation) are first shown in \Cref{fig: example/homogenized/compare_beta/beta_1/stress_strain,fig: example/homogenized/compare_beta/beta_1/energy}.  Without any coalescence dissipation, there is no degradation of the fracture toughness.
The sum of the elastic energy, the plastic energy and the fracture energy equals the accumulated power expenditure due to external work.

\Cref{fig: example/homogenized/compare_beta/beta_0.5/stress_strain,fig: example/homogenized/compare_beta/beta_0.5/energy} show the stress strain curve and the corresponding energy balance when $C = \beta = 0.5$ and $\varepsilon_0 = 0.001 \ll \plasticstrain_c$.
In this case, $\Gc$ and $\psi_c$ are scaled such that $\beta\Gc$ and $\beta\psi_c$ remains the same for comparison purposes.
The degradation function $g^c(\plasticstrain)$ for the fracture toughness decreases as the plastic strain increases (\Cref{fig: example/homogenized/compare_beta/beta_0.5/stress_strain}), resulting in a final effective fracture toughness: $\lim_{\plasticstrain \to \infty} \widehat{\Gc} = \beta \Gc$.
The corresponding energy balance (\Cref{fig: example/homogenized/compare_beta/beta_0.5/energy}) is virtually indistinguishable from the case \textit{without} the fracture dissipation.

\begin{figure}[htb!]
  \centering
  \begin{subfigure}[b]{0.4\textwidth}
    \centering
    \tikzsetnextfilenamesafe{example/homogenized/compare_C_epsilon0/C/stress_strain}
    \includegraphics{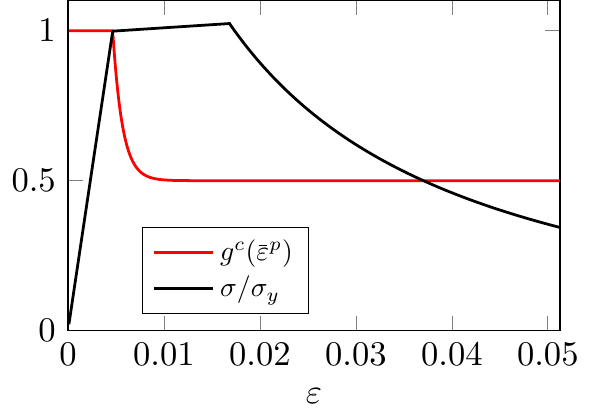}
    \caption{}
    \label{fig: example/homogenized/compare_C_epsilon0/C/stress_strain}
  \end{subfigure}
  \begin{subfigure}[b]{0.4\textwidth}
    \centering
    \tikzsetnextfilenamesafe{example/homogenized/compare_C_epsilon0/C/energy}
    \includegraphics{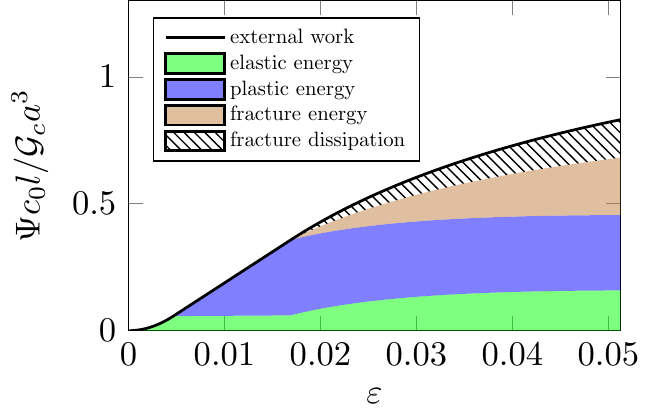}
    \vspace{-12pt}
    \caption{}
    \label{fig: example/homogenized/compare_C_epsilon0/C/energy}
  \end{subfigure}

  \begin{subfigure}[b]{0.4\textwidth}
    \centering
    \tikzsetnextfilenamesafe{example/homogenized/compare_C_epsilon0/epsilon0/stress_strain}
    \includegraphics{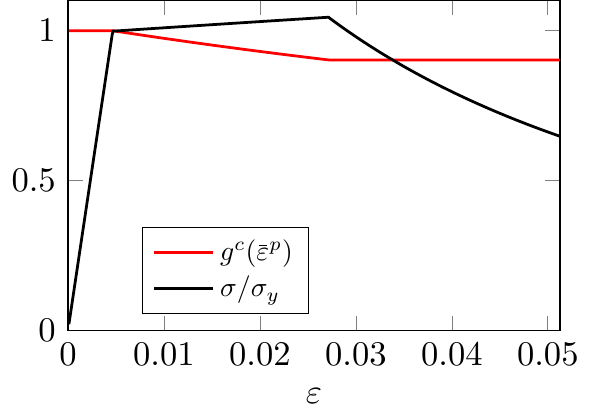}
    \caption{}
    \label{fig: example/homogenized/compare_C_epsilon0/epsilon0/stress_strain}
  \end{subfigure}
  \begin{subfigure}[b]{0.4\textwidth}
    \centering
    \tikzsetnextfilenamesafe{example/homogenized/compare_C_epsilon0/epsilon0/energy}
    \includegraphics{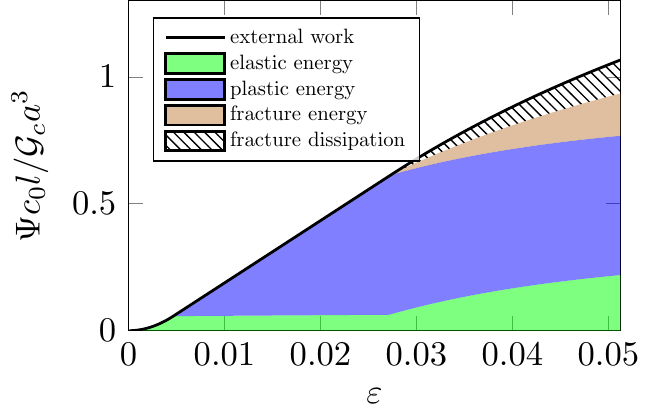}
    \vspace{-12pt}
    \caption{}
    \label{fig: example/homogenized/compare_C_epsilon0/epsilon0/energy}
  \end{subfigure}
  \caption{(a, c) the stress strain curve and the degradation for the fracture toughness and (b, d) the corresponding energy conservation. In (a-b), $C=0.3$, $\beta=0.5$, $\varepsilon=0.001$, and $\dfrac{\psi_cc_0a}{\Gc} = 0.0021$. In (c-d),  $C=\beta=0.5$, $\varepsilon_0 = 0.1$ and $\dfrac{\psi_cc_0a}{\Gc} = 0.0012$. }
  \label{fig: example/homogenized/compare_C_epsilon0}
\end{figure}

\begin{remark}[on the fracture dissipation]
  \vspace{-0.5em}
  Recall from \eqref{eq: dual fracture potential} that the fracture dissipation consists of the coalescence dissipation as well as the dissipative portion of the fracture process:
  \begin{align*}
    \lambda(\dot{d}, \plasticstrain; \beta, \varepsilon_0) = - (1-\beta)\dfrac{\Gc}{c_0l}\alpha_{,d}(d)\left( 1-e^{-\plasticstrain/\varepsilon_0} \right)\dot{d} + (1-C) \dfrac{\Gc}{c_0 l}\alpha_{, d}(d)\dot{d}.
  \end{align*}
  Before the onset of damage, i.e. $\dot{d} = 0$, the rate of the cumulative fracture dissipation density is zero. After damage initiation, if $\plasticstrain_c \gg \varepsilon_0$, and hence $e^{-\plasticstrain_c/\varepsilon_0} \approx 0$, the rate of the cumulative fracture dissipation density is approximately
  \begin{align*}
    \lambda(\dot{d}, \plasticstrain; \beta, \varepsilon_0) \approx (\beta - C) \dfrac{\Gc}{c_0l}\alpha_{,d}(d) \dot{d}.
  \end{align*}
  If $C = \beta$, the rate of the cumulative fracture dissipation density further reduces to
  \begin{align*}
    \lambda(\dot{d}, \plasticstrain; \beta, \varepsilon_0) \approx 0,
  \end{align*}
  and therefore the fracture dissipation shown in \Cref{fig: example/homogenized/compare_beta/beta_0.5/energy} is negligible.

  It follows immediately that a non-negligible amount of fracture dissipation will arise if $C < \beta$.
  In the other case, if the effective fracture toughness is not completely degraded by the plastic strain before damage initiation, i.e. $0 < e^{-\plasticstrain_c/\varepsilon_0} < 1$ and $g^c(\plasticstrain_c) > \beta$, it follows that
  \begin{align*}
    \lambda(\dot{d}, \plasticstrain; \beta, \varepsilon_0) & = - (1-\beta)\dfrac{\Gc}{c_0l}\alpha_{,d}(d)\left( 1-e^{-\plasticstrain/\varepsilon_0} \right)\dot{d} + (1-C) \dfrac{\Gc}{c_0 l}\alpha_{, d}(d)\dot{d}     \\
                                                           & > - (1-\beta)\dfrac{\Gc}{c_0l}\alpha_{,d}(d)\left( 1-e^{-\plasticstrain/\varepsilon_0} \right)\dot{d} + (1-\beta) \dfrac{\Gc}{c_0 l}\alpha_{, d}(d)\dot{d} \\
                                                           & = (1-\beta) \dfrac{\Gc}{c_0 l}\alpha_{, d}(d)\dot{d} e^{-\plasticstrain/\varepsilon_0} > 0.
  \end{align*}
\end{remark}

Following the remark, in the case of $C = 0.3$ and $\beta = 0.5$, the fracture dissipation in the energy conservation is of the same order of magnitude as other contributions, and is indicated using dashed lines (\Cref{fig: example/homogenized/compare_C_epsilon0/C/energy}).
It bears emphasis that the presence of the fracture dissipation does not alter the stress-strain response in this homogeneous setting (\Cref{fig: example/homogenized/compare_C_epsilon0/C/stress_strain}). On the other hand, using a larger $\varepsilon_0 = 0.1$ (together with a smaller critical fracture energy) also results in a non-negligible fracture dissipation (\Cref{fig: example/homogenized/compare_C_epsilon0/epsilon0/energy}).

\subsection{A nonhomogeneous example: uniaxial load-displacement curves}
\label{s: example/nonhomogeneous}

In the homogenized example (\Cref{s: example/homogenized}), we have demonstrated that the use of a linear crack geometric function enables an unperturbed elastic-plastic behavior, and that the Lorentz-type degradation functions lead to a regularization-length-independent critical fracture strength. In this section, we demonstrate that as a consequence of these modeling choices the underlying softening response is also independent of the regularization-length. In particular, the force-displacement response converges as the regularization-length is decreased.

\begin{figure}[htb!]
  \centering
  \begin{overpic}[scale=0.25]{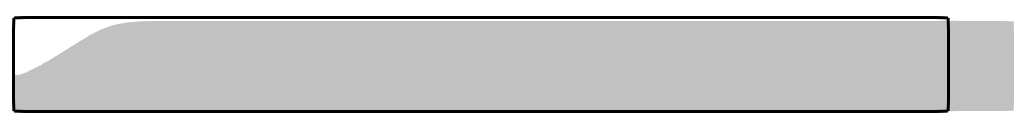}
    \put(45, 13){$L_0$}
    \put(-5, 5){$W_0$}
  \end{overpic}
  \caption{The geometry (black outline) and the final deformed configuration of the bar.}
  \label{fig: example/nonhomogeneous/schematics}
\end{figure}

Consider a two-dimensional domain $\body = [0, L_0] \times [0, W_0]$ with $W_0 / L_0 = 0.1$. The size of the bar is chosen such that no snap-back occurs in the force-displacement response for the values of regularization-length considered. The mesh is sufficiently refined such that the necking due to plastic flow is properly resolved. The domain is uniformly discretized with \texttt{QUAD4} elements. Plane-strain conditions are assumed to hold.

With an estimate of the plastic zone size $r_p = E\Gc/(3 \pi (1-\nu^2) \sigma_y^2)$, the dimensionless parameters are given as
\begin{align}
  \dfrac{W_0}{L_0} = 0.1, \quad \dfrac{\sigma_y}{E} = 0.00476, \quad \dfrac{H}{E} = 0.02, \quad \dfrac{L_0}{r_p} = 1.38, \quad \dfrac{\psi_c c_0 L_0}{\Gc} =
  \begin{cases}
    0.4, \quad \text{E-P-D} \\
    4, \quad \text{E-P-PD}
  \end{cases}.
\end{align}

The critical fracture energy $\psi_c$ is chosen to allow for plastic deformation prior to damage softening. A sequence of values of the phase-field regularization-length are investigated to demonstrate convergence. The mesh is uniformly refined such that $l/h = 8$ remains constant and the phase-field bandwidth is always sufficiently resolved.

\begin{figure}[htb!]
  \centering
  \begin{subfigure}{0.45\textwidth}
    \centering
    \tikzsetnextfilenamesafe{example/nonhomogeneous/compare_l/EPPD}
    \includegraphics{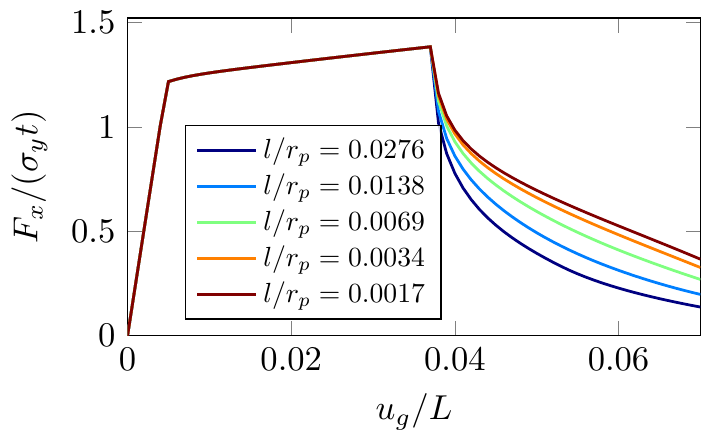}
    \caption{E-P-PD}
    \label{fig: example/nonhomogeneous/compare_l/EPPD}
  \end{subfigure}
  \begin{subfigure}{0.45\textwidth}
    \centering
    \tikzsetnextfilenamesafe{example/nonhomogeneous/compare_l/EPD_beta_1}
    \includegraphics{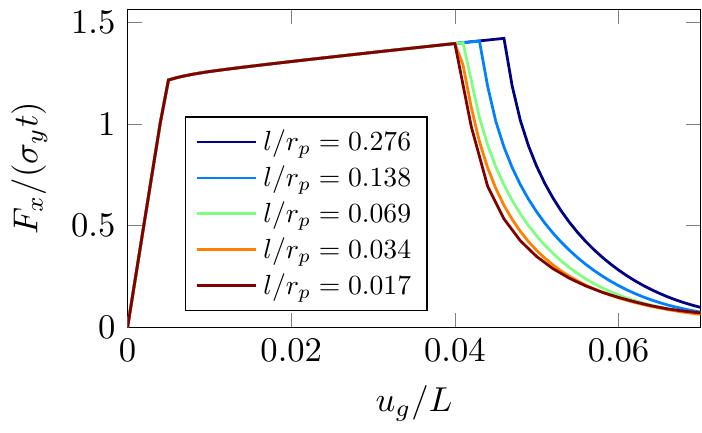}
    \caption{E-P-D with $\beta = 1$}
    \label{fig: example/nonhomogeneous/compare_l/EPD_beta_1}
  \end{subfigure}

  \begin{subfigure}{0.45\textwidth}
    \centering
    \tikzsetnextfilenamesafe{example/nonhomogeneous/compare_l/EPD_beta_0.8}
    \includegraphics{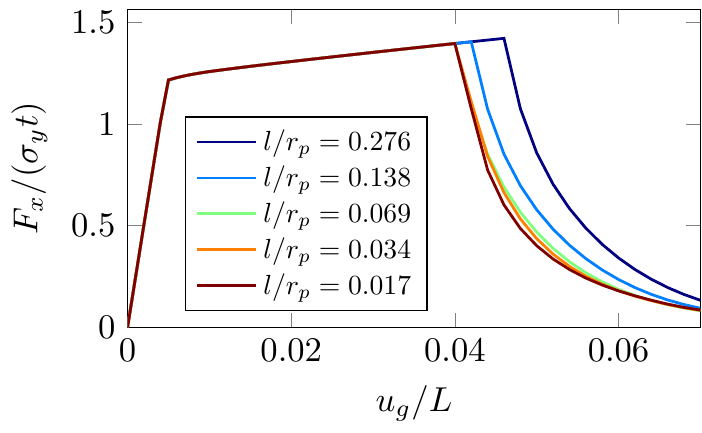}
    \caption{E-P-D with $\beta = 0.8$}
    \label{fig: example/nonhomogeneous/compare_l/EPD_beta_0.8}
  \end{subfigure}
  \begin{subfigure}{0.45\textwidth}
    \centering
    \tikzsetnextfilenamesafe{example/nonhomogeneous/compare_l/EPD_beta_0.6}
    \includegraphics{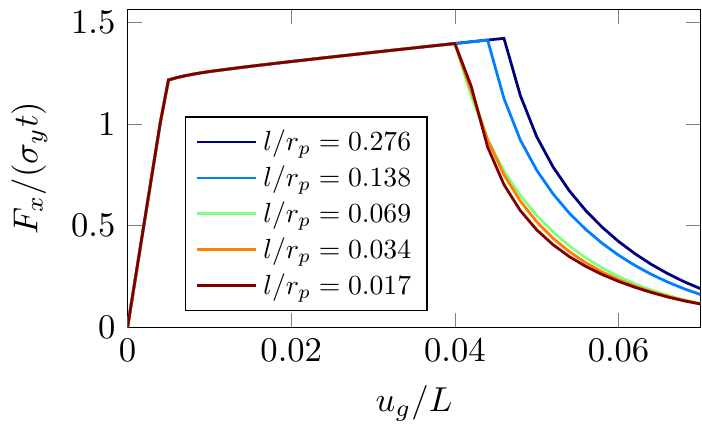}
    \caption{E-P-D with $\beta = 0.6$}
    \label{fig: example/nonhomogeneous/compare_l/EPD_beta_0.6}
  \end{subfigure}
  \caption{Normalized load-displacement curves obtained for (a) the E-P-PD model and (b-d) the E-P-D model with different values of $\beta$. Unit thickness is assumed.}
  \label{fig: example/nonhomogeneous/compare_l}
\end{figure}

The bar is placed in tension through a prescribed horizontal displacement at the right end. Symmetry conditions are applied along the left and the bottom of the bar.  To summarize, the boundary conditions on the displacement field are
\begin{align*}
  u_x(x = 0) = 0, \quad u_x(x = L_0) = u_g, \quad u_y(y = 0) = 0.
\end{align*}
The boundary conditions are illustrated in \Cref{fig: example/nonhomogeneous/schematics} through the difference between the reference and final configurations. A numerical perturbation on the order of machine precision is introduced at $x=0$ to ensure localization at the left end. The total reaction force $F_x$ on the right edge is plotted as a function of the prescribed displacement $u_g$ (\Cref{fig: example/nonhomogeneous/compare_l}).

The convergence of the E-P-PD model and the E-P-D model are illustrated in \Cref{fig: example/nonhomogeneous/compare_l/EPPD,fig: example/nonhomogeneous/compare_l/EPD_beta_1}.
Two additional sets of load-displacement curves are generated for $\beta = 0.8$ and $\beta = 0.6$ in \Cref{fig: example/nonhomogeneous/compare_l/EPD_beta_0.8,fig: example/nonhomogeneous/compare_l/EPD_beta_0.6}. Once the regularization length is sufficiently small, the curves are virtually indistinguishable.

\subsection{Three-point bending}
\label{s: example/3pb}

With modeling choices motivated and demonstrated in \Cref{s: example/homogenized,s: example/nonhomogeneous}, the model is now applied to simulate a three-point bending experiment for a specimen composed of an aluminum alloy \cite{kubik2019ductile}. In particular, we focus attention on the comparison of the E-P-PD model and the proposed E-P-D model with coalescence dissipation.

\begin{figure}[!htb]
  \centering
  \tikzsetnextfilenamesafe{example/3pb/mesh}
  \includegraphics{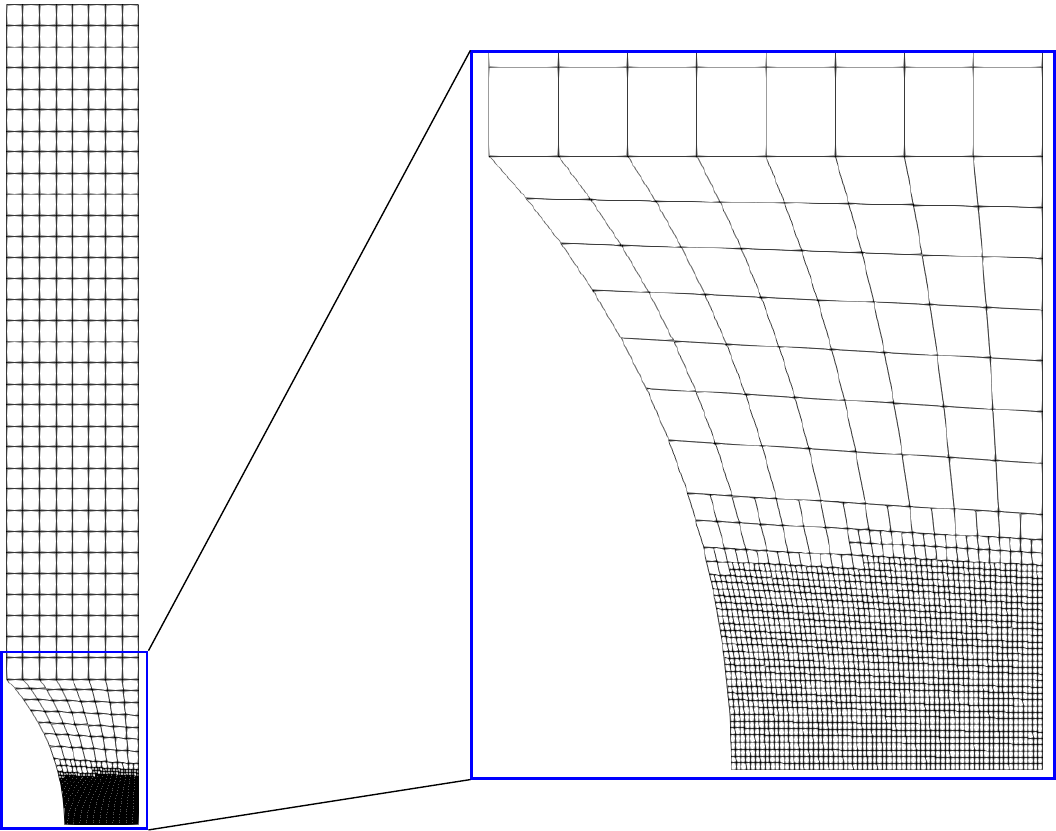}
  \caption{The mesh used for calibration and the zoomed-in view of the refined region.}
  \label{fig: example/3pb/mesh}
\end{figure}

First, material properties and model parameters are calibrated against a tension test of a notched cylindrical specimen (R13 notch). Details of the specimen geometry, experimental setup and force-displacement data are provided in \cite{kubik2018notched}. Our calculations employ a cylindrical coordinate system, and symmetry is assumed about both the centerline and the half-plane of the specimen. The finite element mesh for this problem is shown in \Cref{fig: example/3pb/mesh}. \texttt{QUAD4} elements are used to discretize the domain with a structured mesh. The maximum and the minumum element sizes are \SI{1}{\milli\meter} and \SI{0.0625}{\milli\meter}, respectively.

\begin{figure}[htb!]
  \centering
  \begin{subfigure}{0.3\textwidth}
    \centering
    \tikzsetnextfilenamesafe{example/3pb/force_disp/EPPD}
    \includegraphics{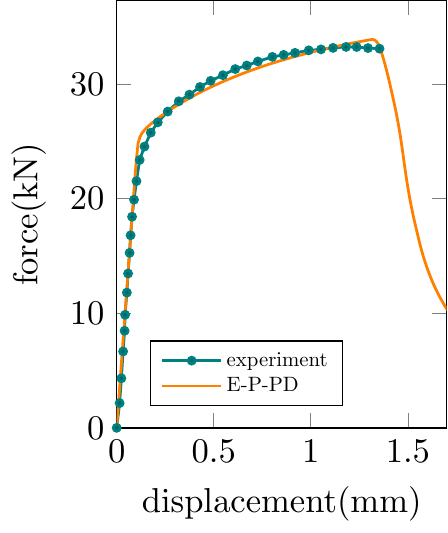}
    \caption{E-P-PD}
  \end{subfigure}
  \begin{subfigure}{0.3\textwidth}
    \centering
    \tikzsetnextfilenamesafe{example/3pb/force_disp/EPD_beta}
    \includegraphics{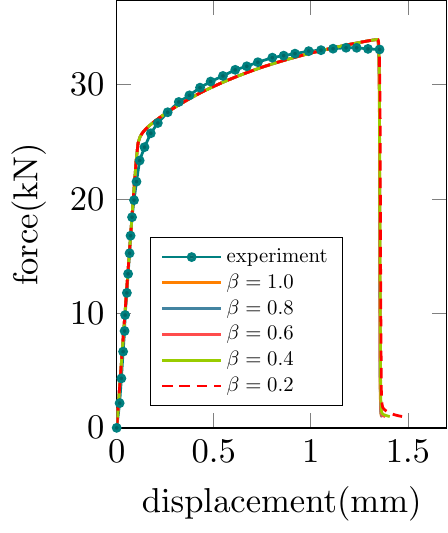}
    \caption{E-P-D with $\varepsilon_0 = 0.05$}
    \label{fig: example/3pb/force_disp/EPD_beta}
  \end{subfigure}
  \begin{subfigure}{0.3\textwidth}
    \centering
    \tikzsetnextfilenamesafe{example/3pb/force_disp/EPD_e0}
    \includegraphics{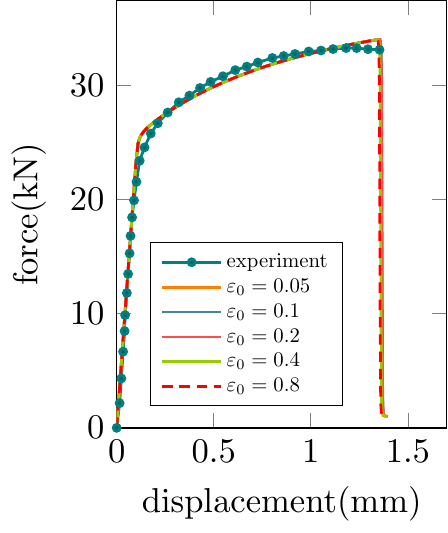}
    \caption{E-P-D with $\beta = 0.5$}
    \label{fig: example/3pb/force_disp/EPD_e0}
  \end{subfigure}
  \caption{Experimental and calibrated force-displacement curves for the cylindrical notched tension test. Experiment data are interpolated from \cite{kubik2018notched}. The calibrated curves correspond to material properties and model parameters summarized in \Cref{tab: example/3pb}.}
  \label{fig: example/3pb/force_disp}
\end{figure}

Calibrated material properties and model parameters are summarized in \Cref{tab: example/3pb}. The E-P-D model is calibrated using a range of $\beta$ and $\varepsilon_0$ values. The corresponding force-displacement curves are shown in \Cref{fig: example/3pb/force_disp}.

We note that a significantly larger critical fracture energy $\psi_c$ is required for the E-P-PD model for the onset of damage to match the experiment. This is expected as the generalized fracture driving energy for the E-P-PD model includes contributions from plasticity, while only the active elastic energy contributes to crack growth in the E-P-D model. According to \eqref{eq: upper bound}, polyconvexity requires a phase-field regularization-length $l \leq \SI{0.01}{\milli\meter}$.

\begin{table}[htb!]
  \centering
  \caption{Summary of the calibrated material properties and model parameters}
  \begin{tabular}{r | c | c | c | c}
    \toprule
    Property/Parameter            & Symbol          & Value(E-P-D)                   & Value(E-P-PD)              & Unit                                       \\
    \midrule
    Young's modulus               & $E$             & \SI{7.25e4}{}                  & \SI{7.25e4}{}              & \SI{}{\mega\pascal}                        \\
    Poisson's ratio               & $\nu$           & 0.33                           & 0.33                       & nondim.                                    \\
    \midrule
    Yield stress                  & $\sigma_y$      & 345                            & 345                        & \SI{}{\mega\pascal}                        \\
    Hardening modulus             & $h$             & 1000                           & 1000                       & \SI{}{\mega\pascal}                        \\
    \midrule
    Fracture toughness            & $\widehat{\Gc}$ & 20                             & 20                         & \SI{}{\milli\joule\per\square\milli\meter} \\
    crack geometric coefficient   & $\xi$           & 1                              & 1                          & nondim.                                    \\
    Critical fracture energy      & $\psi_c$        & 3.7                            & 205                        & \SI{}{\milli\joule\per\cubic\milli\meter}  \\
    Regularization length         & $l$             & 0.1                            & 0.01                       & \SI{}{\milli\meter}                        \\
    \midrule
    Interaction coefficient       & $\beta$         & \{0.1, 0.2, 0.4, 0.6, 0.8, 1\} & 1                          & nondim.                                    \\
    Characteristic plastic strain & $\varepsilon_0$ & \{0.05, 0.1, 0.2, 0.4, 0.8\}   & -                          & nondim.                                    \\
    Elastic degradation function  & $g^e$           & $g^\text{Lorentz}(d; p=1)$     & $g^\text{Lorentz}(d; p=1)$ & nondim.                                    \\
    Plastic degradation function  & $g^p$           & 1                              & $g^\text{Lorentz}(d; p=1)$ & nondim.                                    \\
    \bottomrule
  \end{tabular}
  \label{tab: example/3pb}
\end{table}

\begin{figure}[!htb]
  \centering
  \begin{subfigure}[b]{0.9\textwidth}
    \centering
    \includegraphics[width=\textwidth,scale=0.5]{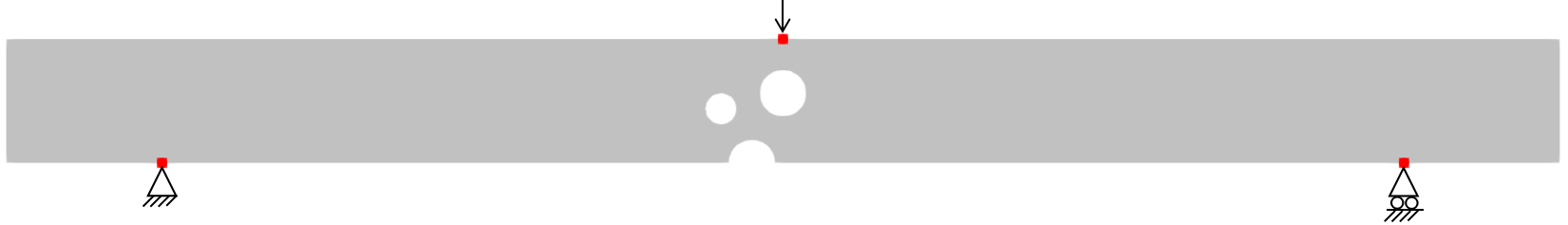}
    \caption{}
  \end{subfigure}
  \begin{subfigure}[b]{0.9\textwidth}
    \centering
    \includegraphics[width=\textwidth,scale=0.5]{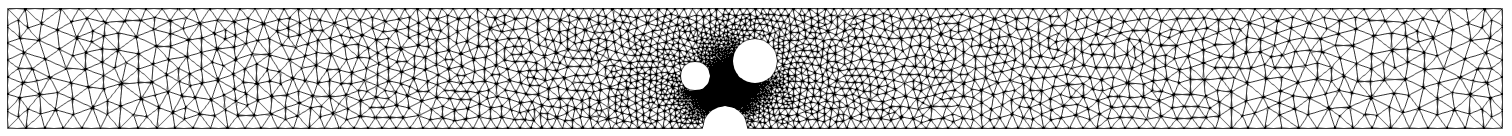}
    \caption{}
    \label{fig: example/3pb/3pb_mesh}
  \end{subfigure}
  \caption{(a) schematics of the three-point bending experiment and (b) the mesh. See \cite{kubik2019ductile} for a detailed drawing with dimensions.}
  \label{fig: example/3pb/3pb}
\end{figure}

The calibrated E-P-D and E-P-PD models are now applied to simulate the three-point bending experiment with the same material. The specimen has a cylindrical notch on the bottom and two cylindrical holes around the middle of the specimen. The two holes are asymmetric about the mid-plane in terms of size and position. Details of the specimen geometry and experimental set up are available in \cite{kubik2019ductile}. Schematics of the boundary conditions and the mesh are shown in \Cref{fig: example/3pb/3pb}.

In the experiment, the crack surface first originates from the half cylindrical notch on the bottom of the specimen, and then grows to connect the notch to the bottom of the left smaller cylindrical hole. After a while, the second crack surface originates from the top of the left smaller hole and grows towards the middle of the upper free surface.

\subsubsection{Two-dimensional results}
\label{s: example/3pb/2D}

\begin{figure}[!htb]
  \centering
  \begin{subfigure}[b]{0.3\textwidth}
    \centering
    \includegraphics[width=\textwidth,scale=0.5]{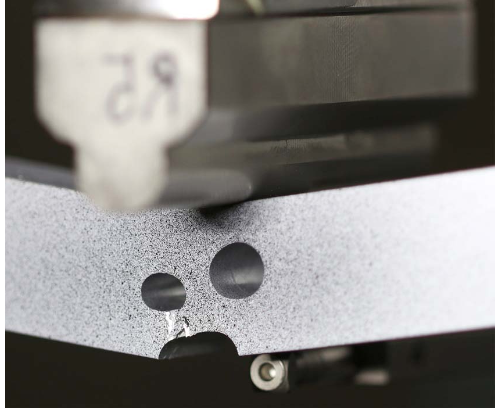}
    \caption{}
  \end{subfigure}
  \hspace{0.1\textwidth}
  \begin{subfigure}[b]{0.35\textwidth}
    \centering
    \includegraphics[width=\textwidth,scale=0.5]{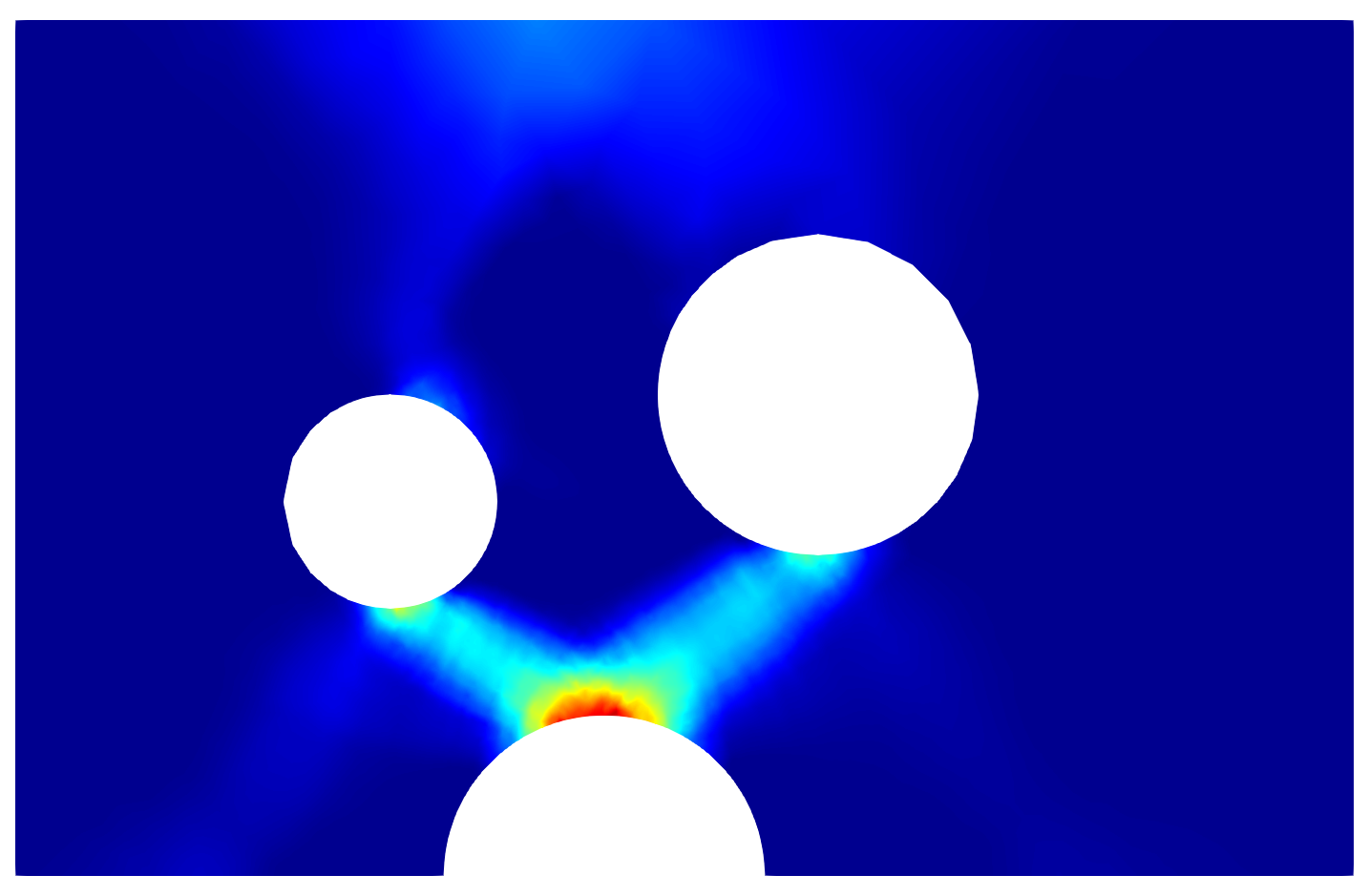}
    \caption{}
  \end{subfigure}
  \begin{subfigure}[b]{0.06\textwidth}
    \centering
    \caption*{$\plasticstrain$}
    \includegraphics[width=\textwidth,scale=0.5]{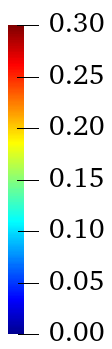}
    \vspace{1em}
  \end{subfigure}
  \label{fig: example/3pb/plastic_strain}
  \caption{(a) a picture of the experiment setup and (b) the effective plastic strain prior to crack initiation.
  }
\end{figure}

First, we focus attention on the initiation and propagation of the first crack, i.e.\ the crack connecting the notch to the left smaller hole. The calculations are performed in two dimensions under plane-strain assumptions. The experimental crack path and the contours of the equivalent plastic strain are shown in \Cref{fig: example/3pb/plastic_strain}. Due to the nature of the linear crack geometric function, both the E-P-PD model and the E-P-D model has the same ductile response up to damage initiation: the equivalent plastic strain localizes around the middle of the bottom cylindrical notch, and two wakes of plastic deformation form at $\ang{45}$ angle towards the lower surfaces of the left smaller cylindrical hole and the right larger cylindrical hole.

It bears emphasis that this three-point bending experiment is particularly challenging for most E-P-D models because the crack trajectories will be qualitatively different in a brittle material and in a ductile material, and it is of common concern that a loosely coupled E-P-D model may fail to predict the correct crack trajectory. In fact, we show that it is important to include the coalescence dissipation to predict the correct crack path.

\begin{figure}[!htb]
  \centering
  \begin{subfigure}[b]{0.3\textwidth}
    \centering
    \includegraphics[width=\textwidth,scale=0.5]{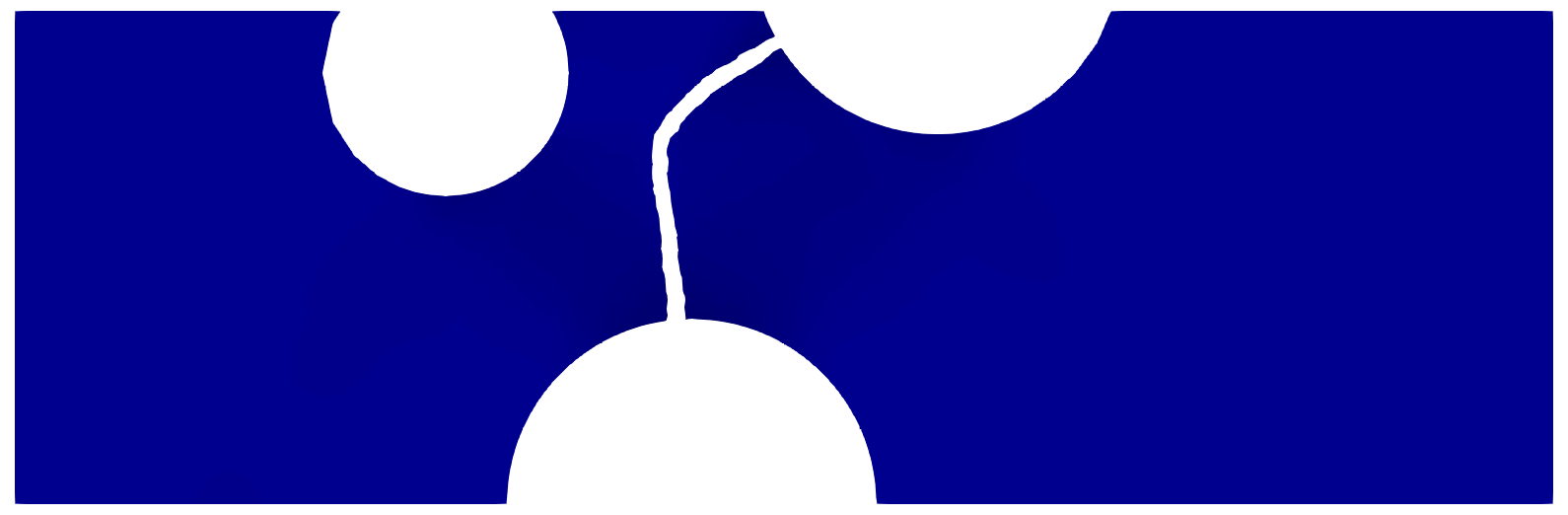}
    \caption{$\varepsilon_0=0.8$}
  \end{subfigure}
  \begin{subfigure}[b]{0.3\textwidth}
    \centering
    \includegraphics[width=\textwidth,scale=0.5]{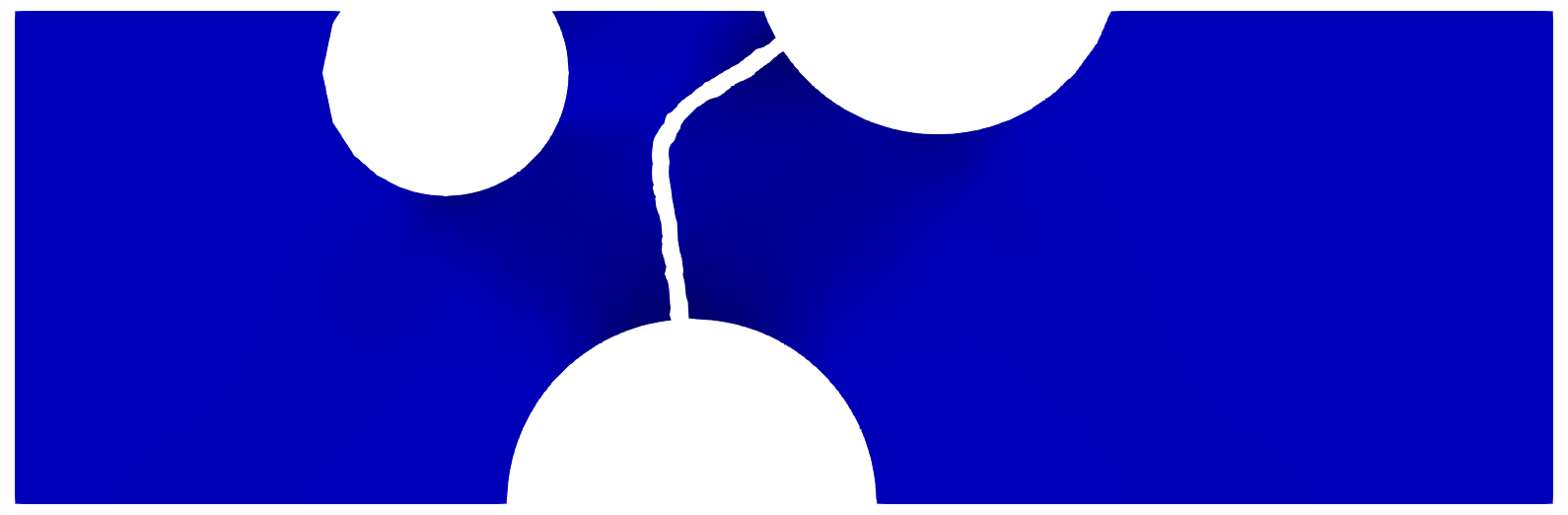}
    \caption{$\varepsilon_0=0.4$}
  \end{subfigure}
  \begin{subfigure}[b]{0.3\textwidth}
    \centering
    \includegraphics[width=\textwidth,scale=0.5]{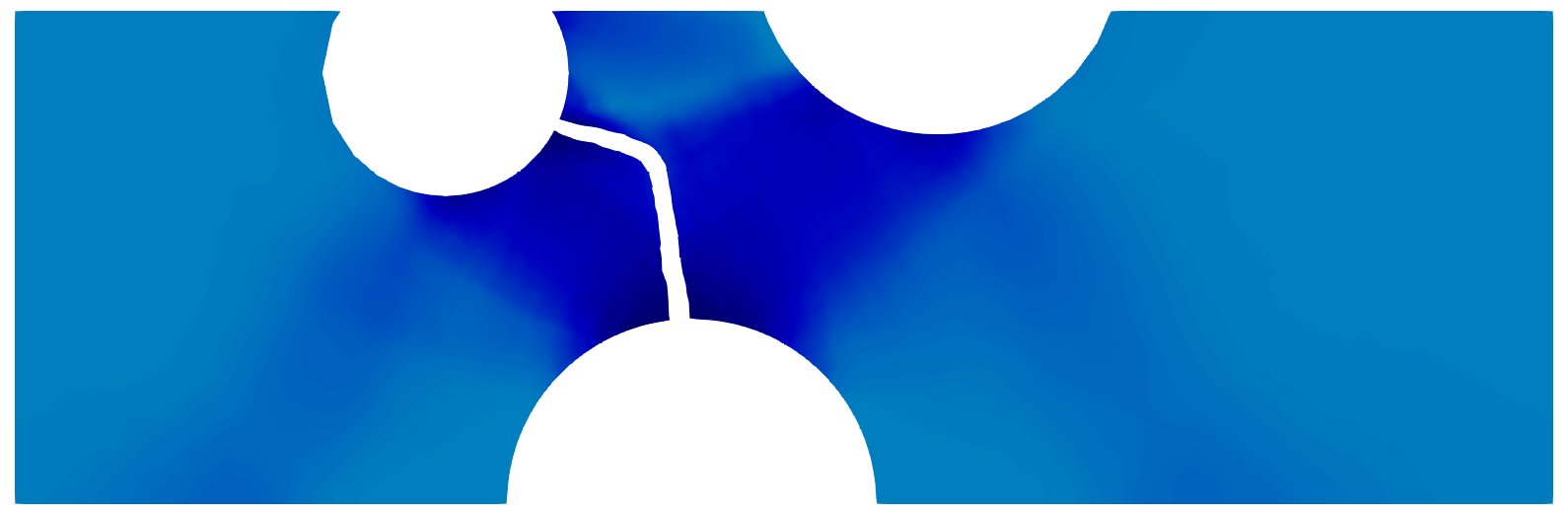}
    \caption{$\varepsilon_0=0.2$}
  \end{subfigure}

  \begin{subfigure}[b]{0.3\textwidth}
    \centering
    \includegraphics[width=\textwidth,scale=0.5]{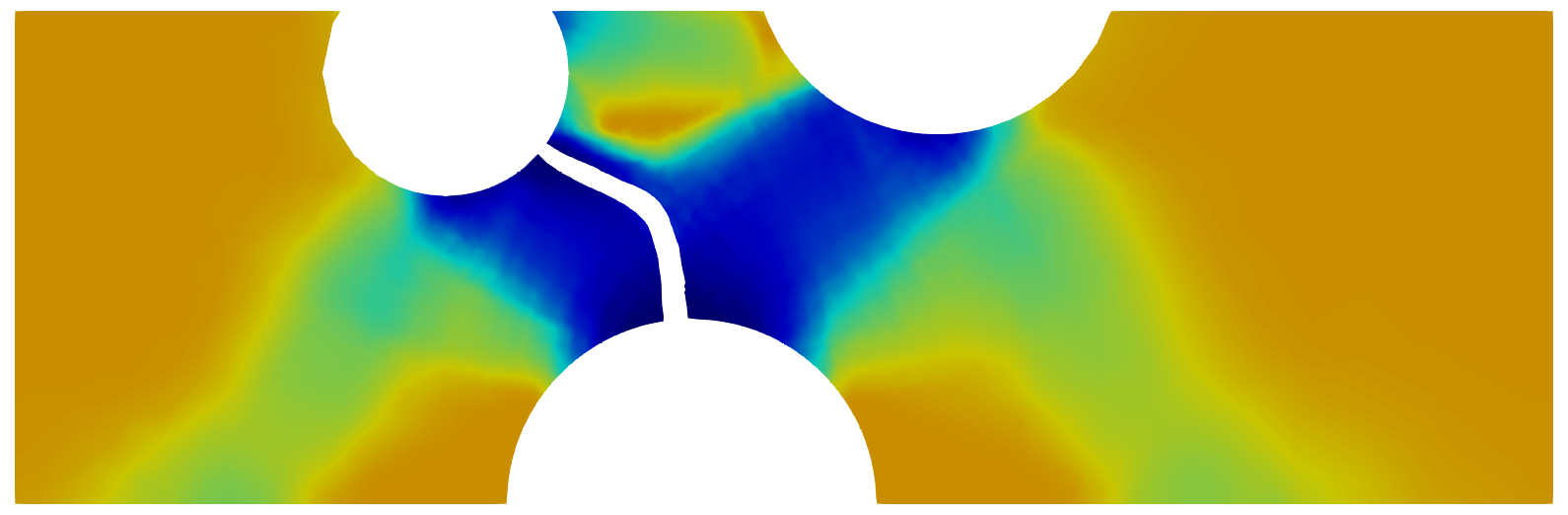}
    \caption{$\varepsilon_0=0.1$}
  \end{subfigure}
  \begin{subfigure}[b]{0.3\textwidth}
    \centering
    \includegraphics[width=\textwidth,scale=0.5]{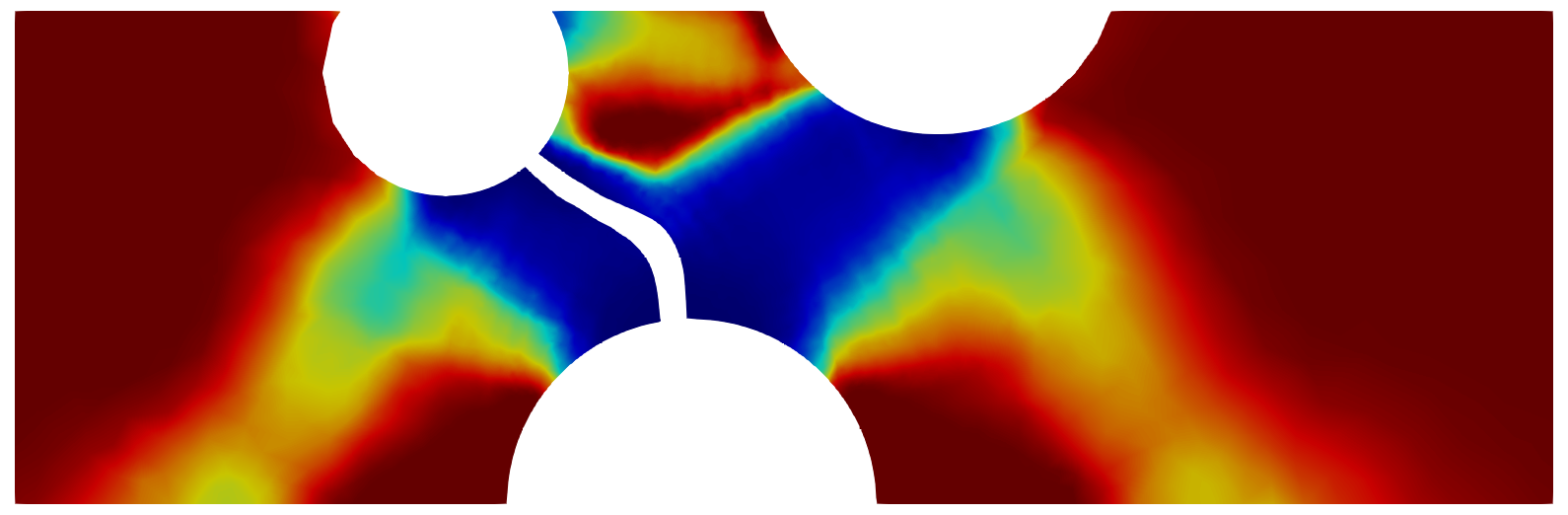}
    \caption{$\varepsilon_0=0.05$}
  \end{subfigure}
  \begin{subfigure}[b]{0.3\textwidth}
    \centering
    \includegraphics[width=\textwidth,scale=0.5]{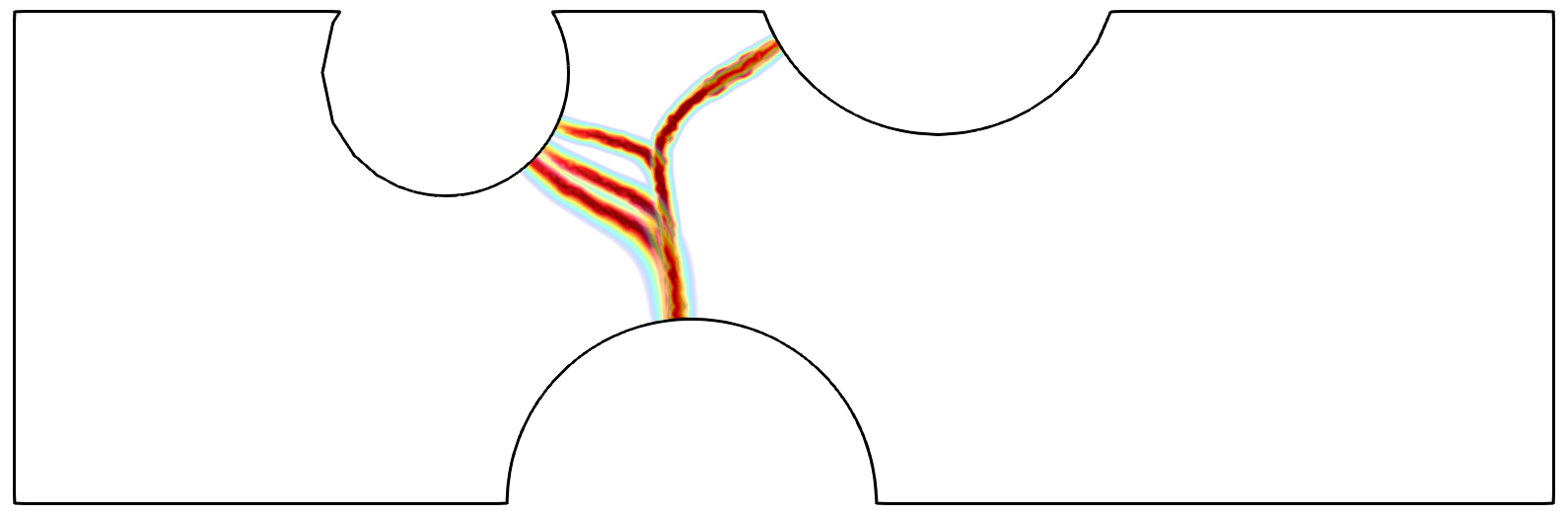}
    \caption{}
  \end{subfigure}
  \caption{(a-e) Contour plots of the effective fracture toughness $\widehat{\Gc}$ with $\beta=0.1$ and different values of $\varepsilon_0$, using the E-P-D model. The portion of the domain with $d < 0.8$ is removed to visualize the crack path. (f) Superposition of the damage contours. All contours are plotted on the reference configuration for the purpose of comparison.}
  \label{fig: example/3pb/2D_comparison_constant_beta}
\end{figure}

\begin{figure}[!htb]
  \centering
  \begin{subfigure}[b]{0.3\textwidth}
    \centering
    \includegraphics[width=\textwidth,scale=0.5]{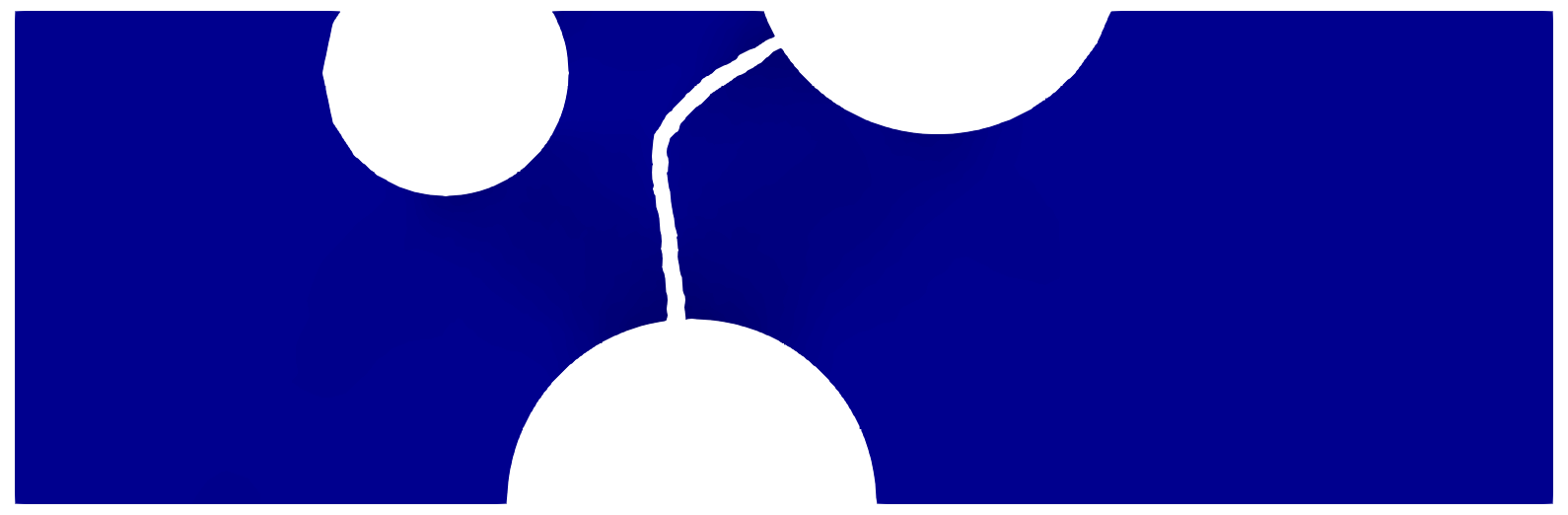}
    \caption{$\beta=1$}
  \end{subfigure}
  \begin{subfigure}[b]{0.3\textwidth}
    \centering
    \includegraphics[width=\textwidth,scale=0.5]{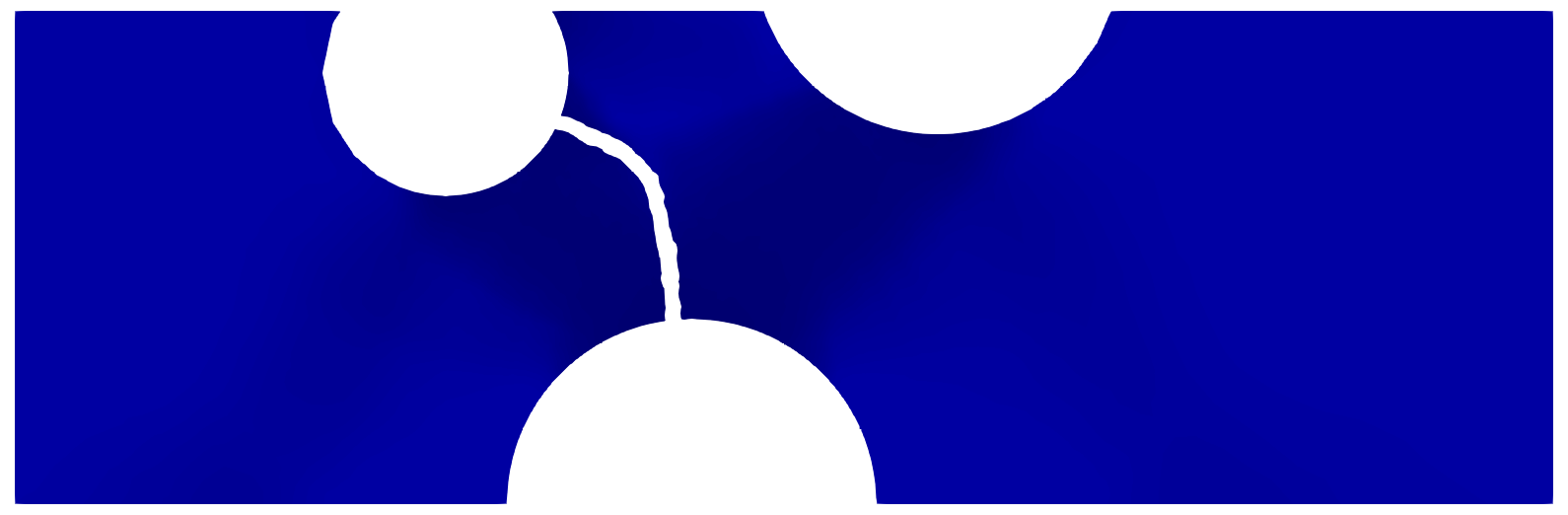}
    \caption{$\beta=0.8$}
  \end{subfigure}
  \begin{subfigure}[b]{0.3\textwidth}
    \centering
    \includegraphics[width=\textwidth,scale=0.5]{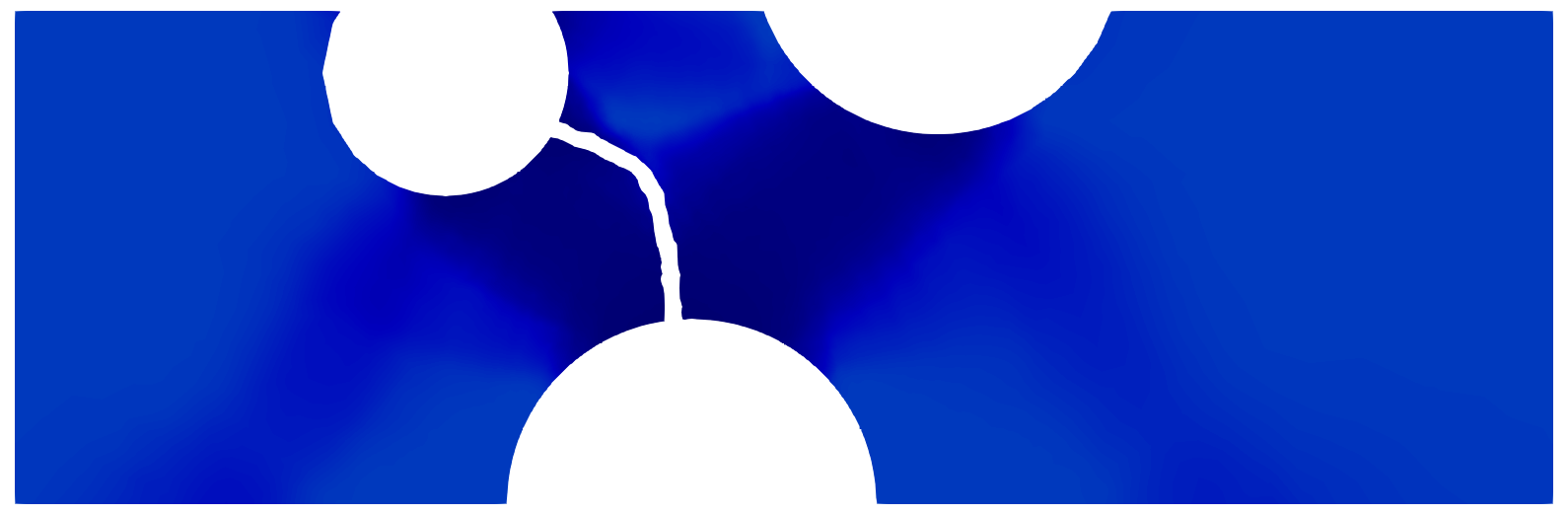}
    \caption{$\beta=0.6$}
  \end{subfigure}

  \begin{subfigure}[b]{0.3\textwidth}
    \centering
    \includegraphics[width=\textwidth,scale=0.5]{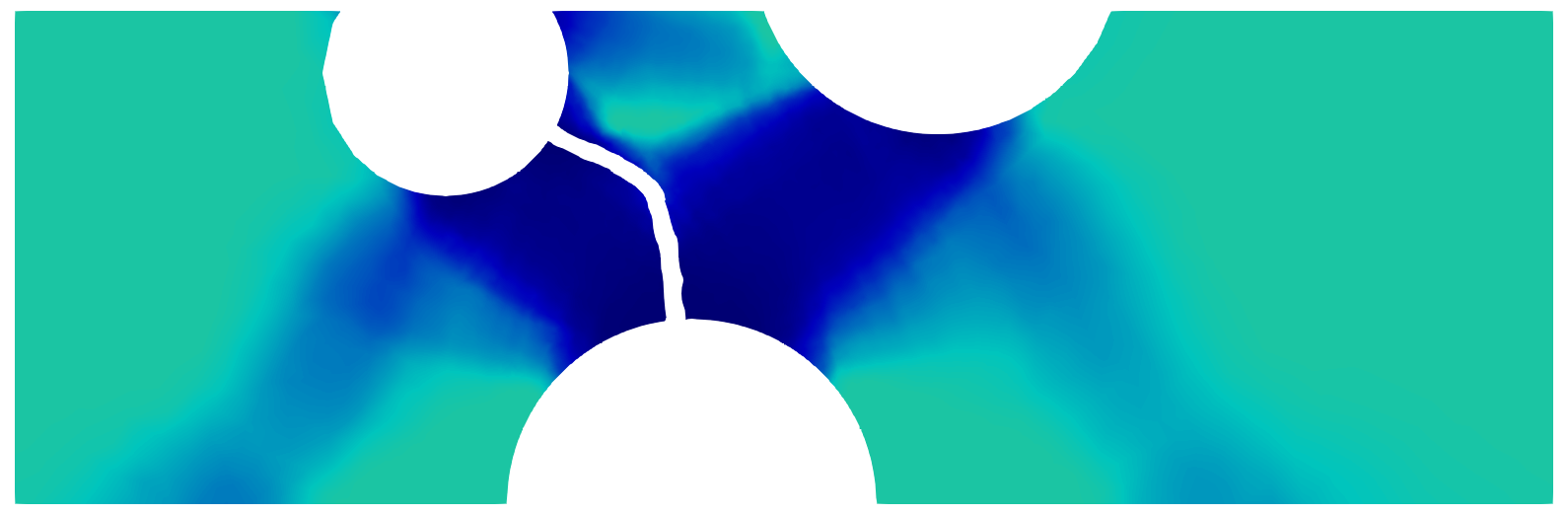}
    \caption{$\beta=0.4$}
  \end{subfigure}
  \begin{subfigure}[b]{0.3\textwidth}
    \centering
    \includegraphics[width=\textwidth,scale=0.5]{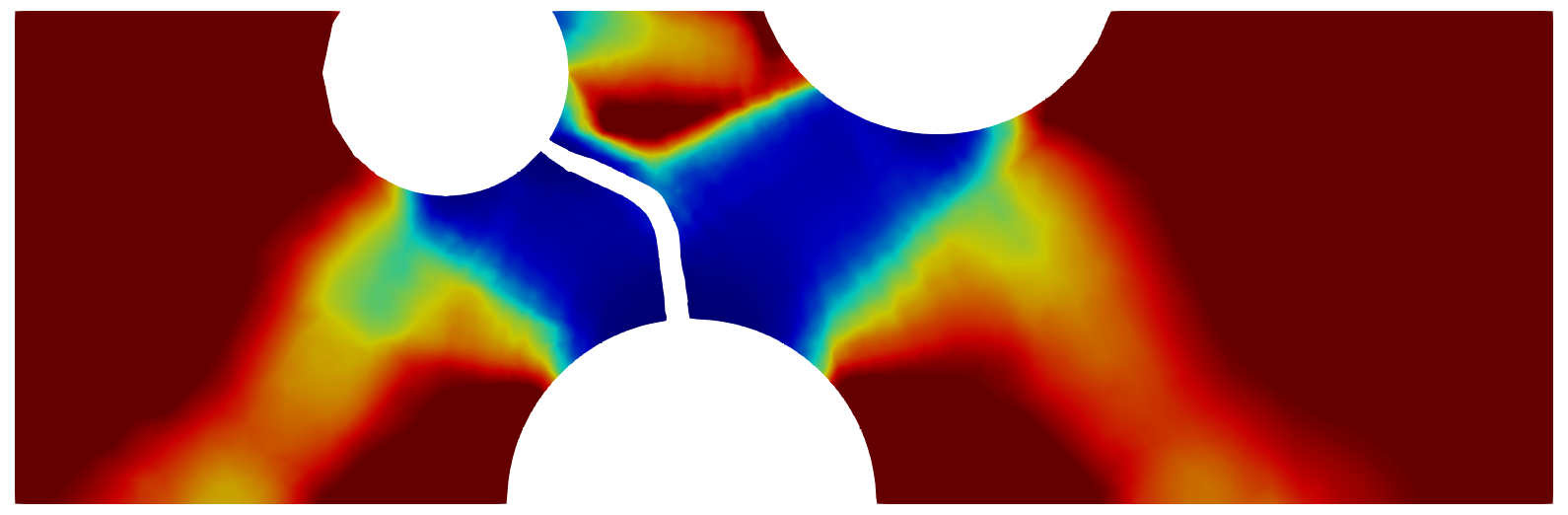}
    \caption{$\beta=0.2$}
  \end{subfigure}
  \begin{subfigure}[b]{0.3\textwidth}
    \centering
    \includegraphics[width=\textwidth,scale=0.5]{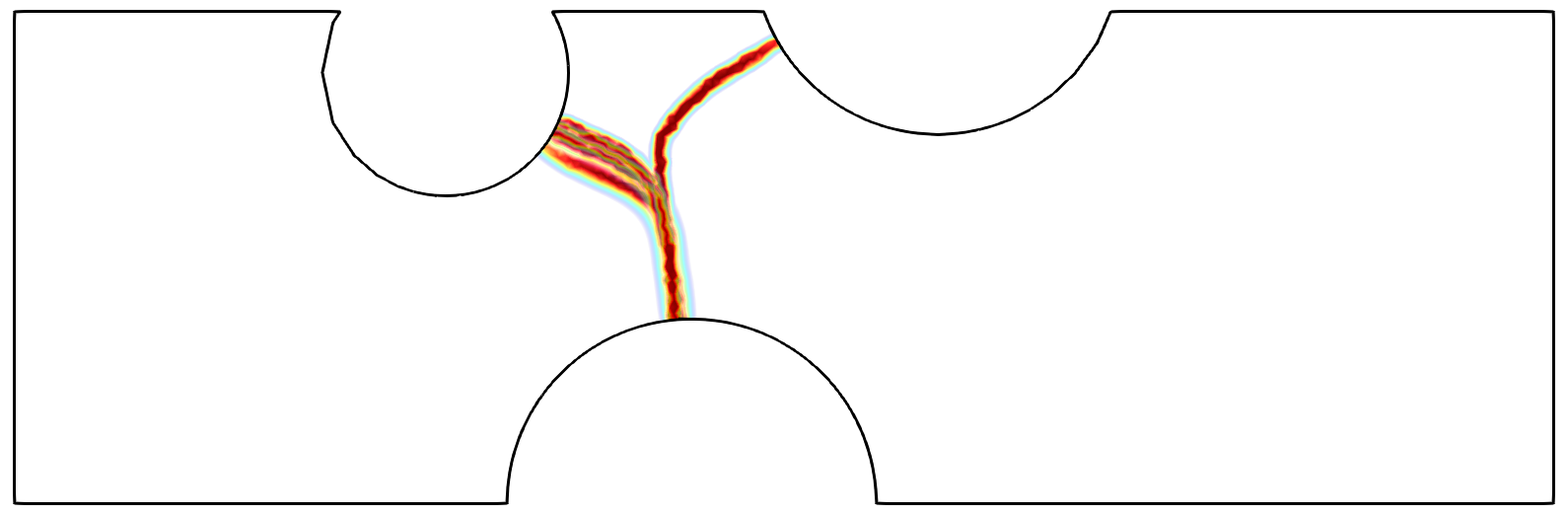}
    \caption{}
  \end{subfigure}
  \caption{(a-e) Contour plots of the effective fracture toughness $\widehat{\Gc}$ with $\varepsilon_0=0.05$ and different values of $\beta$. Domain with $d < 0.8$ is removed to visualize the crack path. (f) Superposition of the damage contours. All contours are plotted on the reference configuration for the purpose of comparision.}
  \label{fig: example/3pb/2D_comparison_constant_e0}
\end{figure}

After damage initiation, crack paths predicted by the E-P-D model and the E-P-PD model differ. For the E-P-D model, the crack paths depends on the parameters chosen for the coalescence dissipation. The effective fracture toughness $\widehat{\Gc}(\plasticstrain;\beta, \varepsilon_0)$ decreases as the plastic strain $\plasticstrain$ increases following a decay characterized by $\beta$ and $\varepsilon_0$. In other words, the spatial variation of the effective fracture toughness conforms with the localization of the plastic strain.
The effect of $\beta$ and $\varepsilon_0$ are illustrated in \Cref{fig: example/3pb/2D_comparison_constant_beta,fig: example/3pb/2D_comparison_constant_e0}.

In \Cref{fig: example/3pb/2D_comparison_constant_beta}, $\beta$ is kept constant and the sensitivity of the fracture path to $\varepsilon_0$ is examined. Smaller values of $\varepsilon_0$ result in faster decay of the fracture toughness as the plastic strain increases. As a result, given the same amount of plastic strain, a smaller $\varepsilon_0$ leads to a higher spatial variation or ``contrast'' in the effective fracture toughness. When $\varepsilon_0 = 0.8$, the contrast is negligible, the effective fracture toughness is approximately homogeneous, and the crack propagates towards the right larger cylindrical hole. When $\varepsilon_0=0.05$, the contrast is high, the fracture toughness is degraded in regions with large plastic strain, and therefore the crack path largely follows the localization of the plastic strain.

Similar arguments hold in \Cref{fig: example/3pb/2D_comparison_constant_e0} where $\varepsilon_0$ is kept constant and the sensitivity of the fracture path to changes in $\beta$ is explored. When $\beta$ is small, the spatial ``contrast'' of the effective fracture toughness is high, and the crack path is affected by the localization of the plastic strain.

\begin{figure}[htb!]
  \centering
  \includegraphics[width=0.3\textwidth,scale=0.5]{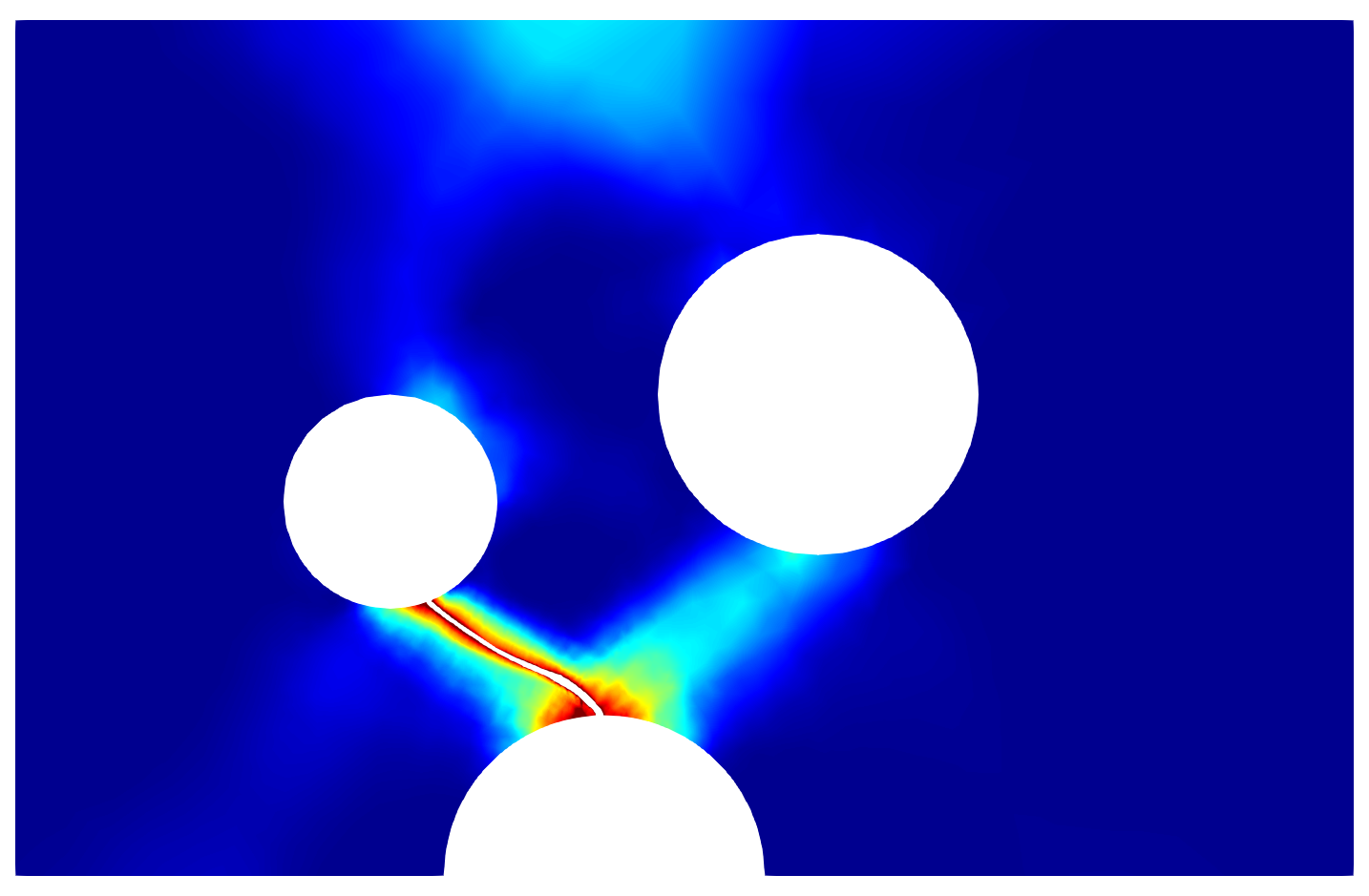}
  \caption{Contour of the equivalent plastic strain obtained using the E-P-PD model. Domain with $d > 0.8$ is removed to visualize the crack path.}
  \label{fig: tpb-2d-eppd}
\end{figure}

For the E-P-PD model, due to the strong coupling between plasticity and fracture, the crack path follows the localization of the plastic strain, and the plastic strain keeps increasing around the crack surface.  The resulting equivalent plastic strain and crack path in this case are shown in \Cref{fig: tpb-2d-eppd}.

\begin{remark}
  \vspace{-0.5em}
  In order to properly resolve the phase-field regularization length $l = \SI{0.01}{\milli\meter}$ for the E-P-PD model, a fairly small element size has to be used, resulting in a system with approximately 12 million degrees of freedom for this two-dimensional problem. In contrast, only 24300 degrees of freedom were used for the E-P-D model.
\end{remark}

\subsubsection{Three-dimensional results}
\label{s: example/3pb/3D}

With a qualitative understanding of the predicted trajectory of the first crack, a subset of the calibrated E-P-D models are now used in three-dimensional simulations of the three-point bending experiment.   The calculated load-deflection curves are compared to experimental measurements.

\begin{figure}[htb!]
  \centering
  \begin{subfigure}{0.45\textwidth}
    \centering
    \tikzsetnextfilenamesafe{example/3pb/load_deflection/constant_beta}
    \includegraphics{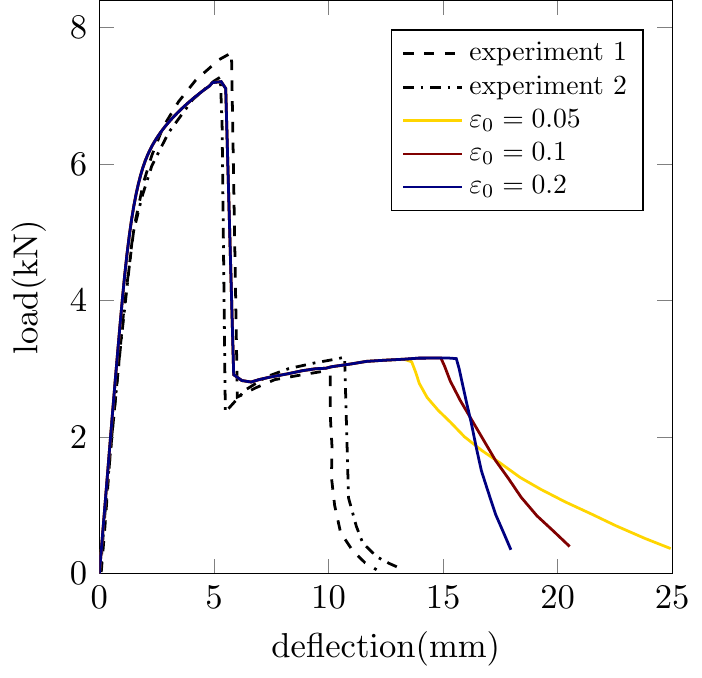}
    \caption{$\beta = 0.1$ and different values of $\varepsilon_0$}
    \label{fig: example/3pb/load_deflection/constant_beta}
  \end{subfigure}
  \begin{subfigure}{0.45\textwidth}
    \centering
    \tikzsetnextfilenamesafe{example/3pb/load_deflection/constant_e0}
    \includegraphics{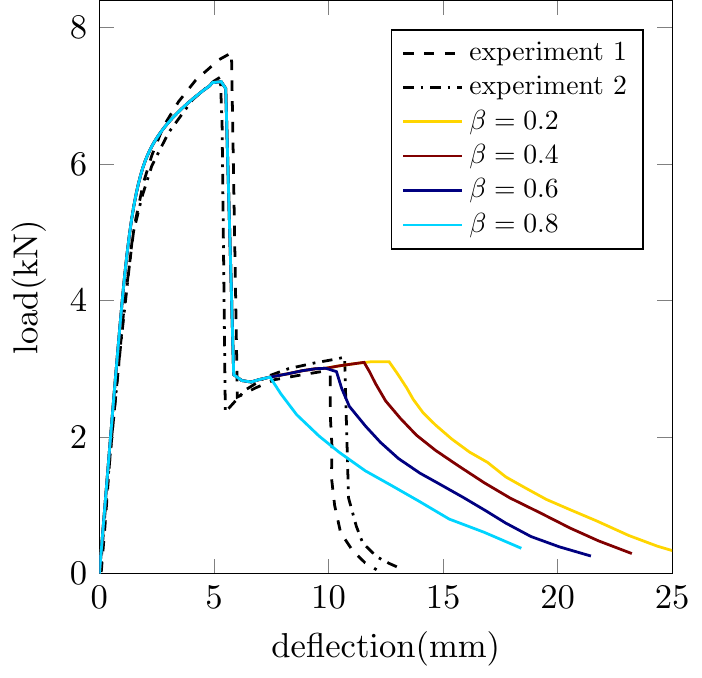}
    \caption{$\varepsilon_0 = 0.05$ and different values of $\beta$}
    \label{fig: example/3pb/load_deflection/constant_e0}
  \end{subfigure}
  \caption{Comparison of load-deflection curves between simulations and experiments for the 3D three-point bend experiment.}
  \label{fig: example/3pb/load_deflection}
\end{figure}

In \cite{kubik2019ductile}, experimental results are reported for two different trials of the three-point bending experiment, both using the same material, geometry, and etc.  Both of the experimental load-deflection curves are shown in \Cref{fig: example/3pb/load_deflection} for the sake of comparison to our simulation results. The experimental curves indicate an initial elastic-plastic response, followed by an abrupt drop in the load, subsequent force increase, and a final abrupt drop.  The two drops in the force-displacement data (occurring for both experiments) correspond to the nucleation and propagation of the crack surfaces, first from the bottom notch to the hole, then from the hole to the top notch.

In the experiments, the first load drop occurs at a deflection around $\SI{5.8}{\milli\meter}$ (experiment 1) and $\SI{5.3}{\milli\meter}$ (experiment 2), and the second load drop occurs at a deflection around $\SI{10.1}{\milli\meter}$ (experiment 1) and $\SI{10.7}{\milli\meter}$ (experiment 2).
In all load-deflection curves predicted by the model-based simulations, the first load drop occurs at a deflection of approximately  $\SI{5.3}{\milli\meter}$, which compares well to the experimental measurements.
However, the prediction of the nucleation and propagation of the second crack is sensitive to model parameters $\beta$ and $\varepsilon_0$. While $\beta = 0.1$ is kept constant (\Cref{fig: example/3pb/load_deflection/constant_beta}), larger $\varepsilon_0$ delays the nucleation of the second crack and leads to a steeper load drop.  While $\varepsilon_0 = 0.05$ is kept constant (\Cref{fig: example/3pb/load_deflection/constant_e0}), larger $\beta$ values accelerate the nucleation of the second crack, but the load drops at approximately the same rate.

\begin{figure}[htb!]
  \centering
  \tikzsetnextfilenamesafe{example/3pb/load_deflection_tuned}
  \includegraphics{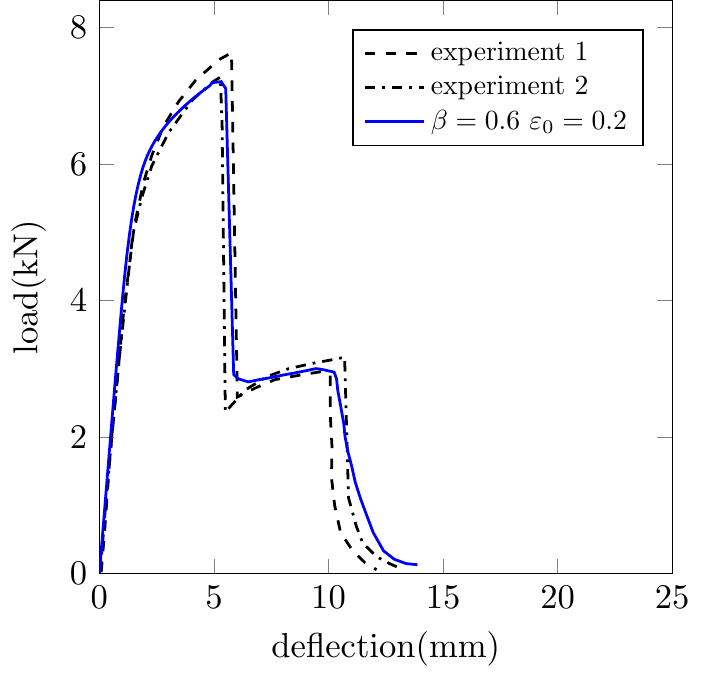}
  \caption{Comparison of load-deflection curves between simulations and experiments with tuned $\beta$ and $\varepsilon_0$.}
  \label{fig: example/3pb/load_deflection_tuned}
\end{figure}

The results shown in \Cref{fig: example/3pb/load_deflection} suggest that choosing $\varepsilon_0=0.2$ and $\beta=0.6$ might result in a force-displacement curve that best matches the experimental data (the combination $\beta=0.6$ and $\varepsilon_0=0.2$ yields a force-displacement response that is practically indistinguishable from those shown in \Cref{fig: example/3pb/force_disp/EPD_beta,fig: example/3pb/force_disp/EPD_e0}).
We performed an additional simulation using this choice of these parameters, and the results are shown in \Cref{fig: example/3pb/load_deflection_tuned}). Indeed, the resulting simulation yields a force-displacement curve that is very close to the experimental result.   Importantly, we find that our model-based simulations yield a better comparison to the experimental load-displacement curves than the various damage models explored in \cite{kubik2019ductile}.

\begin{figure}[!htb]
  \centering
  \begin{subfigure}{0.33\textwidth}
    \centering
    \includegraphics[width=\textwidth,scale=0.5]{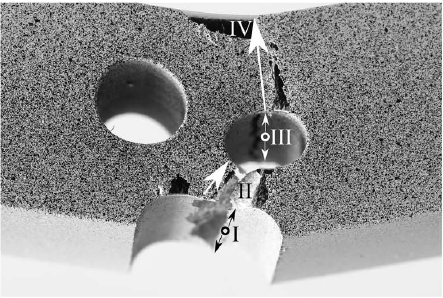}
    \caption{experiment}
    \label{fig: example/3pb/1234/experiment}
  \end{subfigure}
  \hfill
  \begin{minipage}{0.6\textwidth}
    \begin{subfigure}{0.45\textwidth}
      \centering
      \includegraphics[width=\linewidth]{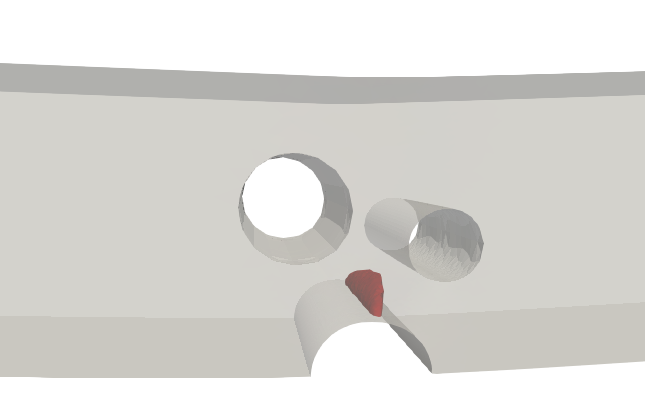}
      \caption{stage I}
      \label{fig: example/3pb/1234/simulation_I}
    \end{subfigure}
    \begin{subfigure}{0.45\textwidth}
      \centering
      \includegraphics[width=\linewidth]{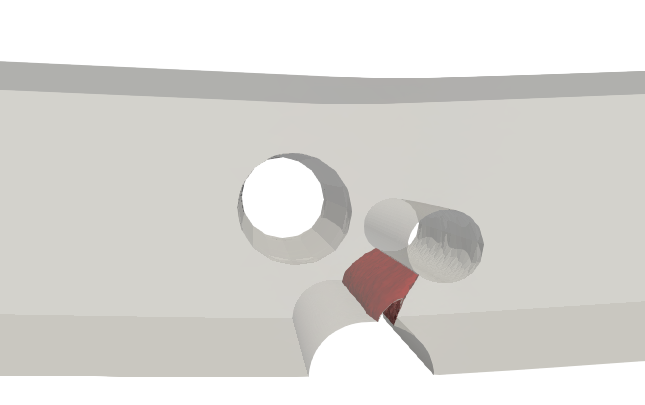}
      \caption{stage II}
      \label{fig: example/3pb/1234/simulation_II}
    \end{subfigure}

    \begin{subfigure}{0.45\textwidth}
      \centering
      \includegraphics[width=\linewidth]{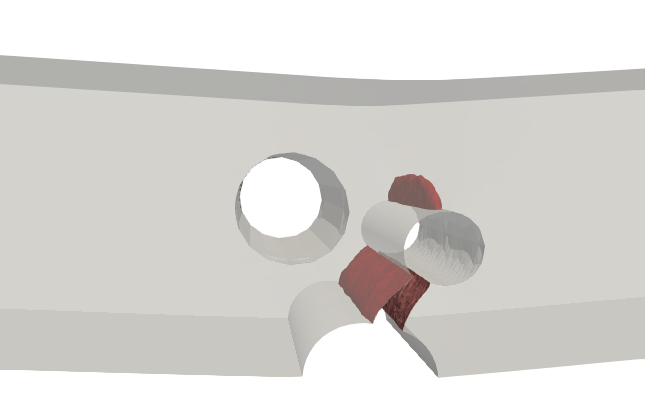}
      \caption{stage III}
      \label{fig: example/3pb/1234/simulation_III}
    \end{subfigure}
    \begin{subfigure}{0.45\textwidth}
      \centering
      \includegraphics[width=\linewidth]{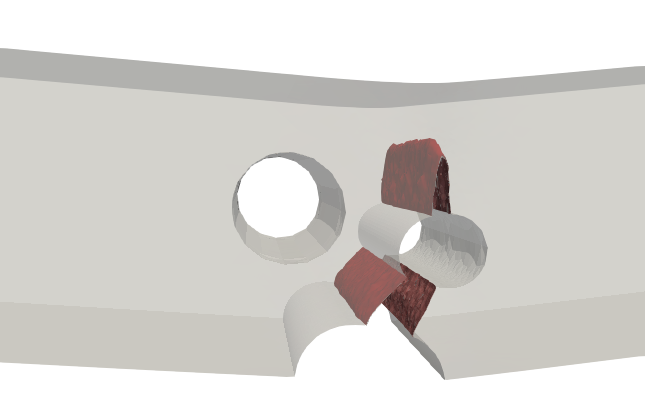}
      \caption{stage IV}
      \label{fig: example/3pb/1234/simulation_IV}
    \end{subfigure}
  \end{minipage}
  \caption{Comparison of crack paths between (a) the experiment and (b-e) the simulation with $\beta = 0.1$ and $\varepsilon_0 = 0.2$. Crack surfaces are highlighted in red.}
  \label{fig: example/3pb/1234}
\end{figure}

While the load-deflection curves differ based on different choices of model parameters, the crack trajectory predictions are relatively insensitive to the model parameters. In the experiments (\Cref{fig: example/3pb/1234/experiment}), the crack nucleation and propagation can be divided into 4 stages. (I) The crack first nucleates around the center of the bottom notch and propagates through the thickness of the specimen. (II) Simultaneously the crack grows slant-wise towards the bottom of the smaller hole. (III) After a period of load increase, the second crack nucleates around the center of the top of the smaller hole and propagates through the thickness towards the free surfaces. (IV) Finally, the crack propagates towards the top free surface of the specimen.
The corresponding stages in the simulation are shown in \Cref{fig: example/3pb/1234/simulation_I,fig: example/3pb/1234/simulation_II,fig: example/3pb/1234/simulation_III,fig: example/3pb/1234/simulation_IV}.

\begin{figure}[!htb]
  \centering
  \begin{subfigure}[b]{0.4\textwidth}
    \centering
    \includegraphics[width=\textwidth,scale=0.5]{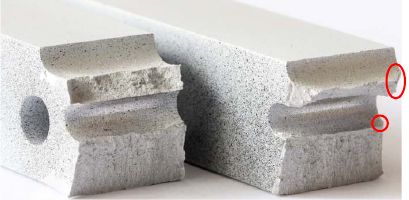}
    \caption{}
  \end{subfigure}
  \begin{subfigure}[b]{0.4\textwidth}
    \centering
    \includegraphics[width=\textwidth,scale=0.5]{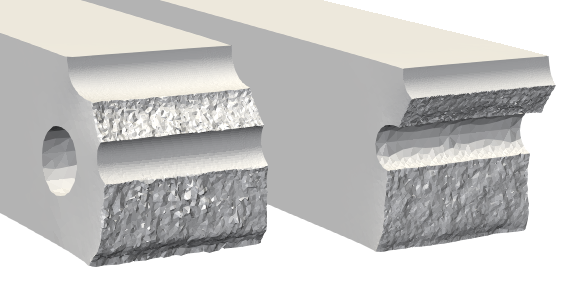}
    \caption{}
  \end{subfigure}
  \caption{Comparison of the final crack surfaces between (a) the experiment and (b) the simulation ($\beta = 0.1$ and $\varepsilon_0 = 0.2$). ``Shear lips'' are circled in red, and the crack surfaces are marked in red.}
  \label{fig: example/3pb/split_comparison}
\end{figure}

The final crack surfaces obtained using $\beta = 0.1$ and $\varepsilon_0 = 0.2$ are shown in \Cref{fig: example/3pb/split_comparison}. In the experiments, ``shear lips'' (circled in red) occur close to the free surfaces of the specimen due to high stress-triaxility. Although such ``shear lips'' are not predicted by the E-P-D model used in this work, the larger-scale crack trajectories agree reasonably well with the experimental observations.

\subsection{Sandia Fracture Challenge Example}
\label{s: example/SFC}

Finally, we now examine the capabilities of the proposed E-P-D model by simulating an experiment from a Sandia Fracture Challenge \cite{boyce2014sandia}.  The specimen is composed of Al-5052 H34.  The hardening model is first calibrated against  tensile test data for a smooth round bar \cite{guo2013experimental}. The calibrated material properties and model constants are summarized in \Cref{tab: example/SFC}. Based on the calibrated tensile test results, the crack should initiate at a critical equivalent plastic strain of $\varepsilon^p_c = 0.8$. The fracture toughness $\Gc$ is then scaled accordingly, such that $\widehat{\Gc}$ matches experimental measurements.

\begin{table}[htb!]
  \centering
  \caption{Summary of the calibrated material properties and model parameters for the Sandia Fracture Challenge specimen}
  \begin{tabular}{r | c | c | c }
    \toprule
    Property/Parameter               & Symbol          & Value           & Unit                                       \\
    \midrule
    Bulk modulus                     & $K$             & \SI{6.495e4}{}  & \SI{}{\mega\pascal}                        \\
    Shear modulus                    & $G$             & \SI{2.4906e4}{} & \SI{}{\mega\pascal}                        \\
    \midrule
    Yield stress                     & $\sigma_y$      & 25              & \SI{}{\mega\pascal}                        \\
    Hardening modulus                & $h$             & 250             & \SI{}{\mega\pascal}                        \\
    \midrule
    Fracture toughness               & $\widehat{\Gc}$ & 13.6            & \SI{}{\milli\joule\per\square\milli\meter} \\
    Fracture dissipation coefficient & $\xi$           & 1               & nondim.                                    \\
    Critical fracture energy         & $\psi_c$        & 1               & \SI{}{\milli\joule\per\cubic\milli\meter}  \\
    Regularization length            & $l$             & 0.3             & \SI{}{\milli\meter}                        \\
    \midrule
    Interaction coefficient          & $\beta$         & 0.5             & nondim.                                    \\
    Characteristic plastic strain    & $\varepsilon_0$ & 0.1             & nondim.                                    \\
    \bottomrule
  \end{tabular}
  \label{tab: example/SFC}
\end{table}

\begin{figure}[!htb]
  \centering
  \tikzsetnextfilenamesafe{example/SFC/schematics}
  \includegraphics{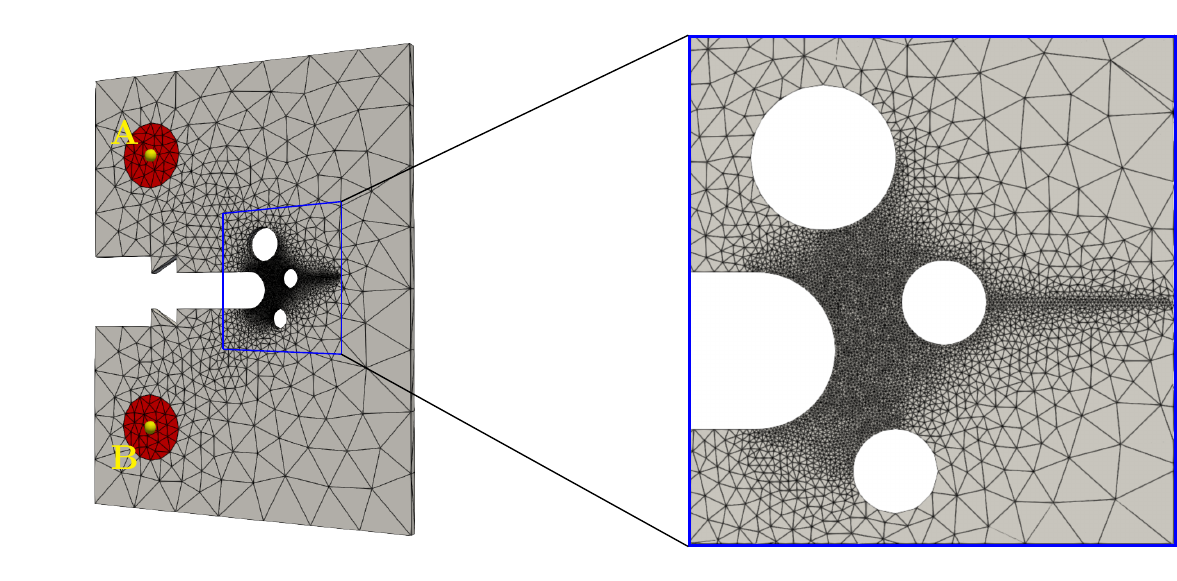}
  \caption{Sandia Fracture Challenge problem geometry and mesh, with zoomed-in view of the refined region. The specimen is loaded vertically with displacement control at lines of nodes A and B marked as yellow spheres.}
  \label{fig: example/SFC/schematics}
\end{figure}

The calibrated model is then applied to simulate crack initiation and growth in the specimen adopted for the Sandia Fracture Challenge. Drawings of the specimen with detailed dimensions are available in \cite{guo2013experimental,ambati_phase-field_2016}. The geometry and the mesh of the specimen is shown in \Cref{fig: example/SFC/schematics}. The three-dimensional geometry is discretized using tetrahedral elements (\texttt{TET4}). The mesh is locally refined near the round notch to sufficiently capture plastic and fracture localizations. It is assumed that the extension of the second crack is known a priori, and the region to the right of the middle hole is also refined to properly resolve the phase-field. The resulting mesh consists of 10831 tetrahedra.

The specimen is loaded in tension via two pins that connect to the specimen through the circular holes whose centers coincide with nodes $A$ and $B$ in \Cref{fig: example/SFC/schematics}.  Rather than apply the loads to the boundaries of the holes or employ a contact algorithm with the pins, here we simplify the model by explicitly meshing the pins and assuming they are perfectly bonded to the specimen at the hole locations.  The material of the pins is    assumed to be elastic with moduli that match the specimen.

\begin{figure}[!htb]
  \centering
  \begin{subfigure}{\textwidth}
    \centering
    \tikzsetnextfilenamesafe{example/SFC/force_disp}
    \includegraphics{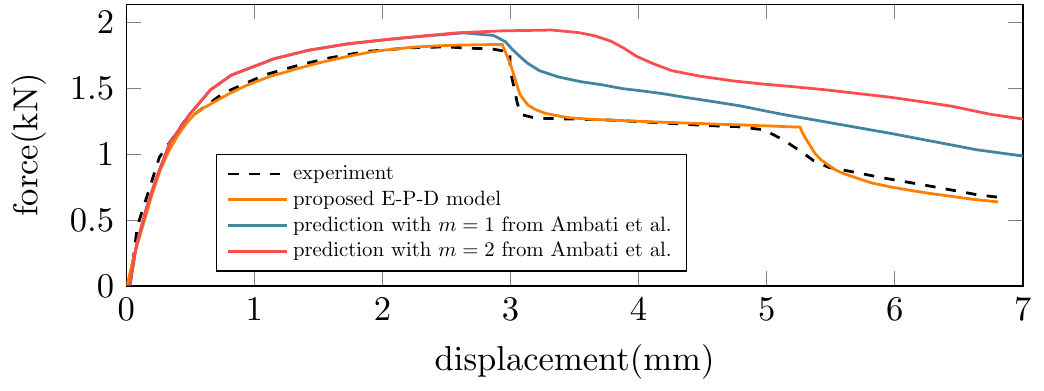}
    \caption{}
    \label{fig: example/SFC/force_disp}
  \end{subfigure}

  \begin{subfigure}{0.17\textwidth}
    \centering
    \includegraphics[width=\textwidth,scale=0.5]{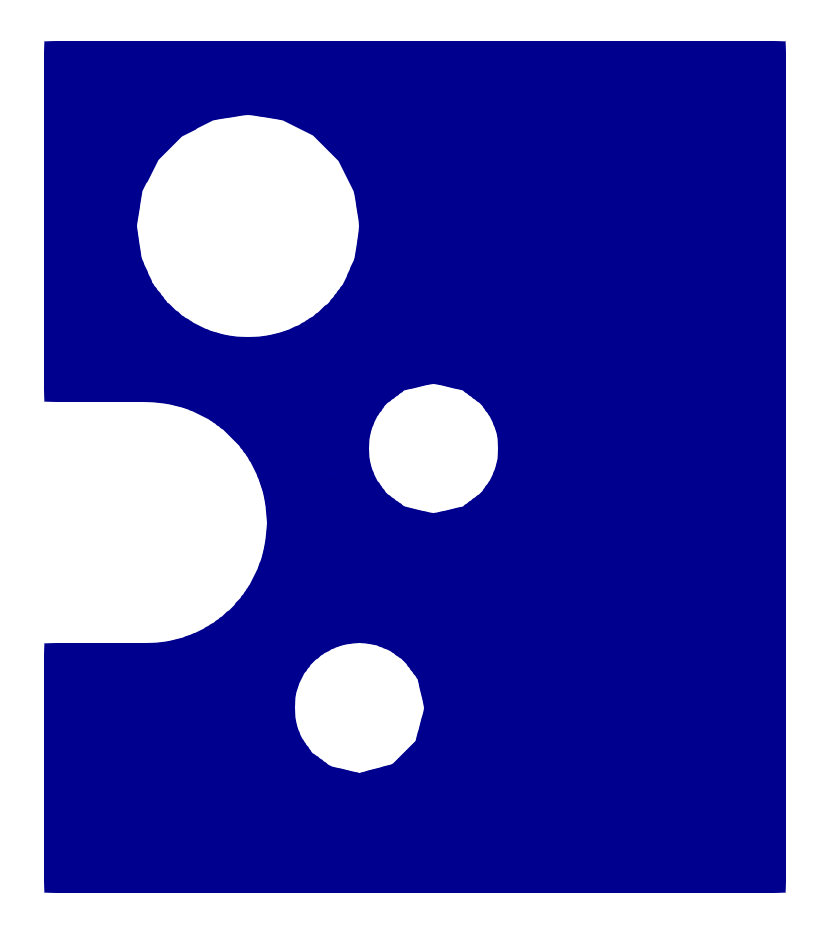}
    \caption{$u = \SI{2.87}{\milli\meter}$}
    \label{fig: example/SFC/d_1}
  \end{subfigure}
  \hspace{0.03\textwidth}
  \begin{subfigure}{0.17\textwidth}
    \centering
    \includegraphics[width=\textwidth,scale=0.5]{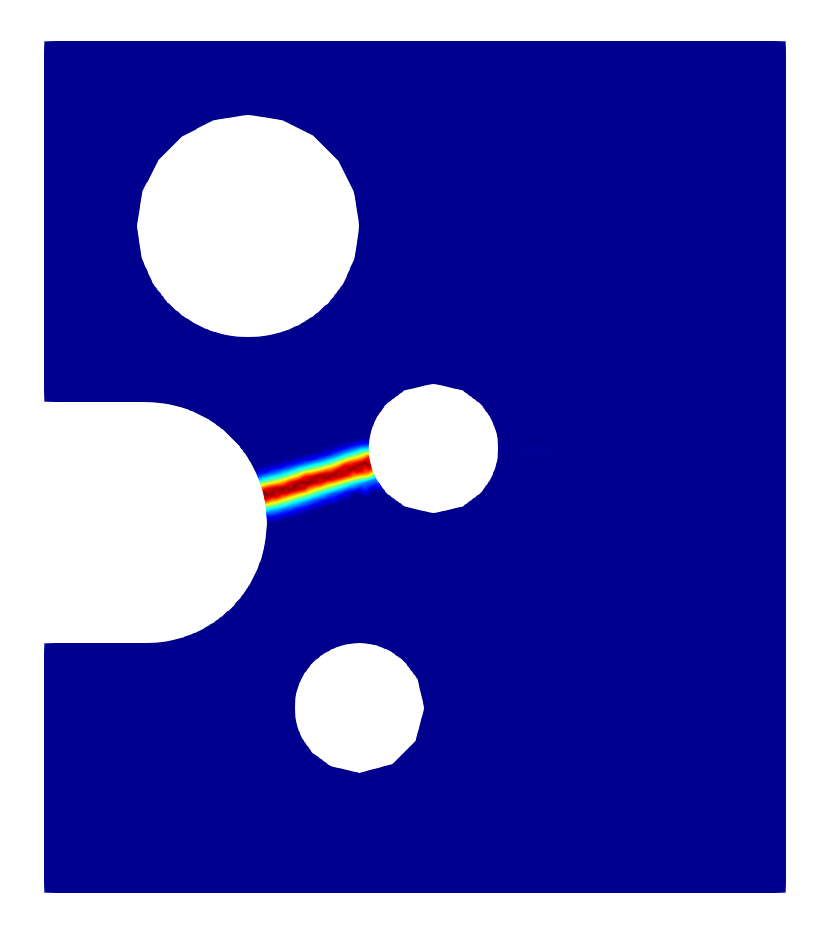}
    \caption{$u = \SI{5.28}{\milli\meter}$}
    \label{fig: example/SFC/d_2}
  \end{subfigure}
  \hspace{0.03\textwidth}
  \begin{subfigure}{0.17\textwidth}
    \centering
    \includegraphics[width=\textwidth,scale=0.5]{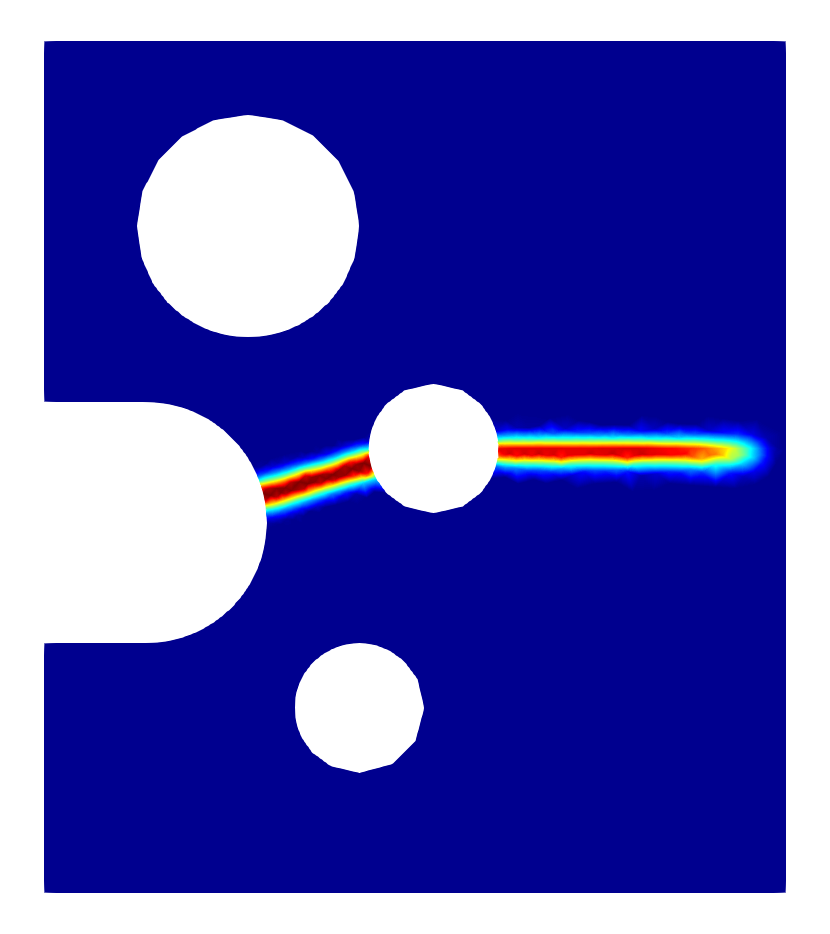}
    \caption{$u = \SI{6.81}{\milli\meter}$}
    \label{fig: example/SFC/d_3}
  \end{subfigure}
  \begin{subfigure}{0.04\textwidth}
    \centering
    \caption*{$d$}
    \includegraphics[width=\textwidth,scale=0.5]{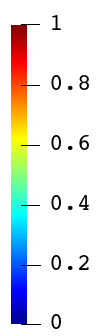}
    \vspace{1em}
  \end{subfigure}

  \begin{subfigure}{0.17\textwidth}
    \centering
    \includegraphics[width=\textwidth,scale=0.5]{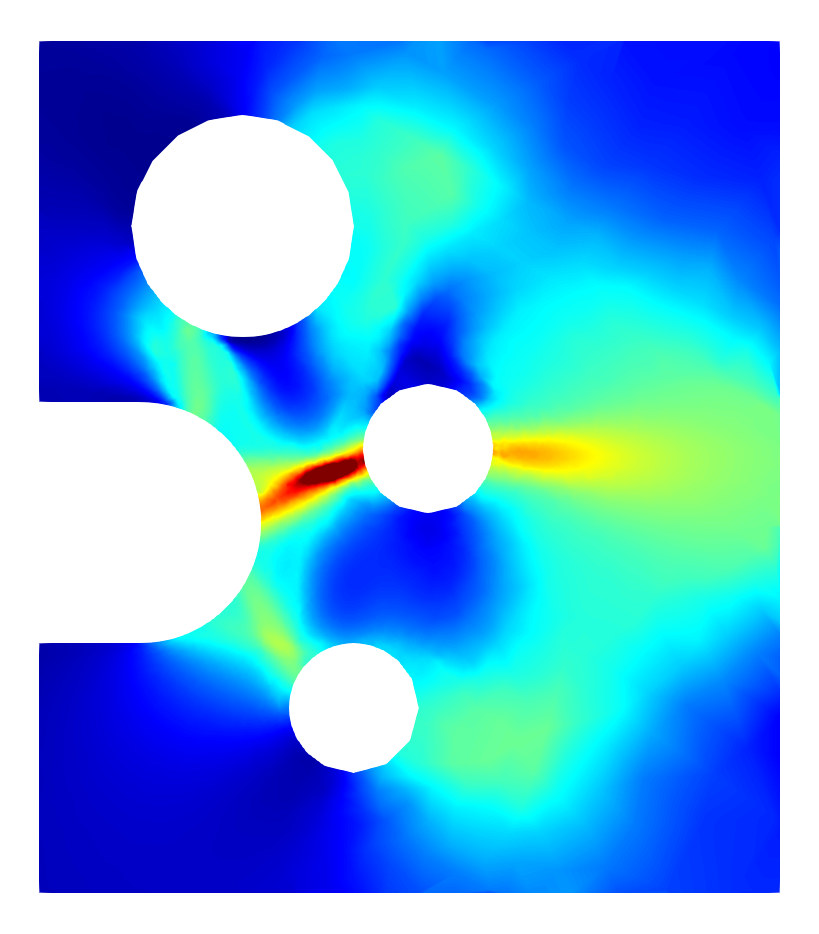}
    \caption{$u = \SI{2.87}{\milli\meter}$}
    \label{fig: example/SFC/We_1}
  \end{subfigure}
  \hspace{0.03\textwidth}
  \begin{subfigure}{0.17\textwidth}
    \centering
    \includegraphics[width=\textwidth,scale=0.5]{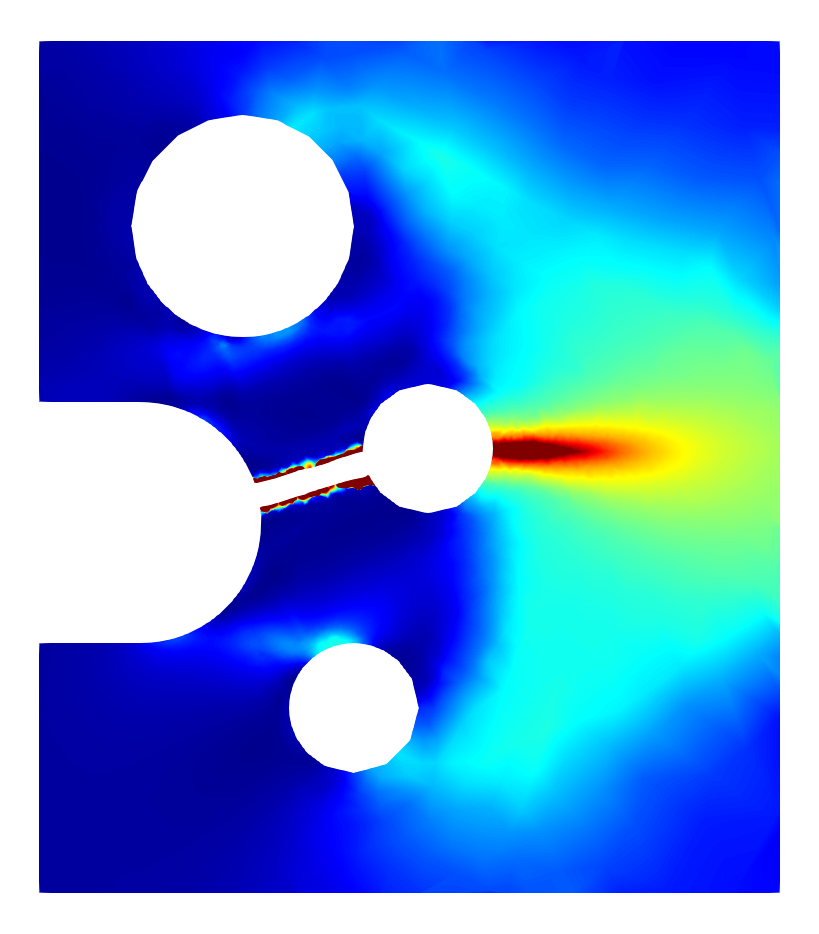}
    \caption{$u = \SI{5.28}{\milli\meter}$}
    \label{fig: example/SFC/We_2}
  \end{subfigure}
  \hspace{0.03\textwidth}
  \begin{subfigure}{0.17\textwidth}
    \centering
    \includegraphics[width=\textwidth,scale=0.5]{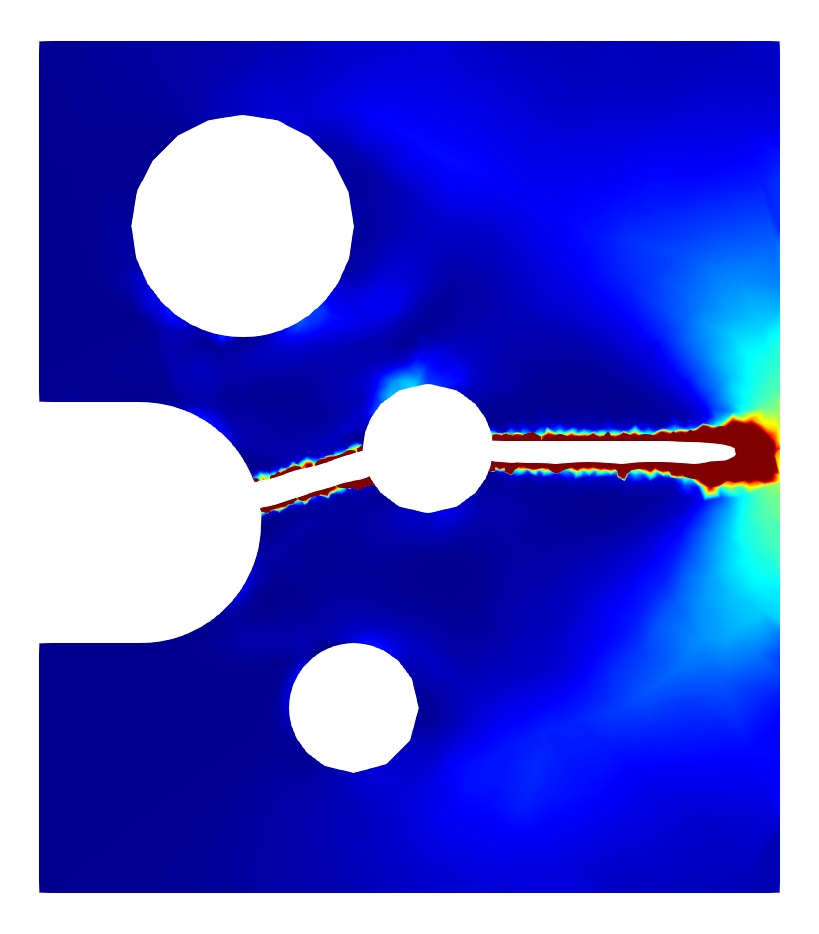}
    \caption{$u = \SI{6.81}{\milli\meter}$}
    \label{fig: example/SFC/We_3}
  \end{subfigure}
  \begin{subfigure}{0.04\textwidth}
    \centering
    \caption*{$W^e$}
    \includegraphics[width=\textwidth,scale=0.5]{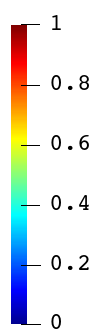}
    \vspace{1em}
  \end{subfigure}

  \begin{subfigure}{0.17\textwidth}
    \centering
    \includegraphics[width=\textwidth,scale=0.5]{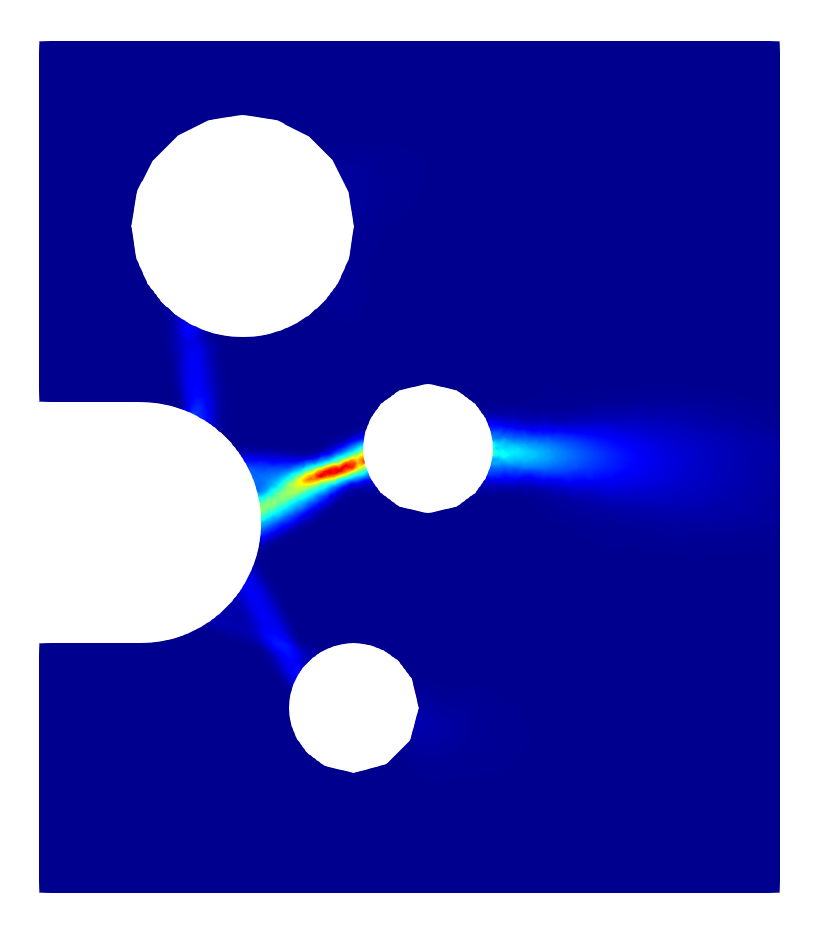}
    \caption{$u = \SI{2.87}{\milli\meter}$}
    \label{fig: example/SFC/Wp_1}
  \end{subfigure}
  \hspace{0.03\textwidth}
  \begin{subfigure}{0.17\textwidth}
    \centering
    \includegraphics[width=\textwidth,scale=0.5]{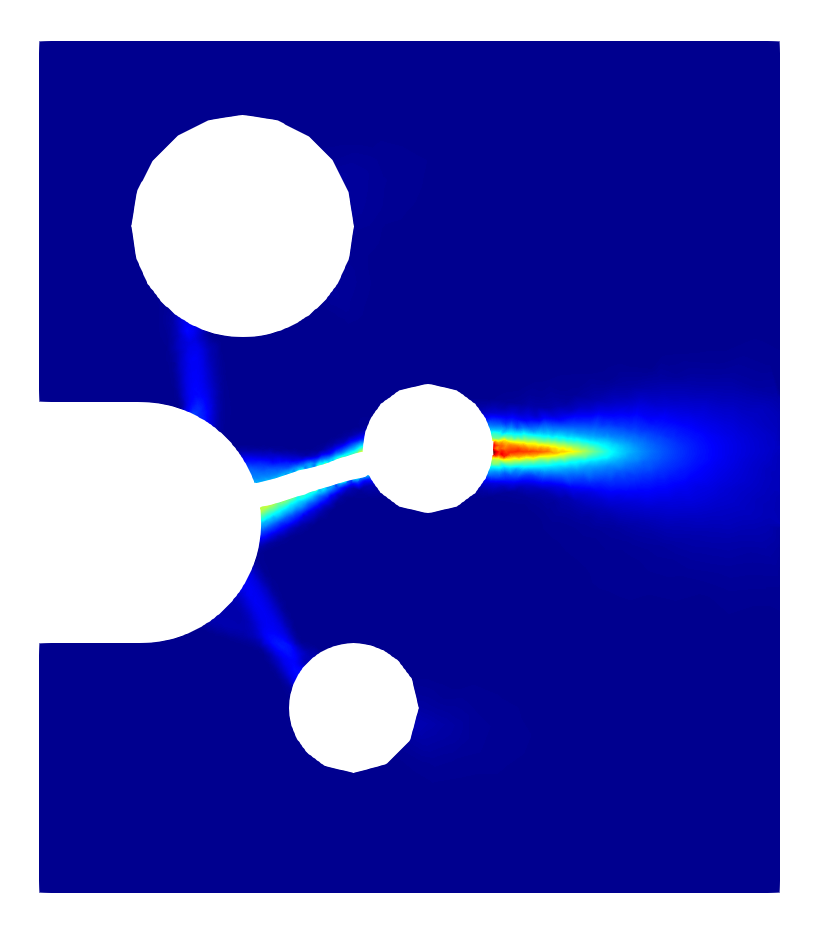}
    \caption{$u = \SI{5.28}{\milli\meter}$}
    \label{fig: example/SFC/Wp_2}
  \end{subfigure}
  \hspace{0.03\textwidth}
  \begin{subfigure}{0.17\textwidth}
    \centering
    \includegraphics[width=\textwidth,scale=0.5]{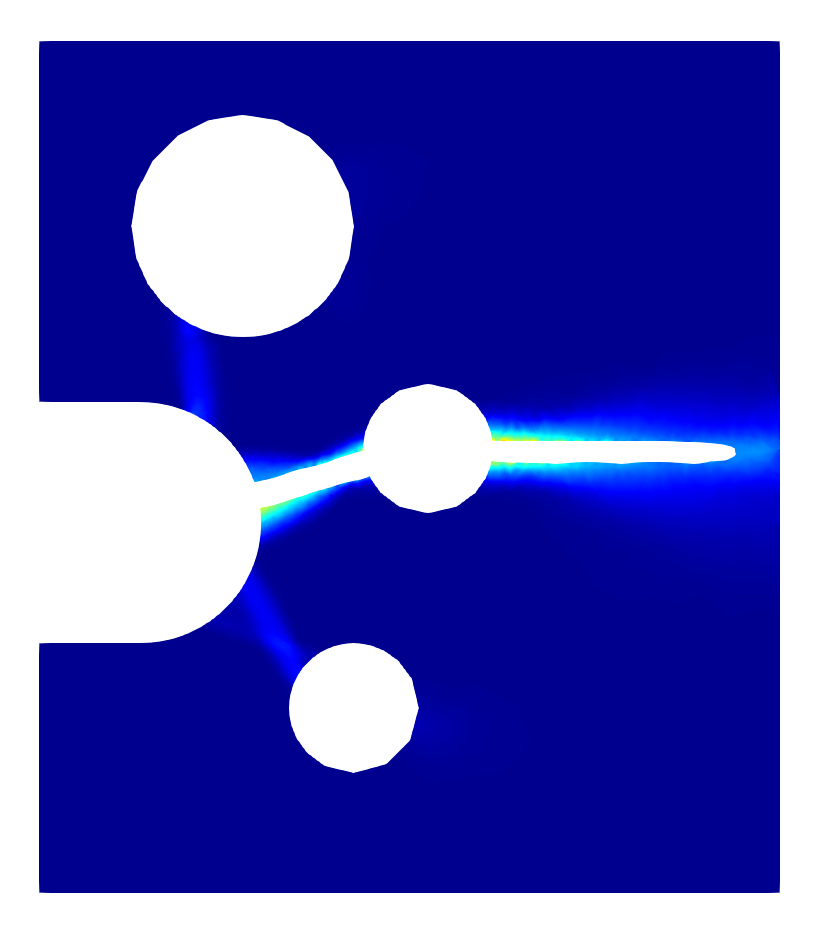}
    \caption{$u = \SI{6.81}{\milli\meter}$}
    \label{fig: example/SFC/Wp_3}
  \end{subfigure}
  \begin{subfigure}{0.04\textwidth}
    \centering
    \caption*{$W^p$}
    \includegraphics[width=\textwidth,scale=0.5]{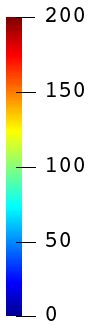}
    \vspace{1em}
  \end{subfigure}

  \begin{subfigure}{0.17\textwidth}
    \centering
    \includegraphics[width=\textwidth,scale=0.5]{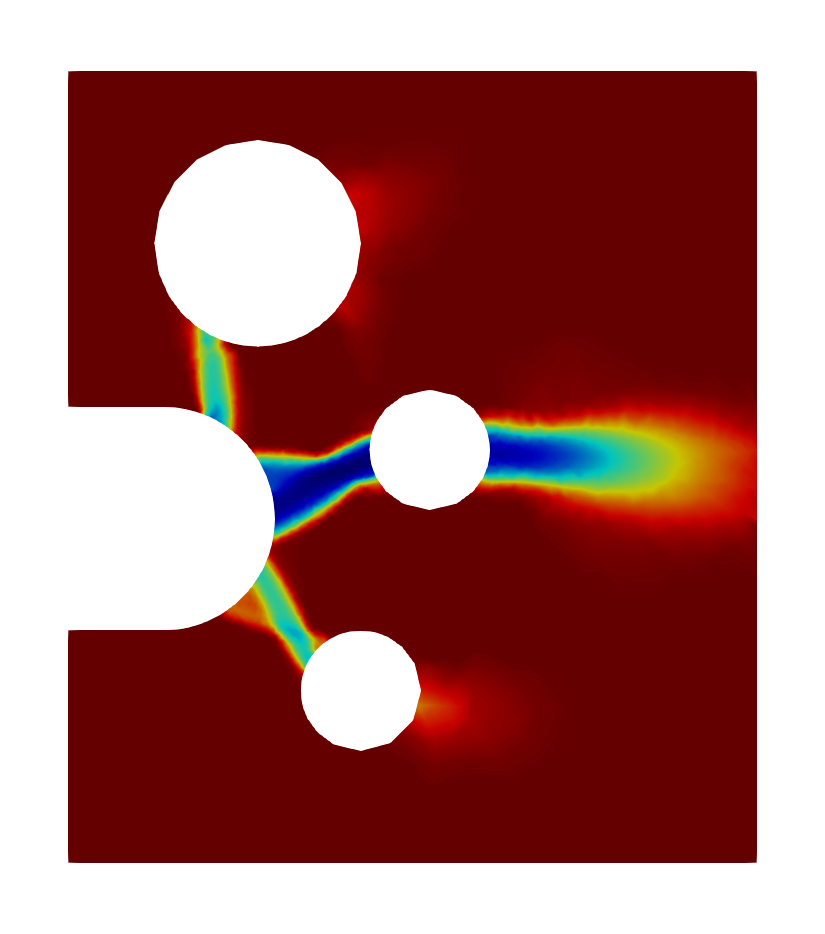}
    \caption{$u = \SI{2.87}{\milli\meter}$}
    \label{fig: example/SFC/gc_1}
  \end{subfigure}
  \hspace{0.03\textwidth}
  \begin{subfigure}{0.17\textwidth}
    \centering
    \includegraphics[width=\textwidth,scale=0.5]{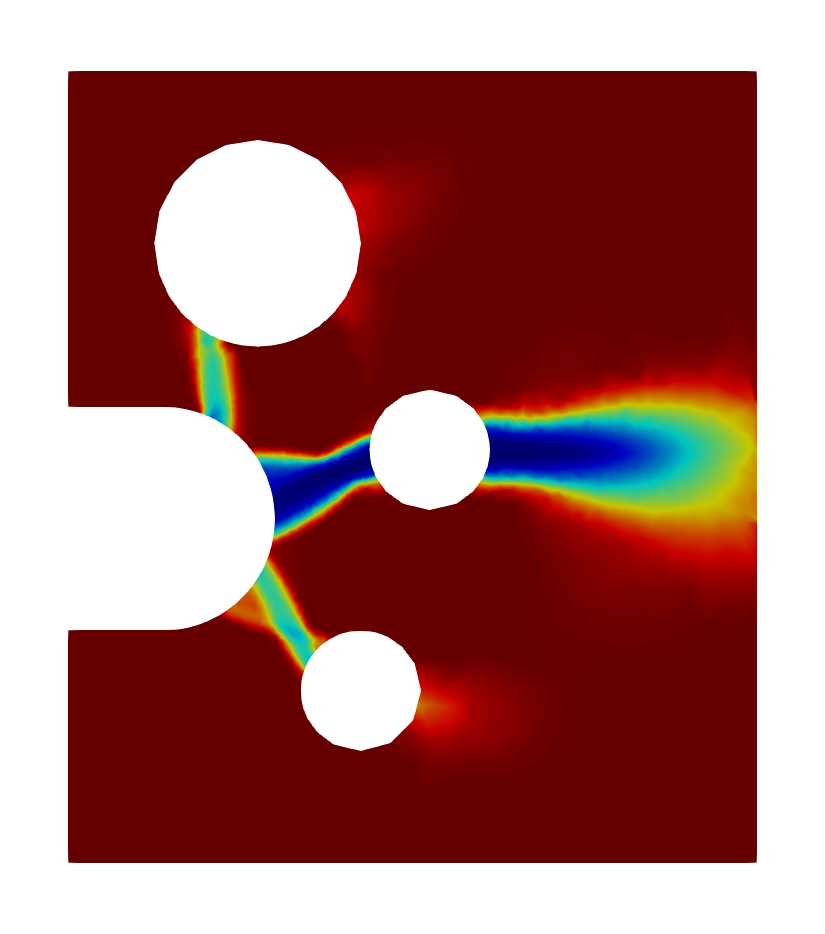}
    \caption{$u = \SI{5.28}{\milli\meter}$}
    \label{fig: example/SFC/gc_2}
  \end{subfigure}
  \hspace{0.03\textwidth}
  \begin{subfigure}{0.17\textwidth}
    \centering
    \includegraphics[width=\textwidth,scale=0.5]{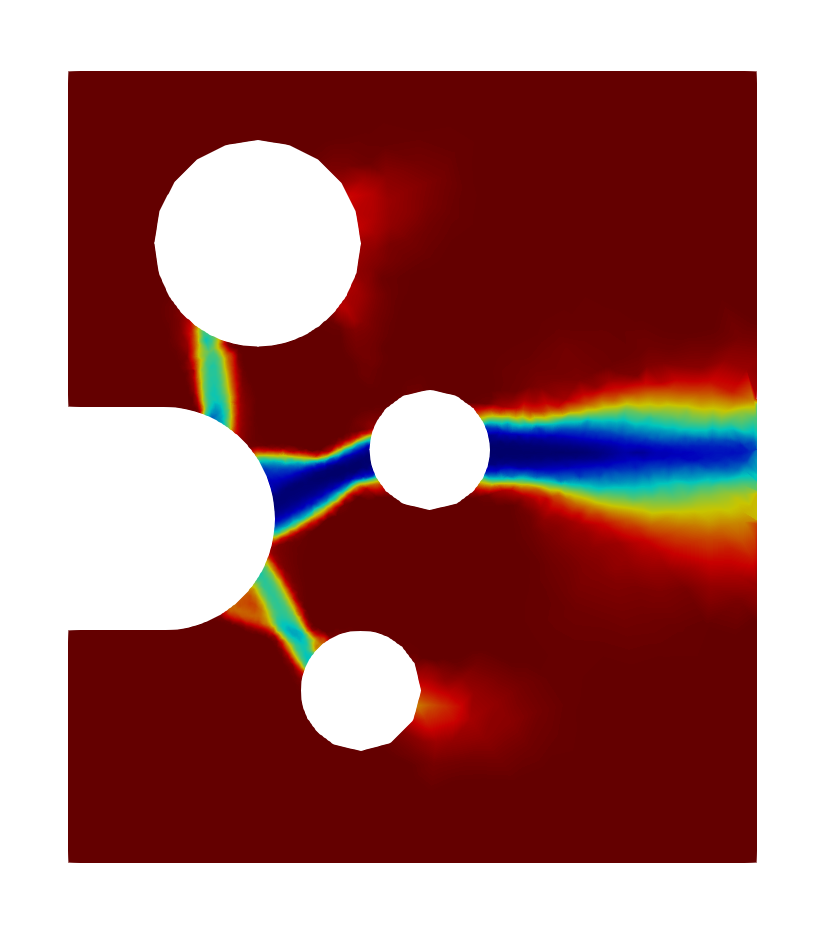}
    \caption{$u = \SI{6.81}{\milli\meter}$}
    \label{fig: example/SFC/gc_3}
  \end{subfigure}
  \begin{subfigure}{0.04\textwidth}
    \centering
    \caption*{$g^c$}
    \includegraphics[width=\textwidth,scale=0.5]{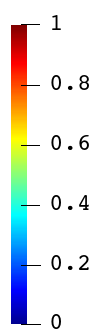}
    \vspace{1em}
  \end{subfigure}
  \caption{The Sandia Fracture Challenge. (a) The force displacement curves from the experiment and the numerical simulations. Contour plots of (b-d) the phase-field $d$, (e-g) the active part of the elastic energy $W^e_\activeenergy$, (h-j) the plastic energy $W^p$, and (k-m) the coalescence degradation function $g^c$ at three different loads. In subfigures (e-j) the domain within the contour of $d=0.8$ is removed to facilitate crack path visualization. }
\end{figure}

The resulting force-displacement curves are shown in \Cref{fig: example/SFC/force_disp}. The response predicted by our proposed E-P-D model indicates an excellent agreement with the experimental measurements. Force-displacement curves from \cite{ambati_phase-field_2016} are also included for comparison.  The proposed E-P-D model clearly provides the best quantitative and qualitative match with the experiment.

Contour plots of the phase-field $d$ are shown in \Cref{fig: example/SFC/d_1,fig: example/SFC/d_2,fig: example/SFC/d_3}. The first crack nucleates on the left side of the middle hole, and then connects with the round notch. After a while, the second crack nucleates on the right side of the middle hole and propagates to the right, with a nearly horizontal orientation.
The active part of the elastic energy is shown in \Cref{fig: example/SFC/We_1,fig: example/SFC/We_2,fig: example/SFC/We_3}, where the stress concentration at the crack tips and a stress release after the first crack can be clearly observed.
The plastic energy is shown in \Cref{fig: example/SFC/Wp_1,fig: example/SFC/Wp_2,fig: example/SFC/Wp_3}, and the coalescence degradation is shown in \Cref{fig: example/SFC/gc_1,fig: example/SFC/gc_2,fig: example/SFC/gc_3}. The fracture toughness is substantially degraded in the regions with high plastic localization.

\section{Conclusion}
\label{s: conclusion}

This work developed a new phase-field model for ductile fracture with finite deformation kinematics and within the confines of a variational framework.   The framework provides two mechanisms for coupling  plasticity and fracture, either using the plastic degradation function (E-P-PD) or through the coalescence dissipation (E-P-D). The proposed E-P-D model using this novel coalescence dissipation was compared with the E-P-PD model from \cite{brandon2020cohesive}.

Several important modeling choices are discussed in \Cref{s: example/homogenized} in a ``homogenized'' setting: A crack geometric function $\alpha(d; \xi > 0)$ allows for an unperturbed elastic-plastic response prior to damage initiation (\Cref{s: example/homogenized/alpha}); a Lorentz-type degradation function separates out the critical fracture energy as an independent material property such that a regularization-length independent finite critical fracture strength can be obtained (\Cref{s: example/homogenized/degradation}); the decrease in the fracture toughness as a function of the increasing plastic strain can be modeled using the proposed coalescence dissipation (\Cref{s: example/homogenized/coalescence}). Convergence of the softening response is numerically demonstrated in \Cref{s: example/nonhomogeneous}.
The force-displacement responses obtained using both the proposed E-P-D model and the E-P-PD model presented in \cite{brandon2020cohesive} become insensitive to the regularization length as the regularization length is sufficiently small, regardless of the amount of coalescence dissipation.

The performance of the proposed E-P-D model is evaluated by simulating a three-point bending experiment in \Cref{s: example/3pb}. The model is first calibrated against a standard notched tension test, and then applied to predict the crack path and load-deflection response of the aluminum specimen in the three-point bending experiment.
The predicted load-deflection curves and crack trajectories are found to be in good agreement with experimental measurements.

\section*{Acknowledgements}
\label{s: acknowledgements}

This research was supported by a research grant to Duke University from Sandia National Laboratories.  Sandia National Laboratories is a multimission laboratory managed and operated by National Technology and Engineering Solutions of Sandia, LLC., a wholly owned subsidiary of Honeywell International, Inc., for the U.S. Department of Energy’s National Nuclear Security Administration, USA under contract DE-NA0003525.

\clearpage

\bibliographystyle{elsarticle-num-names}
\bibliography{bib.bib}

\begin{thebibliography}{48}
\expandafter\ifx\csname natexlab\endcsname\relax\def\natexlab#1{#1}\fi
\providecommand{\url}[1]{\texttt{#1}}
\providecommand{\href}[2]{#2}
\providecommand{\path}[1]{#1}
\providecommand{\DOIprefix}{doi:}
\providecommand{\ArXivprefix}{arXiv:}
\providecommand{\URLprefix}{URL: }
\providecommand{\Pubmedprefix}{pmid:}
\providecommand{\doi}[1]{\href{http://dx.doi.org/#1}{\path{#1}}}
\providecommand{\Pubmed}[1]{\href{pmid:#1}{\path{#1}}}
\providecommand{\bibinfo}[2]{#2}
\ifx\xfnm\relax \def\xfnm[#1]{\unskip,\space#1}\fi
\bibitem[{Francfort and Marigo(1998)}]{Francfort98}
\bibinfo{author}{G.~Francfort}, \bibinfo{author}{J.~Marigo},
\newblock \bibinfo{title}{Revisiting brittle fracture as an energy minimization
  problem},
\newblock \bibinfo{journal}{Journal of the Mechanics and Physics of Solids}
  \bibinfo{volume}{46} (\bibinfo{year}{1998}) \bibinfo{pages}{1319--1342}.
\bibitem[{Bourdin et~al.(2000)Bourdin, Francfort, and Marigo}]{Bourdin2000}
\bibinfo{author}{B.~Bourdin}, \bibinfo{author}{G.~Francfort},
  \bibinfo{author}{J.~Marigo},
\newblock \bibinfo{title}{Numerical experiments in revisited brittle fracture},
\newblock \bibinfo{journal}{Journal of the Mechanics and Physics of Solids}
  \bibinfo{volume}{48} (\bibinfo{year}{2000}) \bibinfo{pages}{797--826}.
\bibitem[{Karma et~al.(2001)Karma, Kessler, and Levine}]{karma_2001}
\bibinfo{author}{A.~Karma}, \bibinfo{author}{D.~A. Kessler},
  \bibinfo{author}{H.~Levine},
\newblock \bibinfo{title}{Phase-field model of mode {III} dynamic fracture},
\newblock \bibinfo{journal}{Physical Review Letters} \bibinfo{volume}{87}
  (\bibinfo{year}{2001}) \bibinfo{pages}{045501--1--4}.
\bibitem[{Karma and Lobkovsky(2004)}]{karma_2004}
\bibinfo{author}{A.~Karma}, \bibinfo{author}{A.~E. Lobkovsky},
\newblock \bibinfo{title}{Unsteady crack motion and branching in a phase-field
  model of brittle fracture},
\newblock \bibinfo{journal}{Physical Review Letters} \bibinfo{volume}{93}
  (\bibinfo{year}{2004}) \bibinfo{pages}{245510}.
\bibitem[{Amor et~al.(2009{\natexlab{a}})Amor, Marigo, and Maurini}]{amor_2009}
\bibinfo{author}{H.~Amor}, \bibinfo{author}{J.~J. Marigo},
  \bibinfo{author}{C.~Maurini},
\newblock \bibinfo{title}{Regularized formulation of the variational brittle
  fracture with unilateral contact: Numerical experiments},
\newblock \bibinfo{journal}{Journal of the Mechanics and Physics of Solids}
  \bibinfo{volume}{57} (\bibinfo{year}{2009}{\natexlab{a}})
  \bibinfo{pages}{1209--1229}.
\bibitem[{Amor et~al.(2009{\natexlab{b}})Amor, Marigo, and
  Maurini}]{AMOR20091209}
\bibinfo{author}{H.~Amor}, \bibinfo{author}{J.-J. Marigo},
  \bibinfo{author}{C.~Maurini},
\newblock \bibinfo{title}{Regularized formulation of the variational brittle
  fracture with unilateral contact: Numerical experiments},
\newblock \bibinfo{journal}{Journal of the Mechanics and Physics of Solids}
  \bibinfo{volume}{57} (\bibinfo{year}{2009}{\natexlab{b}})
  \bibinfo{pages}{1209 -- 1229}.
\bibitem[{Miehe et~al.(2010{\natexlab{a}})Miehe, Welschinger, and
  Hofacker}]{miehe_2010_p1}
\bibinfo{author}{C.~Miehe}, \bibinfo{author}{F.~Welschinger},
  \bibinfo{author}{M.~Hofacker},
\newblock \bibinfo{title}{Thermodynamically consistent phase-field models of
  fracture: Variational principles and multi-field {F}{E} implementations},
\newblock \bibinfo{journal}{International Journal for Numerical Methods in
  Engineering} \bibinfo{volume}{83} (\bibinfo{year}{2010}{\natexlab{a}})
  \bibinfo{pages}{1273--1311}.
\bibitem[{Miehe et~al.(2010{\natexlab{b}})Miehe, Hofacker, and
  Welschinger}]{miehe_2010_p2}
\bibinfo{author}{C.~Miehe}, \bibinfo{author}{M.~Hofacker},
  \bibinfo{author}{F.~Welschinger},
\newblock \bibinfo{title}{A phase field model for rate-independent crack
  propagation: Robust algorithmic implementation based on operator splits},
\newblock \bibinfo{journal}{Computer Methods in Applied Mechanics and
  Engineering} \bibinfo{volume}{199} (\bibinfo{year}{2010}{\natexlab{b}})
  \bibinfo{pages}{2765--2778}.
\bibitem[{May et~al.(2015)May, Vignollet, and De~Borst}]{may2015numerical}
\bibinfo{author}{S.~May}, \bibinfo{author}{J.~Vignollet},
  \bibinfo{author}{R.~De~Borst},
\newblock \bibinfo{title}{A numerical assessment of phase-field models for
  brittle and cohesive fracture: $\gamma$-convergence and stress oscillations},
\newblock \bibinfo{journal}{European Journal of Mechanics-A/Solids}
  \bibinfo{volume}{52} (\bibinfo{year}{2015}) \bibinfo{pages}{72--84}.
\bibitem[{Negri(2020)}]{negri2020gamma}
\bibinfo{author}{M.~Negri},
\newblock \bibinfo{title}{$\gamma$-convergence for high order phase field
  fracture: continuum and isogeometric formulations},
\newblock \bibinfo{journal}{Computer Methods in Applied Mechanics and
  Engineering} \bibinfo{volume}{362} (\bibinfo{year}{2020})
  \bibinfo{pages}{112858}.
\bibitem[{Borden(2012)}]{borden2012isogeometric}
\bibinfo{author}{M.~J. Borden}, \bibinfo{title}{Isogeometric analysis of
  phase-field models for dynamic brittle and ductile fracture}, Ph.D. thesis,
  The University of Texas at Austin, \bibinfo{year}{2012}.
\bibitem[{Lorentz et~al.(2011)Lorentz, Cuvilliez, and
  Kazymyrenko}]{lorentz2011convergence}
\bibinfo{author}{E.~Lorentz}, \bibinfo{author}{S.~Cuvilliez},
  \bibinfo{author}{K.~Kazymyrenko},
\newblock \bibinfo{title}{Convergence of a gradient damage model toward a
  cohesive zone model},
\newblock \bibinfo{journal}{Comptes Rendus M{\'e}canique} \bibinfo{volume}{339}
  (\bibinfo{year}{2011}) \bibinfo{pages}{20--26}.
\bibitem[{Lorentz(2017)}]{lorentz2017nonlocal}
\bibinfo{author}{E.~Lorentz},
\newblock \bibinfo{title}{A nonlocal damage model for plain concrete consistent
  with cohesive fracture},
\newblock \bibinfo{journal}{International Journal of Fracture}
  \bibinfo{volume}{207} (\bibinfo{year}{2017}) \bibinfo{pages}{123--159}.
\bibitem[{Wu(2017)}]{wu2017unified}
\bibinfo{author}{J.-Y. Wu},
\newblock \bibinfo{title}{A unified phase-field theory for the mechanics of
  damage and quasi-brittle failure},
\newblock \bibinfo{journal}{Journal of the Mechanics and Physics of Solids}
  \bibinfo{volume}{103} (\bibinfo{year}{2017}) \bibinfo{pages}{72--99}.
\bibitem[{Geelen et~al.(2019)Geelen, Liu, Hu, Tupek, and
  Dolbow}]{geelen2019phase}
\bibinfo{author}{R.~J. Geelen}, \bibinfo{author}{Y.~Liu},
  \bibinfo{author}{T.~Hu}, \bibinfo{author}{M.~R. Tupek},
  \bibinfo{author}{J.~E. Dolbow},
\newblock \bibinfo{title}{A phase-field formulation for dynamic cohesive
  fracture},
\newblock \bibinfo{journal}{Computer Methods in Applied Mechanics and
  Engineering} \bibinfo{volume}{348} (\bibinfo{year}{2019})
  \bibinfo{pages}{680--711}.
\bibitem[{Hu et~al.(2020)Hu, Guilleminot, and Dolbow}]{HuGary2020}
\bibinfo{author}{T.~Hu}, \bibinfo{author}{J.~Guilleminot},
  \bibinfo{author}{J.~E. Dolbow},
\newblock \bibinfo{title}{A phase-field model of fracture with frictionless
  contact and random fracture properties: Application to thin-film fracture and
  soil dessication},
\newblock \bibinfo{journal}{Computer Methods in Applied Mechanics and
  Engineering} \bibinfo{volume}{368} (\bibinfo{year}{2020})
  \bibinfo{pages}{113106}.
\bibitem[{Talamini et~al.(2020)Talamini, Tupek, Stershic, Hu, Foulk~III,
  Ostien, and Dolbow}]{brandon2020cohesive}
\bibinfo{author}{B.~Talamini}, \bibinfo{author}{M.~R. Tupek},
  \bibinfo{author}{A.~J. Stershic}, \bibinfo{author}{T.~Hu},
  \bibinfo{author}{J.~W. Foulk~III}, \bibinfo{author}{J.~T. Ostien},
  \bibinfo{author}{J.~E. Dolbow},
\newblock \bibinfo{title}{A cohesive phase-field model for ductile fracture},
\newblock \bibinfo{journal}{Journal of the Mechanics and Physics of Solids}
  (\bibinfo{year}{2020}).
\bibitem[{Alessi et~al.(2014)Alessi, Marigo, and Vidoli}]{alessi_gradient_2014}
\bibinfo{author}{R.~Alessi}, \bibinfo{author}{J.-J. Marigo},
  \bibinfo{author}{S.~Vidoli},
\newblock \bibinfo{title}{Gradient {Damage} {Models} {Coupled} with
  {Plasticity} and {Nucleation} of {Cohesive} {Cracks}},
\newblock \bibinfo{journal}{Archive for Rational Mechanics and Analysis}
  \bibinfo{volume}{214} (\bibinfo{year}{2014}) \bibinfo{pages}{575--615}.
\bibitem[{Alessi et~al.(2015)Alessi, Marigo, and Vidoli}]{alessi_gradient_2015}
\bibinfo{author}{R.~Alessi}, \bibinfo{author}{J.-J. Marigo},
  \bibinfo{author}{S.~Vidoli},
\newblock \bibinfo{title}{Gradient damage models coupled with plasticity:
  Variational formulation and main properties},
\newblock \bibinfo{journal}{Mechanics of Materials}  (\bibinfo{year}{2015})
  \bibinfo{pages}{351--367}.
\bibitem[{Alessi et~al.(2018)Alessi, Marigo, Maurini, and
  Vidoli}]{alessi_coupling_2018}
\bibinfo{author}{R.~Alessi}, \bibinfo{author}{J.-J. Marigo},
  \bibinfo{author}{C.~Maurini}, \bibinfo{author}{S.~Vidoli},
\newblock \bibinfo{title}{Coupling damage and plasticity for a phase-field
  regularisation of brittle, cohesive and ductile fracture: {One}-dimensional
  examples},
\newblock \bibinfo{journal}{International Journal of Mechanical Sciences}
  \bibinfo{volume}{149} (\bibinfo{year}{2018}) \bibinfo{pages}{559--576}.
\bibitem[{Ambati et~al.(2015)Ambati, Gerasimov, and
  Lorenzis}]{ambati_phase-field_2015}
\bibinfo{author}{M.~Ambati}, \bibinfo{author}{T.~Gerasimov},
  \bibinfo{author}{L.~D. Lorenzis},
\newblock \bibinfo{title}{Phase-field modeling of ductile fracture},
\newblock \bibinfo{journal}{Computational Mechanics} \bibinfo{volume}{55}
  (\bibinfo{year}{2015}) \bibinfo{pages}{1017--1040}.
\bibitem[{Ambati et~al.(2016)Ambati, Kruse, and
  De~Lorenzis}]{ambati_phase-field_2016}
\bibinfo{author}{M.~Ambati}, \bibinfo{author}{R.~Kruse},
  \bibinfo{author}{L.~De~Lorenzis},
\newblock \bibinfo{title}{A phase-field model for ductile fracture at finite
  strains and its experimental verification},
\newblock \bibinfo{journal}{Computational Mechanics} \bibinfo{volume}{57}
  (\bibinfo{year}{2016}) \bibinfo{pages}{149--167}.
\bibitem[{Miehe et~al.(2016)Miehe, Aldakheel, and Raina}]{miehe_phase_2016}
\bibinfo{author}{C.~Miehe}, \bibinfo{author}{F.~Aldakheel},
  \bibinfo{author}{A.~Raina},
\newblock \bibinfo{title}{Phase field modeling of ductile fracture at finite
  strains: {A} variational gradient-extended plasticity-damage theory},
\newblock \bibinfo{journal}{International Journal of Plasticity}
  \bibinfo{volume}{84} (\bibinfo{year}{2016}) \bibinfo{pages}{1--32}.
\bibitem[{Borden et~al.(2016)Borden, Hughes, Landis, Anvari, and
  Lee}]{borden_phase-field_2016}
\bibinfo{author}{M.~J. Borden}, \bibinfo{author}{T.~J. Hughes},
  \bibinfo{author}{C.~M. Landis}, \bibinfo{author}{A.~Anvari},
  \bibinfo{author}{I.~J. Lee},
\newblock \bibinfo{title}{A phase-field formulation for fracture in ductile
  materials: Finite deformation balance law derivation, plastic degradation,
  and stress triaxiality effects},
\newblock \bibinfo{journal}{Computer Methods in Applied Mechanics and
  Engineering} \bibinfo{volume}{312} (\bibinfo{year}{2016})
  \bibinfo{pages}{130--166}.
\bibitem[{Borden et~al.(2017)Borden, Hughes, Landis, Anvari, and
  Lee}]{borden_phase-field_2017}
\bibinfo{author}{M.~J. Borden}, \bibinfo{author}{T.~J.~R. Hughes},
  \bibinfo{author}{C.~M. Landis}, \bibinfo{author}{A.~Anvari},
  \bibinfo{author}{I.~J. Lee},
\newblock \bibinfo{title}{Phase-{Field} {Formulation} for {Ductile}
  {Fracture}},
\newblock in: \bibinfo{booktitle}{Advances in {Computational} {Plasticity}: {A}
  {Book} in {Honour} of {D}. {Roger} {J}. {Owen}}, volume~\bibinfo{volume}{46},
  \bibinfo{publisher}{Springer International Publishing}, \bibinfo{year}{2017},
  p. \bibinfo{pages}{443}.
\bibitem[{Alessi et~al.(2017)Alessi, Ambati, Gerasimov, Vidoli, and
  De~Lorenzis}]{alessi_comparison_2017}
\bibinfo{author}{R.~Alessi}, \bibinfo{author}{M.~Ambati},
  \bibinfo{author}{T.~Gerasimov}, \bibinfo{author}{S.~Vidoli},
  \bibinfo{author}{L.~De~Lorenzis},
\newblock \bibinfo{title}{Comparison of {Phase}-{Field} {Models} of {Fracture}
  {Coupled} with {Plasticity}},
\newblock in: \bibinfo{booktitle}{Advances in {Computational} {Plasticity}: {A}
  {Book} in {Honour} of {D}. {Roger} {J}. {Owen}}, volume~\bibinfo{volume}{46},
  \bibinfo{publisher}{Springer International Publishing}, \bibinfo{year}{2017},
  p. \bibinfo{pages}{443}.
\bibitem[{Simo(1988)}]{simo1988framework}
\bibinfo{author}{J.~C. Simo},
\newblock \bibinfo{title}{A framework for finite strain elastoplasticity based
  on maximum plastic dissipation and the multiplicative decomposition: Part i.
  continuum formulation},
\newblock \bibinfo{journal}{Computer methods in applied mechanics and
  engineering} \bibinfo{volume}{66} (\bibinfo{year}{1988})
  \bibinfo{pages}{199--219}.
\bibitem[{Ortiz and Stainier(1999)}]{ortiz_variational_1999}
\bibinfo{author}{M.~Ortiz}, \bibinfo{author}{L.~Stainier},
\newblock \bibinfo{title}{The variational formulation of viscoplastic
  constitutive updates},
\newblock \bibinfo{journal}{Computer Methods in Applied Mechanics and
  Engineering} \bibinfo{volume}{171} (\bibinfo{year}{1999})
  \bibinfo{pages}{419--444}.
\bibitem[{Pham and Marigo(2013)}]{pham2013onset}
\bibinfo{author}{K.~Pham}, \bibinfo{author}{J.-J. Marigo},
\newblock \bibinfo{title}{From the onset of damage to rupture: construction of
  responses with damage localization for a general class of gradient damage
  models},
\newblock \bibinfo{journal}{Continuum Mechanics and Thermodynamics}
  \bibinfo{volume}{25} (\bibinfo{year}{2013}) \bibinfo{pages}{147--171}.
\bibitem[{Ambrosio and Tortorelli(1990)}]{ambrosio1990approximation}
\bibinfo{author}{L.~Ambrosio}, \bibinfo{author}{V.~M. Tortorelli},
\newblock \bibinfo{title}{Approximation of functional depending on jumps by
  elliptic functional via t-convergence},
\newblock \bibinfo{journal}{Communications on Pure and Applied Mathematics}
  \bibinfo{volume}{43} (\bibinfo{year}{1990}) \bibinfo{pages}{999--1036}.
\bibitem[{Miehe et~al.(2015)Miehe, Hofacker, Sch{\"a}nzel, and
  Aldakheel}]{miehe2015phase}
\bibinfo{author}{C.~Miehe}, \bibinfo{author}{M.~Hofacker},
  \bibinfo{author}{L.-M. Sch{\"a}nzel}, \bibinfo{author}{F.~Aldakheel},
\newblock \bibinfo{title}{Phase field modeling of fracture in multi-physics
  problems. part ii. coupled brittle-to-ductile failure criteria and crack
  propagation in thermo-elastic--plastic solids},
\newblock \bibinfo{journal}{Computer Methods in Applied Mechanics and
  Engineering} \bibinfo{volume}{294} (\bibinfo{year}{2015})
  \bibinfo{pages}{486--522}.
\bibitem[{Rodriguez et~al.(2016)Rodriguez, Matinmanesh, Phull, Schemitsch,
  Zalzal, Clarkin, Papini, and Towler}]{rodriguez2016silica}
\bibinfo{author}{O.~Rodriguez}, \bibinfo{author}{A.~Matinmanesh},
  \bibinfo{author}{S.~Phull}, \bibinfo{author}{E.~H. Schemitsch},
  \bibinfo{author}{P.~Zalzal}, \bibinfo{author}{O.~M. Clarkin},
  \bibinfo{author}{M.~Papini}, \bibinfo{author}{M.~R. Towler},
\newblock \bibinfo{title}{Silica-based and borate-based, titania-containing
  bioactive coatings characterization: critical strain energy release rate,
  residual stresses, hardness, and thermal expansion},
\newblock \bibinfo{journal}{Journal of functional biomaterials}
  \bibinfo{volume}{7} (\bibinfo{year}{2016}) \bibinfo{pages}{32}.
\bibitem[{Chowdhury et~al.(2019)Chowdhury, Wise, Ganesh, and
  Gillespie~Jr}]{chowdhury2019effects}
\bibinfo{author}{S.~C. Chowdhury}, \bibinfo{author}{E.~A. Wise},
  \bibinfo{author}{R.~Ganesh}, \bibinfo{author}{J.~W. Gillespie~Jr},
\newblock \bibinfo{title}{Effects of surface crack on the mechanical properties
  of silica: A molecular dynamics simulation study},
\newblock \bibinfo{journal}{Engineering Fracture Mechanics}
  \bibinfo{volume}{207} (\bibinfo{year}{2019}) \bibinfo{pages}{99--108}.
\bibitem[{Vo et~al.(2020)Vo, He, Blum, Damone, and Newell}]{vo2020molecular}
\bibinfo{author}{T.~Vo}, \bibinfo{author}{B.~He}, \bibinfo{author}{M.~Blum},
  \bibinfo{author}{A.~Damone}, \bibinfo{author}{P.~Newell},
\newblock \bibinfo{title}{Molecular scale insight of pore morphology relation
  with mechanical properties of amorphous silica using reaxff},
\newblock \bibinfo{journal}{Computational Materials Science}
  \bibinfo{volume}{183} (\bibinfo{year}{2020}) \bibinfo{pages}{109881}.
\bibitem[{Yin and Kaliske(2020)}]{yin2020ductile}
\bibinfo{author}{B.~Yin}, \bibinfo{author}{M.~Kaliske},
\newblock \bibinfo{title}{A ductile phase-field model based on degrading the
  fracture toughness: Theory and implementation at small strain},
\newblock \bibinfo{journal}{Computer Methods in Applied Mechanics and
  Engineering} \bibinfo{volume}{366} (\bibinfo{year}{2020})
  \bibinfo{pages}{113068}.
\bibitem[{Borden et~al.(2016)Borden, Hughes, Landis, Anvari, and
  Lee}]{borden2016phase}
\bibinfo{author}{M.~J. Borden}, \bibinfo{author}{T.~J. Hughes},
  \bibinfo{author}{C.~M. Landis}, \bibinfo{author}{A.~Anvari},
  \bibinfo{author}{I.~J. Lee},
\newblock \bibinfo{title}{A phase-field formulation for fracture in ductile
  materials: Finite deformation balance law derivation, plastic degradation,
  and stress triaxiality effects},
\newblock \bibinfo{journal}{Computer Methods in Applied Mechanics and
  Engineering} \bibinfo{volume}{312} (\bibinfo{year}{2016})
  \bibinfo{pages}{130--166}.
\bibitem[{Ambati et~al.(2016)Ambati, Kruse, and De~Lorenzis}]{ambati2016phase}
\bibinfo{author}{M.~Ambati}, \bibinfo{author}{R.~Kruse},
  \bibinfo{author}{L.~De~Lorenzis},
\newblock \bibinfo{title}{A phase-field model for ductile fracture at finite
  strains and its experimental verification},
\newblock \bibinfo{journal}{Computational Mechanics} \bibinfo{volume}{57}
  (\bibinfo{year}{2016}) \bibinfo{pages}{149--167}.
\bibitem[{Heister et~al.(2015)Heister, Wheeler, and Wick}]{heister2015primal}
\bibinfo{author}{T.~Heister}, \bibinfo{author}{M.~F. Wheeler},
  \bibinfo{author}{T.~Wick},
\newblock \bibinfo{title}{A primal-dual active set method and
  predictor-corrector mesh adaptivity for computing fracture propagation using
  a phase-field approach},
\newblock \bibinfo{journal}{Computer Methods in Applied Mechanics and
  Engineering} \bibinfo{volume}{290} (\bibinfo{year}{2015})
  \bibinfo{pages}{466--495}.
\bibitem[{Balay et~al.(2019)Balay, Abhyankar, Adams, Brown, Brune, Buschelman,
  Dalcin, Dener, Eijkhout, Gropp, Karpeyev, Kaushik, Knepley, May, McInnes,
  Mills, Munson, Rupp, Sanan, Smith, Zampini, Zhang, and
  Zhang}]{petsc-web-page}
\bibinfo{author}{S.~Balay}, \bibinfo{author}{S.~Abhyankar},
  \bibinfo{author}{M.~F. Adams}, \bibinfo{author}{J.~Brown},
  \bibinfo{author}{P.~Brune}, \bibinfo{author}{K.~Buschelman},
  \bibinfo{author}{L.~Dalcin}, \bibinfo{author}{A.~Dener},
  \bibinfo{author}{V.~Eijkhout}, \bibinfo{author}{W.~D. Gropp},
  \bibinfo{author}{D.~Karpeyev}, \bibinfo{author}{D.~Kaushik},
  \bibinfo{author}{M.~G. Knepley}, \bibinfo{author}{D.~A. May},
  \bibinfo{author}{L.~C. McInnes}, \bibinfo{author}{R.~T. Mills},
  \bibinfo{author}{T.~Munson}, \bibinfo{author}{K.~Rupp},
  \bibinfo{author}{P.~Sanan}, \bibinfo{author}{B.~F. Smith},
  \bibinfo{author}{S.~Zampini}, \bibinfo{author}{H.~Zhang},
  \bibinfo{author}{H.~Zhang}, \bibinfo{title}{{PETS}c {W}eb page},
  \bibinfo{howpublished}{\url{https://www.mcs.anl.gov/petsc}},
  \bibinfo{year}{2019}. \URLprefix \url{https://www.mcs.anl.gov/petsc}.
\bibitem[{Neto et~al.(2005)Neto, Pires, and Owen}]{neto2005f}
\bibinfo{author}{E.~D.~S. Neto}, \bibinfo{author}{F.~A. Pires},
  \bibinfo{author}{D.~Owen},
\newblock \bibinfo{title}{F-bar-based linear triangles and tetrahedra for
  finite strain analysis of nearly incompressible solids. part i: formulation
  and benchmarking},
\newblock \bibinfo{journal}{International Journal for Numerical Methods in
  Engineering} \bibinfo{volume}{62} (\bibinfo{year}{2005})
  \bibinfo{pages}{353--383}.
\bibitem[{Hu(2020{\natexlab{a}})}]{raccoon}
\bibinfo{author}{T.~Hu}, \bibinfo{title}{{RACCOON}},
  \bibinfo{year}{2020}{\natexlab{a}}. \URLprefix
  \url{https://github.com/hugary1995/raccoon}.
\bibitem[{Hu(2020{\natexlab{b}})}]{raccoon_doc}
\bibinfo{author}{T.~Hu}, \bibinfo{title}{{RACCOON} documentation},
  \bibinfo{year}{2020}{\natexlab{b}}. \URLprefix
  \url{https://hugary1995.github.io/raccoon}.
\bibitem[{Permann et~al.(2020)Permann, Gaston, Andr{\v{s}}, Carlsen, Kong,
  Lindsay, Miller, Peterson, Slaughter, Stogner et~al.}]{permann2020moose}
\bibinfo{author}{C.~J. Permann}, \bibinfo{author}{D.~R. Gaston},
  \bibinfo{author}{D.~Andr{\v{s}}}, \bibinfo{author}{R.~W. Carlsen},
  \bibinfo{author}{F.~Kong}, \bibinfo{author}{A.~D. Lindsay},
  \bibinfo{author}{J.~M. Miller}, \bibinfo{author}{J.~W. Peterson},
  \bibinfo{author}{A.~E. Slaughter}, \bibinfo{author}{R.~H. Stogner}, et~al.,
\newblock \bibinfo{title}{Moose: Enabling massively parallel multiphysics
  simulation},
\newblock \bibinfo{journal}{SoftwareX} \bibinfo{volume}{11}
  (\bibinfo{year}{2020}) \bibinfo{pages}{100430}.
\bibitem[{Benson and Munson(2006)}]{benson2006flexible}
\bibinfo{author}{S.~J. Benson}, \bibinfo{author}{T.~S. Munson},
\newblock \bibinfo{title}{Flexible complementarity solvers for large-scale
  applications},
\newblock \bibinfo{journal}{Optimization Methods and Software}
  \bibinfo{volume}{21} (\bibinfo{year}{2006}) \bibinfo{pages}{155--168}.
\bibitem[{Kub{\'\i}k et~al.(2019)Kub{\'\i}k, {\v{S}}ebek, Zapletal,
  Petru{\v{s}}ka, and N{\'a}vrat}]{kubik2019ductile}
\bibinfo{author}{P.~Kub{\'\i}k}, \bibinfo{author}{F.~{\v{S}}ebek},
  \bibinfo{author}{J.~Zapletal}, \bibinfo{author}{J.~Petru{\v{s}}ka},
  \bibinfo{author}{T.~N{\'a}vrat},
\newblock \bibinfo{title}{Ductile failure predictions for the three-point
  bending test of a complex geometry made from aluminum alloy},
\newblock \bibinfo{journal}{Journal of Engineering Materials and Technology}
  \bibinfo{volume}{141} (\bibinfo{year}{2019}).
\bibitem[{Kub{\'\i}k et~al.(2018)Kub{\'\i}k, {\v{S}}ebek, and
  Petru{\v{s}}ka}]{kubik2018notched}
\bibinfo{author}{P.~Kub{\'\i}k}, \bibinfo{author}{F.~{\v{S}}ebek},
  \bibinfo{author}{J.~Petru{\v{s}}ka},
\newblock \bibinfo{title}{Notched specimen under compression for ductile
  failure criteria},
\newblock \bibinfo{journal}{Mechanics of Materials} \bibinfo{volume}{125}
  (\bibinfo{year}{2018}) \bibinfo{pages}{94--109}.
\bibitem[{Boyce et~al.(2014)Boyce, Kramer, Fang, Cordova, Neilsen, Dion,
  Kaczmarowski, Karasz, Xue, Gross et~al.}]{boyce2014sandia}
\bibinfo{author}{B.~L. Boyce}, \bibinfo{author}{S.~L. Kramer},
  \bibinfo{author}{H.~E. Fang}, \bibinfo{author}{T.~E. Cordova},
  \bibinfo{author}{M.~K. Neilsen}, \bibinfo{author}{K.~Dion},
  \bibinfo{author}{A.~K. Kaczmarowski}, \bibinfo{author}{E.~Karasz},
  \bibinfo{author}{L.~Xue}, \bibinfo{author}{A.~J. Gross}, et~al.,
\newblock \bibinfo{title}{The sandia fracture challenge: blind round robin
  predictions of ductile tearing},
\newblock \bibinfo{journal}{International Journal of Fracture}
  \bibinfo{volume}{186} (\bibinfo{year}{2014}) \bibinfo{pages}{5--68}.
\bibitem[{Guo(2013)}]{guo2013experimental}
\bibinfo{author}{J.~Guo}, \bibinfo{title}{An experimental and numerical
  investigation on damage evolution and ductile fracture mechanism of aluminum
  alloy}, Ph.D. thesis, PhD dissertation, The University of Tokushima,
  \bibinfo{year}{2013}.

\end{thebibliography}

\end{document}